\documentclass[fleqn,usenatbib]{mnras}
\usepackage{newtxtext,newtxmath}
\usepackage{longtable}
\usepackage{tabularx}
\usepackage{verbatim}
\usepackage[normalem]{ulem}

\hypersetup{linkcolor=red}          % red equations

\usepackage[dvipsnames]{xcolor}

\def\equationautorefname~#1\null{Equation~(#1)\null}

\defcitealias{Zhu_et_al.(2023a)}{Paper I}
\defcitealias{Zhu_et_al.(2023b)}{Paper III}

\usepackage{scalerel,tikz}
\usetikzlibrary{svg.path}
\definecolor{orcidlogocol}{HTML}{A6CE39}
\tikzset{orcidlogo/.pic={
 \fill[orcidlogocol] svg{M256,128c0,70.7-57.3,128-128,128C57.3,256,0,198.7,0,128C0,57.3,57.3,0,128,0C198.7,0,256,57.3,256,128z};
 \fill[white] svg{M86.3,186.2H70.9V79.1h15.4v48.4V186.2z}
 svg{M108.9,79.1h41.6c39.6,0,57,28.3,57,53.6c0,27.5-21.5,53.6-56.8,53.6h-41.8V79.1z M124.3,172.4h24.5c34.9,0,42.9-26.5,42.9-39.7c0-21.5-13.7-39.7-43.7-39.7h-23.7V172.4z}
 svg{M88.7,56.8c0,5.5-4.5,10.1-10.1,10.1c-5.6,0-10.1-4.6-10.1-10.1c0-5.6,4.5-10.1,10.1-10.1C84.2,46.7,88.7,51.3,88.7,56.8z};
}}
\newcommand\orcidicon[1]{\href{https://orcid.org/#1}{\mbox{\scalerel*{
\begin{tikzpicture}[yscale=-1,transform shape]
\pic{orcidlogo};
\end{tikzpicture}
}{|}}}}

\title[Stellar population vs. kinematics in MaNGA]{MaNGA DynPop -- II. Global stellar population, gradients, and star-formation histories from integral-field spectroscopy of 10K galaxies: link with galaxy rotation, shape, and total-density gradients}

\author[S. Lu et al.]
{Shengdong Lu\orcidicon{0000-0002-6726-9499}$^{1}$\thanks{E-mail: \url{lushengdong93@icloud.com}},
Kai Zhu\orcidicon{0000-0002-2583-2669}$^{2,3,4}$\thanks{E-mail: \url{kaizhu@nao.cas.cn}},
Michele Cappellari\orcidicon{0000-0002-1283-8420}$^{5}$,
Ran Li\orcidicon{0000-0003-3899-0612}$^{2,3,4}$,
Shude Mao\orcidicon{0000-0001-8317-2788}$^{1}$,
Dandan Xu$^{1}$
\\
\\
$^{1}$Department of Astronomy, Tsinghua University, Beijing 100084, China\\
$^{2}$National Astronomical Observatories, Chinese Academy of Sciences, 20A Datun Road, Chaoyang District, Beijing 100101, China\\
$^{3}$Institute for Frontiers in Astronomy and Astrophysics, Beijing Normal University, Beijing 102206, China\\
$^{4}$School of Astronomy and Space Science, University of Chinese Academy of Sciences, Beijing 100049, China\\
$^{5}$Sub-department of Astrophysics, Department of Physics, University of Oxford, Denys Wilkinson Building, Keble Road, Oxford, OX1 3RH, UK
}
\date{Accepted 2023 September 5}
\pubyear{2023}

\begin{document}
\label{firstpage}
\pagerange{\pageref{firstpage}--\pageref{lastpage}}
\maketitle
\begin{abstract}
This is the second paper of the MaNGA DynPop series, which analyzes the global stellar population, radial gradients, and non-parametric star-formation history of $\sim 10$K galaxies from the MaNGA Survey final data release 17 (DR17) and relates them with dynamical properties of galaxies. We confirm the correlation between the stellar population properties and the stellar velocity dispersion $\sigma_{\rm e}$, but also find that younger galaxies are more metal-poor at fixed $\sigma_{\rm e}$. Stellar age, metallicity, and mass-to-light ratio $M_{\ast}/L$ all decrease with increasing galaxy rotation, while their radial gradients become more negative (i.e. lower value at the outskirts). The exception is the slow rotators, which also appear to have significantly negative metallicity gradients, confirming the mass-metallicity gradient correlation. Massive disk galaxies in the green valley, on the $(\sigma_{\rm e},\rm Age)$ plane, show the most negative age and metallicity gradients, consistent with their old central bulges surrounded by young star-forming disks and metal-poor gas accretion. Galaxies with high $\sigma_{\rm e}$, steep total mass-density slope, low dark matter fraction, high $M_{\ast}/L$, and high metallicity have the highest star-formation rate at earlier times, and are currently quenched. We also discover a population of low-mass star-forming galaxies with low rotation but physically distinct from the massive slow rotators. A catalogue of these stellar population properties is provided publicly.
\end{abstract}

\begin{keywords}
catalogues – galaxies: abundances – galaxies: evolution – galaxies: formation – galaxies: stellar content
\end{keywords}

\section{Introduction}
Galaxies trace the large-scale structure in the Universe. Thus, their evolution with cosmic time is a key question for the study of the Universe. Evolution of galaxies is a complicated process, involving multiple components, including stars, gas, dark matter, active galactic nuclei (AGNs), as well as the surrounding environments of the galaxies. Therefore, although many efforts have been made to understand the evolution of galaxies, it still remains one of the great challenges of modern astrophysics (see \citealt{Somerville2015,Naab2017} for reviews).

Stellar population of galaxies encodes the information of galaxy mass assembly and chemical enrichment histories, which is thus essential for understanding the evolution of galaxies. Early studies typically made use of galaxy colour to classify different stellar populations and chemical compositions in galaxies (see \citealt{Baade_et_al.(1963),Tinsley_et_al.(1980)} for reviews), which, however, suffers from the degeneracies between age and metallicity (e.g. \citealt{Worthey(1994)}), as well as between the star-formation history (SFH) and dust attenuation effect (e.g. \citealt{Silva_et_al.(1998),Devriendt_et_al.(1999),Pozzetti_et_al.(2000)}). The latter is even stronger for late-type galaxies (LTGs), which have more dust than early-type galaxies (ETGs).

To break the degeneracies, people try to fit some spectral absorption features that have different sensitivities on age, metallicity, and dust attenuation effect, e.g. the Lick system \citep{Worthey(1994)}, using a linear combination of single stellar population (SSP) templates. SSP models combine stellar evolution theory in the form of an initial mass function (IMF; e.g. \citealt{Salpeter1955,Chabrier(2003),Kroupa(2001)}), stellar tracks and/or isochrones (e.g. \citealt{Bruzual_et_al.(1993),Vazdekis_et_al.(1996),Conroy_et_al.(2009)}), and stellar libraries that are either empirical or theoretical or a combination (e.g. \citealt{Maraston_et_al.(1998),Walcher_et_al.(2009),Vazdekis_et_al.(2010),Vazdekis_et_al.(2016),Conroy_et_al.(2018)}) to predict the spectrum of a stellar population with given age and metallicity. Another way to study the stellar population of galaxies is the full-spectrum fitting, instead of fitting specific absorption features. Existing fitting methods include {\sc ppxf} \citep{Cappellari_et_al.(2004),Cappellari2017,Cappellari(2023)}, {\sc starlight} \citep{Cid_Fernandes_et_al.(2005)}, {\sc stecmap} \citep{Ocvirk_et_al.(2006a),Ocvirk_et_al.(2006b)}, {\sc vespa} \citep{Tojeiro_et_al.(2007)}, {\sc fit3D} \citep{Sanchez_et_al.(2016a),Sanchez_et_al.(2016b),Lacerda_et_al.(2022)}, and {\sc firefly} \citep{Wilkinson_et_al.(2017)}.

With the methods above, many efforts have been made to study the stellar population and star-formation history of galaxies. \citet{Kauffmann_et_al.(2003)} used the $4000\,\Angstrom$ break strength and the Balmer absorption-line index H$\delta_{\rm A}$ to study the star formation histories, dust attenuation, and stellar masses of $\sim 120000$ galaxies from the Sloan Digital Sky Survey \citep{Stoughton_et_al.(2002)}. \citet{Gallazzi_et_al.(2005)} studied the connection between metallicity, age and stellar mass in nearby galaxies with the similar method but applies a Bayesian approach. \citet{Thomas2010} obtained luminosity-weighted ages, metallicities and $\alpha$/Fe ratios from the Lick absorption line indices for 3360 early-type galaxies and found that the scaling relations between galaxy stellar population properties and dynamical properties (i.e. galaxy mass and velocity dispersion) are not sensitive to environment densities. More recently, \citet{ScholzDaz_et_al.(2023)} made use of the absorption feature method to study the dependence of star-formation histories of galaxies on halo mass. With full-spectrum fitting, \citet{Cappellari_et_al.(2013b)} and \citet{McDermid_et_al.(2015)} studied the correlation between stellar population properties and dynamical properties of 260 early-type galaxies in ATLAS$^{3\rm D}$ \citep{Cappellari_et_al.(2011)} and found that stellar population properties show systematic variation on the mass-size plane, roughly along the direction of velocity dispersion of galaxies. \citet{Scott_et_al.(2017)} extended this study to $\sim 1300$ galaxies from the SAMI survey \citep{Bryant_et_al.(2015)} and \citet{Li_et_al.(2018)} to $\sim 2000$ galaxies with different morphologies (including both early-type and late-type galaxies) in MaNGA \citep{Bundy_et_al.(2015)} and also found similar trends.

Besides, with the development of large-scale integral field unit (IFU) surveys, such as SAURON \citep{Bacon_et_al.(2001),de_Zeeuw_et_al.(2002)}, ATLAS$^{\rm 3D}$ \citep{Cappellari_et_al.(2011)}, CALIFA \citep{Sanchez_et_al.(2012)}, SAMI \citep{Bryant_et_al.(2015)}, and MaNGA \citep{Bundy_et_al.(2015)}, one can not only study the global stellar population properties, but also analyze the spatially resolved stellar population properties, with which the population profiles and gradients can be studied. For example, \citet{Kuntschner2010} analyzed the absorption line strength maps of 48 early-type galaxies from the SAURON survey to study their stellar population gradients and found that the age gradient depends on dynamical mass, with the scatter being smaller at the low mass end. \citet{GarciaBenito2017} studied the gradient of mass-weighted age of 661 galaxies with various morphologies from the CALIFA survey and found that negative mass-weighted age gradient is seen for all the samples investigated and the gradient increases (becomes flatter) with increasing stellar mass surface density. \citet{Goddard2017} applied the stellar population analysis to a representative sample of 721 galaxies from the MaNGA survey and concluded that galaxy mass is the main driver of the variation of stellar population gradients among different galaxies; the galaxy environment seems to have no obvious influence on the stellar population gradients, consistent with the findings of \citet{Zheng_et_al.(2017)} with 1105 galaxies from the MaNGA survey and \citet{Santucci_et_al.(2020)} with 96 passive central galaxies and their satellite counterparts from the SAMI survey. The mass dependence of stellar population radial profiles (gradients) is also seen in \citet{Belfiore2018}, in which low-mass star-forming galaxies appear to have nearly flat specific star-formation rate (sSFR) profiles, while the more massive star-forming galaxies show a significant decrease in sSFR at the galaxy center. With a larger sample of $\sim 2000$ galaxies with various morphologies from MaNGA, \citet{Li_et_al.(2018)} studied the distributions of stellar population gradients on the mass-size plane and found that the variation of stellar population gradients have complicated dependence on dynamical parameters of galaxies and their variation cannot be accurately captured by mass alone. Interested readers are referred to \citet{Sanchez(2020)} for a review on the spatially resolved spectroscopic properties of galaxies (see sec. 4.7 for stellar population gradients).

With the end of the MaNGA survey, the spatially resolved spectra of over 10000 unique nearby galaxies have been successfully obtained \citep{MaNGA_dr17}. \citet{firefly} (known as the {\sc firefly} catalogue) and \citet{pipe3d} (known as the {\sc pipe3d} catalogue) have provided the catalogues of both the global and spatially resolved stellar population properties for the full sample of MaNGA, which have been widely used in many studies on the analyses of stellar content of galaxies. For example, \citet{Barrera-Ballesteros_et_al.(2022)} studied the stellar population radial profiles of $\sim 10000$ MaNGA galaxies with the {\sc pypipe3d} analysis pipeline \citep{Lacerda_et_al.(2022)}. \citet{Boardman_et_al.(2022),Boardman_et_al.(2023)} studied the driving factors of gas metallicity gradients, combining the {\sc pipe3d} catalogue and the gas metallicity maps derived with the MaNGA DAP (Data Analysis Pipeline; \citealt{Belfiore_et_al.(2019),Westfall_et_al.(2019)}) emission line fluxes with the \citet{Marino_et_al.(2013)} calibrator. Apart from using the existing catalogues, \citet{Bernardi_et_al.(2023a),Bernardi_et_al.(2023b)} studied the influence of stellar initial mass function (IMF) on the estimates of stellar population properties (e.g. age, metallicity, and stellar mass-to-light ratio) and further the influence on estimating the half-stellar-mass radius using the full MaNGA sample.

With a series of papers of the MaNGA DynPop (Dynamics and stellar Population) project, we aim to not only provide catalogues of dynamical and stellar population properties for the final sample of the MaNGA survey ($\sim 10000$ nearby galaxies; \citealt{MaNGA_dr17}) as a legacy product, but also to study the formation and evolution of galaxies with the combination of the dynamical and stellar population properties. In Paper I \citep{Zhu_et_al.(2023a)}, we performed the Jeans Anisotropic Modelling (JAM; \citealt{Cappellari2008,Cappellari2020}) on the full sample of MaNGA survey and obtained their quality-assessed dynamical properties. This work is the second paper of the MaNGA DynPop series. The goal of this work is to provide a catalogue of stellar population properties, gradients, and star-formation history properties for all the $\sim 10000$ MaNGA galaxies. With these properties, we will also make analyses of the scaling relations of stellar populations and the correlation of population gradients and star-formation history versus structural and dynamical properties of galaxies. The readers are also referred to Paper III \citep{Zhu_et_al.(2023b)} for a study of multiple dynamical scaling relations of these galaxies, and Paper IV \citep{Wang_et_al.(2023)} for a study of the density profiles from galaxies to clusters, combining the stellar dynamical modelling and weak lensing.

The paper is organized as follows. In \autoref{sec:methodology}, we describe the MaNGA data used for this work (\autoref{sec:manga}), the structural and dynamical properties of MaNGA galaxies (\autoref{sec:jam_props}), and stellar population synthesis method (\autoref{sec:sp}). In \autoref{sec:global_sp}, we present the one-dimensional correlation between global stellar population properties and velocity dispersion of galaxies (\autoref{sec:1d_sp_relation}) and the variation of global stellar population properties in two-dimensional planes (\autoref{sec:2d_sp_relation}). \autoref{sec:sp_gradients} presents the correlation between stellar population profiles/gradients and other galaxy properties. The analysis of star-formation histories of MaNGA galaxies is given in \autoref{sec:sfh}. Finally, we summarize our findings in \autoref{sec:conclusion}. Throughout the paper, we assume a flat Universe with $\Omega_{\rm m} = 0.307$ and $H_0 = 67.7\,\mathrm{km\cdot s^{-1}\cdot Mpc^{-1}}$ \citep{Planck2016}.

\section{Data and Methodology}
\label{sec:methodology}
\subsection{MaNGA data product}
\label{sec:manga}
MaNGA (Mapping Nearby Galaxies at Apache Point Observatory; \citealt{Bundy_et_al.(2015)}) is currently the largest integral field units (IFU) spectroscope survey in the world, targeting $\sim 10000$ nearby galaxies in the redshift range $0.01<z<0.15$ \citep{Yan_et_al.(2016b),Wake_et_al.(2017)}. The spectra cover a simultaneous wavelength range from $3600\,\Angstrom$ to $10300\,\Angstrom$, with a spectral resolution $R\sim 2000$ (\citealt{Drory_et_al.(2015)}). They are produced by the Data Reduction Pipeline (DRP; \citealt{Law_et_al.(2016)}), which calibrates and reduces the raw data from MaNGA observation. The observation covers a spatial range from $1.5R_{\rm e}$ to $2.5R_{\rm e}$ (where $R_{\rm e}$ is the effective radius of galaxies) for each observed galaxy, conducting the ``Primary+'' and ``Secondary'' samples ($\sim 67\%$ and $\sim 33\%$, respectively; \citealt{Bundy_et_al.(2015)}). Readers are referred to the following papers for more details on the MaNGA instrumentation \citep{Drory_et_al.(2015)}, observing strategy \citep{Law_et_al.(2015)}, spectrophotometric calibration \citep{Smee_et_al.(2013),Yan_et_al.(2016a)}, and survey execution and initial data quality \citep{Yan_et_al.(2016b)}. The MaNGA project is now finished and the complete data have been released (\citealt{MaNGA_dr17}).

\subsection{Structural and dynamical properties of galaxies}
\label{sec:jam_props}
In this work, we mainly use the structural and dynamical properties of MaNGA galaxies from the first paper of this series (Paper I; \citealt{Zhu_et_al.(2023a)}). In \citetalias{Zhu_et_al.(2023a)}, we perform the Jeans Anisotropic Modelling (JAM; \citealt{Cappellari2008,Cappellari2020}) to the complete sample of MaNGA survey ($\sim 10000$ nearby galaxies) and obtain their quality-assessed dynamical and structural properties, using the {\sc python} version of JAM, {\sc JamPy}\footnote{Version 6.3.3, available from \url{https://pypi.org/project/jampy/}}. Below, we briefly introduce the parameters used in this work:
\begin{enumerate}
    \item ellipticity ($\epsilon$): $\epsilon$ is the ellipticity of the elliptical half-light isophotes of the galaxies, obtained
    with {\sc find\_galaxy} routine in the the {\sc python} software, {\sc MgeFit}\footnote{Version 5.0.14, available from \url{https://pypi.org/project/mgefit/}} by \citet{Cappellari(2002)}.
    
    \item galaxy size ($R_{\rm e}^{\rm maj}$ and $R_{\rm e}$): $R_{\rm e}^{\rm maj}$ is the semi-major axis of the elliptical half-light isophotes of the galaxies. $R_{\rm e}$ satisfies $\pi R_{\rm e}^2  = A$, where $A$ is the area of the elliptical half-light isophotes of galaxies. The galaxy sizes here are derived from the Multi-Gaussian Expansion (MGE) models \citep{Emsellem_et_al.(1994),Cappellari(2002)} with the {\sc MgeFit} software and are scaled by a factor of 1.35 to approximately match the popular galaxy sizes obtained from extrapolated photometry (see fig.~7 of \citealt{Cappellari_et_al.(2013b)}).
    
    \item galaxy velocity dispersion ($\sigma_{\rm e}$): $\sigma_{\rm e}$ used here is calculated as the square-root of the luminosity-weighted average second moments of the velocity within the elliptical half-light isophote, defined as:
    \begin{equation}
    \label{eq:sigma_e}
    \sigma_{\rm e} = \sqrt{\frac{\sum_{k}F_{k}(V_{k}^2+\sigma_{k}^2)}{\sum_{k}F_{k}}},
    \end{equation}
    where $V_{k}$ and $\sigma_{k}$ are the mean line-of-sight velocity and dispersion in the $k$-th IFU spaxel, respectively, and $F_{k}$ is the flux in the $k$-th spaxel. The summation goes over all the spaxels within the elliptical half-light isophote. Velocities and dispersion here are obtained by the MaNGA Data Analysis Pipeline (DAP; \citealt{Belfiore_et_al.(2019),Westfall_et_al.(2019)}).

     \item spin parameter of galaxies ($\lambda_{R_{\rm e}}$): $\lambda_{R_{\rm e}}$ is calculated as \citep{Emsellem_et_al.(2007)}:
    \begin{equation}
        \lambda_{R_{\rm e}} = \frac{\sum_{k} F_k R_k\left|V_k\right|}{\sum_{k} F_k R_k \sqrt{V_k^2+\sigma_k^2}},
    \end{equation}
    where $R_k$ is the distance to the galaxy center of $k-$th spaxel. The other parameters are the same as \autoref{eq:sigma_e}. $\lambda_{R_{\rm e}}$ is also calculated within the elliptical half-light isophotes of galaxies.

    \item galaxy $r-$band luminosity ($L_r$): $L_r$ is the total SDSS $r-$band \citep{Stoughton_et_al.(2002)} luminosity, derived by analytically integrating the MGE parametrization of the galaxy SDSS $r-$band surface brightness distribution.

    \item galaxy mass ($M_{\rm JAM}$): $M_{\rm JAM}$ provides an approximation to the total stellar mass of a galaxy derived from stellar dynamics. More specifically, $M_{\rm JAM}\equiv L_r\times(M/L_r)_{\rm e}$, where $(M/L_r)_{\rm e}$ is the total (including stars and dark matter) mass-to-light ratio within a {\em sphere} of radius $r=R_{\rm e}$ from JAM. Given the generally small amounts of dark matter within $R_{\rm e}$, which is typically on the order of 10\% (see \citetalias{Zhu_et_al.(2023b)} of this series), the total  $(M/L_r)_{\rm e}$ is close to the stellar mass-to-light ratio $(M_\ast/L_r)_{\rm e}$ for most galaxies. For this reason, $M_{\rm JAM}\approx L_r\times(M_{\ast}/L_r)_{\rm e}=M_{\ast}$. When dark matter becomes important, $M_{\rm JAM}$ will overestimate $M_{\ast}$, but for most galaxies one expects the dynamical measurement to be more robust than the mass inferred from stellar population, due to the influence of possible IMF variations among galaxies \citep{Dokkum2010,Cappellari_et_al.(2012),Li_et_al.(2017)}, as well as due to the effect of outshining, which makes it difficult to measure accurate masses of very young stellar populations \citep[e.g][]{Maraston2010,Sorba_et_al.(2015)}. The readers are referred to \citet{Smith2020} for a review on IMF variations. For the reasons above, we adopt $M_{\rm JAM}$ as the estimator of the galaxies stellar masses.
    
    \item galaxy total mass density slope ($\gamma_{\rm tot}$): $\gamma_{\rm tot}$ is the mass-weighted total density slope within a sphere of effective radius (e.g. \citealt{Dutton_and_Treu(2014),Li_et_al.(2019)}). The total density slope of galaxies are obtained by JAM.

    \item dark matter fraction ($f_{\rm DM}$): $f_{\rm DM}$ used here is calculated as the dark matter fraction within a sphere of radius $R_{\rm e}$.
\end{enumerate}

We note here that \citetalias{Zhu_et_al.(2023a)} provides structural and dynamical properties of MaNGA galaxies from multiple JAM models with different assumptions on dark halos and velocity ellipsoids. In this work, we take the total mass-to-light ratio, $(M/L_r)_{\rm e}$, from the JAM with a mass-follows-light model (i.e. the total galaxy mass is assumed to be proportional to the $r-$band luminosity of the galaxy) and a cylindrical-aligned velocity ellipsoid \citep{Cappellari2008}, following the practice of \citet{Cappellari_et_al.(2013b)}. The total mass-to-light ratio in this model is set to be constant across the whole single galaxy and thus does not vary with the aperture size. The other JAM-related parameters (i.e. the total mass density slope and the dark matter fraction) are from the results of the JAM modelling with a constant stellar mass-to-light ratio, a spherical NFW \citep{Navarro_et_al.(1997)} dark halo, and a cylindrical-aligned velocity ellipsoid. We note here that the total mass-to-light ratio (or equivalently, $M_{\rm JAM}$) and the total mass density slope of galaxies would not change too much under various model assumptions, while the dark matter fraction changes dramatically with model assumptions. Our readers are referred to \citetalias{Zhu_et_al.(2023a)} for more details on the discrepancies among different models. In addition, \citetalias{Zhu_et_al.(2023a)} provides quality flags ($\rm Qual$) of all the 10000 galaxies for the users with different purposes. Galaxies with $\rm Qual=-1$ are those with unreliable/irregular kinematic
maps, for which no kinematic properties can be trusted. Thus, in this work, we only take the galaxies with $\rm Qual>-1$ (9147 unique galaxies out of 10296 observations) to make analyses, in order to balance the data reliability and the sample size. The readers are referred to table~2 of \citetalias{Zhu_et_al.(2023a)} for more details of data quality flags. We note, however, the above selection criterion is only applied when making statistical analyses, while the catalogue of stellar population properties and star-formation histories will contain all the 10296 observations, among which 10010 are unique galaxies.

\subsection{Stellar population synthesis}
\label{sec:sp}
\subsubsection{Stellar libraries}
\label{sec:libraries}
In this study, we make use of the Penalized Pixel-Fitting method ({\sc ppxf}\footnote{Version 8.2.3, available from \url{https://pypi.org/project/ppxf/}. Also see \citet{Ge_et_al.(2018)} for the comparison between the {\sc ppxf} software and the {\sc starlight} software \citep{Cid_Fernandes_et_al.(2005)}.}) by \citet{Cappellari_et_al.(2004)} as upgraded in \citet{Cappellari2017,Cappellari(2023)} to carry out the spectrum fitting. For comparison, we adopt three different single stellar population (SSP) models in this work:
\begin{enumerate}
    \item the {\sc fsps}\footnote{Available from \url{https://github.com/cconroy20/fsps}.} model \citep{Conroy_et_al.(2009),Conroy_et_al.(2010)}, which allows one to compute SSP models for a specified set of parameters (e.g. age, metallicty, initial mass function etc.). In this work, we adopt 43 ages linearly spaced by 0.1 dex in $\lg\rm Age$ (years) from 6 to 10.2 (i.e. from 1 Myr to 15.85 Gyr) and 9 equally-spaced metallicities $[Z/H] = $ [-1.75, -1.5, -1.25, -1, -0.75, -0.5, -0.25, 0, 0.25].
    \item the {\sc galaxev}\footnote{Available from \url{http://www.bruzual.org/bc03/}.} model, which provides a Fortran code (version 2020) to generate SSP models under a set of parameters as {\sc fsps}. Here, we adopt the same age grid as {\sc fsps}, while use a different metallicity grid with $[Z/H]=$ [-2.34, -1.74, -0.73, -0.42, 0, 0.47].
    \item the {\sc miles}-based \citep{Sanchez-Blazquez_et_al.(2006)} SPS models of \citet{Vazdekis_et_al.(2010)}\footnote{Available from \url{http://miles.iac.es}.} (hereafter, the Vazdekis model), for which we adopt 25 linearly spaced $\lg\rm Age$ from 7.8 to 10.2, with the same step of 0.1 dex (i.e. from 63.1 Myr to 15.85 Gyr) and 7 metallicities $[Z/H] = $[-2.32, -1.71, -1.31, -0.71, -0.40, 0, 0.22].
\end{enumerate}

\begin{figure}
\centering
\includegraphics[width=1\columnwidth]{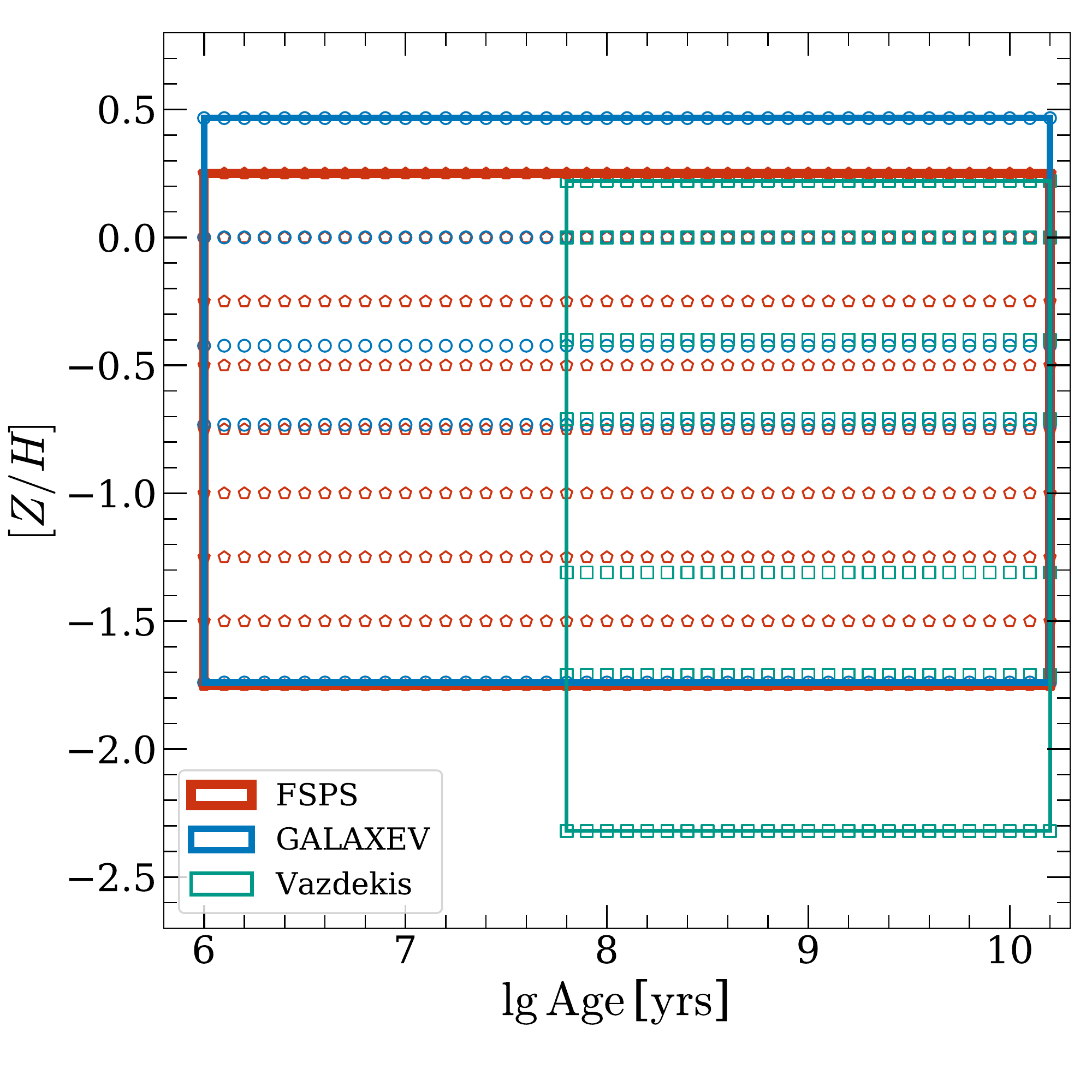}
\caption{Sketch map of age and metallicity grids of the {\sc fsps} (red), {\sc galaxev} (blue), and Vazdekis (green) models.}
\label{fig:libs}
\end{figure}

For all models we adopt the same \citet{Salpeter1955} IMF. We use the default Padova \citep{Girardi_et_al.(2000),Marigo_et_al.(2008)} evolutionary isochrone for the Vazdekis and {\sc galaxev} models and the MIST isochrones \citep{Choi2016} for the {\sc fsps} model. To better illustrate the age and metallicity grids of each library, we present them in \autoref{fig:libs}. In general, the {\sc galaxev} and {\sc fsps} models have the same age and similar metallicity range, while the Vazdekis model lacks low-age templates and has a lower metallicity boundary. In this work, we take the {\sc fsps} model as our default setting and only make comparisons between the models when necessary (see \autoref{sec:1d_sp_relation} for explanations).

\subsubsection{Fitting procedure}
\label{sec:dust}
Photons emitted by a distant galaxy will experience two dust extinction processes: one from the dust of the Milky Way (MW), and the other from the distant galaxy itself. The second effect is theoretically stronger for edge-on late-type galaxies. In this work, we consider these two processes separately. Firstly, we correct the MW dust attenuation effect using a \citet{Calzetti_et_al.(2000)} reddening curve, with the $\rm E(B-V)$ values of MW for the corresponding sky regions of our samples from the NASA Sloan Atlas (NSA)
catalogue\footnote{\url{http://nsatlas.org/data}} \citep{Blanton_et_al.(2011)}. After correcting the dust attenuation effect of MW, the corrected spectra are taken as the input for {\sc ppxf} fitting. The dust attenuation effect from the target galaxy is then evaluated by setting the \texttt{dust} keyword in the software, assuming a two-parameter attenuation curve, i.e. $(A_V,\delta)$, where $A_V$ is the attenuation of the spectrum at $\lambda = 5500\,\Angstrom$ ($V-$band) and $\delta$ is the UV slope of the spectrum (see \citealt[][sec.~3.7]{Cappellari(2023)} for details). 

\begin{figure*}
\centering
\includegraphics[width=2\columnwidth]{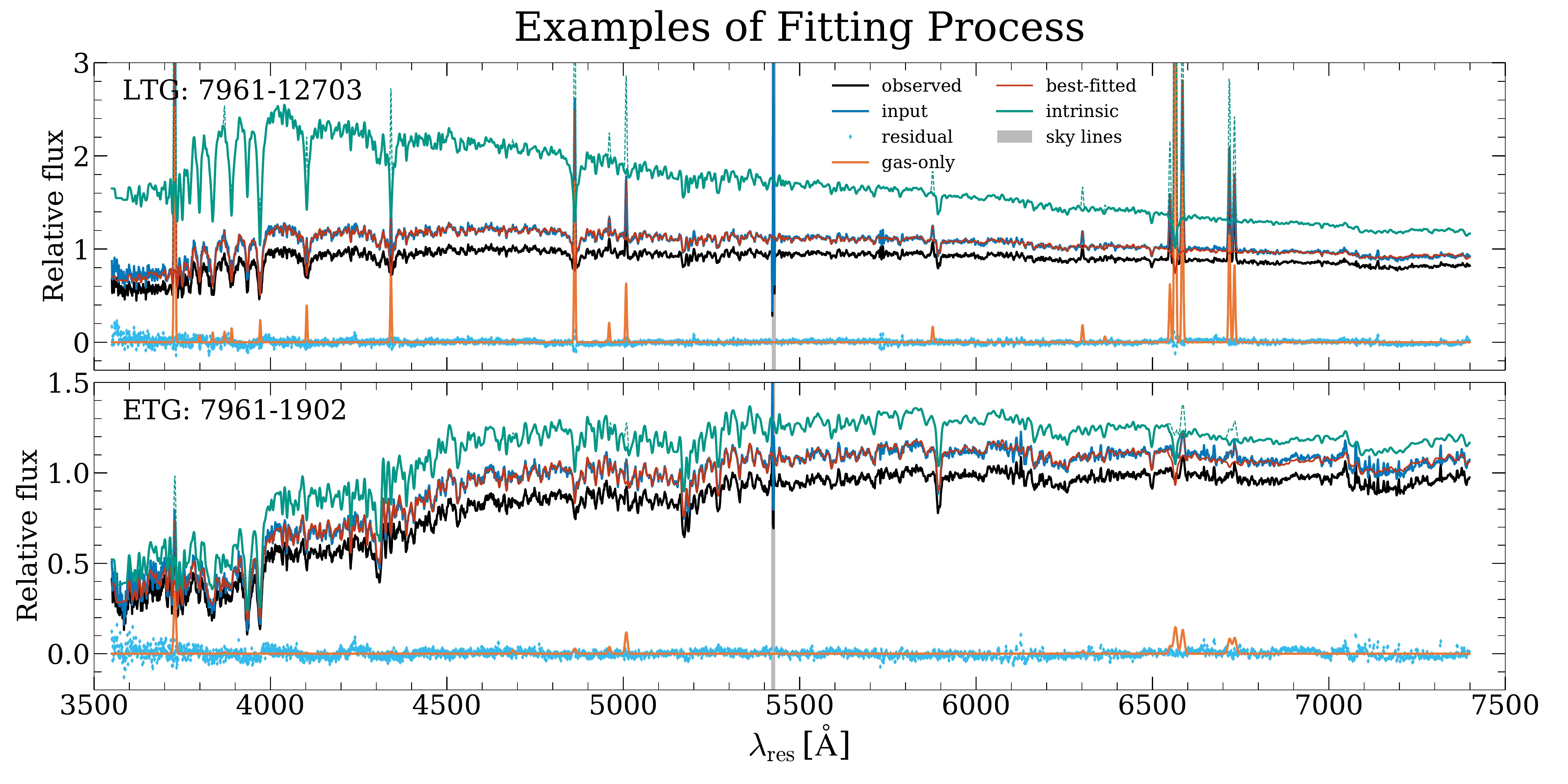}
\caption{Two examples (LTG: 7961-12703; ETG: 7961-1902) of spectra at different fitting stages. In each panel, the black curve is the observed spectrum (including stellar continuum and gas emission lines) obtained from MaNGA DRP datacube; the blue curve is the observed spectrum with the Milky Way dust attenuation effect corrected, which is taken as the input of {\sc ppxf} fitting (including stellar continuum and gas emission lines); the red curve is the best-fitted spectrum given by {\sc ppxf}, with the solid curve representing the best-fitted stellar continuum and the dashed curve representing the best-fitted gas emission lines; the green curve is the intrinsic spectrum of the galaxy, which is derived by the combination of best-fitted templates (including stellar and gas emission line templates, indicating by solid and dashed curves, respectively) with velocity and dispersion convolved (no dust attenuation). The orange curve is the best-fitted gas-only spectrum and the cyan diamonds are the fitting residuals. The grey shaded region is the region of sky lines, which is masked and not fitted.}
\label{fig:example_spec}
\end{figure*}

To better demonstrate this process, we present \autoref{fig:example_spec}, where we show the spectra of two example galaxies (both are the stacked spectra within half-light isophotes; a star-forming late-type galaxy and a passive early-type galaxy) at different stages of this fitting process: the observed spectra from DRP datacube are shown by the black curves; after correcting the MW dust attenuation effect, we get the blue curves, which are then taken as the input spectra for {\sc ppxf} fitting. 

During the spectrum fitting, we simultaneously fit the stellar populations and the kinematics of the stellar continuum and the gas emission lines by assuming a Gaussian distribution and free velocities and dispersions of the stellar component and the gas emission lines. The gas emission lines included in our fitting are listed in \citet{Belfiore_et_al.(2019)} (see their table 1), which are taken as the default setting of the {\sc ppxf} software. During the fitting, we do not fit the multiplicative polynomials by setting \texttt{mdegree=-1} in the {\sc ppxf} software. We emphasize here that in our analysis, we conducted a simultaneously fitting of kinematics and stellar population properties, rather than fitting them separately. We confirm that this simultaneous fitting procedure had minimal impact on the stellar population properties. The fitting range is this work is $3550\,\Angstrom-7400\,\Angstrom$. For all stellar population properties, we do not adopt regularization (i.e. \texttt{regul}$=0$), while for star-formation history-related parameters, we adopt \texttt{regul}$=100$ to obtain smoother template weight distributions (see \autoref{sec:sfh} for more details).

As shown in \autoref{fig:example_spec}, the red curves are the best-fitted results of the given spectra, which are the linear combinations of the best-fitted templates with best-fitted velocities and velocity dispersions convolved and the best-fitted dust attenuation effect (of the investigated galaxy) considered. The green curves are the best-fitted ``intrinsic'' spectra of the distant galaxies, without undergoing any dust extinction processes. As can be clearly seen, the late-type galaxy (LTG) has obviously stronger gas emission (the orange curves) and more significant dust attenuation effect from its own dust than the early-type galaxy (ETG), which is consistent with our previous understanding that LTGs typically have more dust than ETGs. 

\begin{figure}
\centering
\includegraphics[width=1\columnwidth]{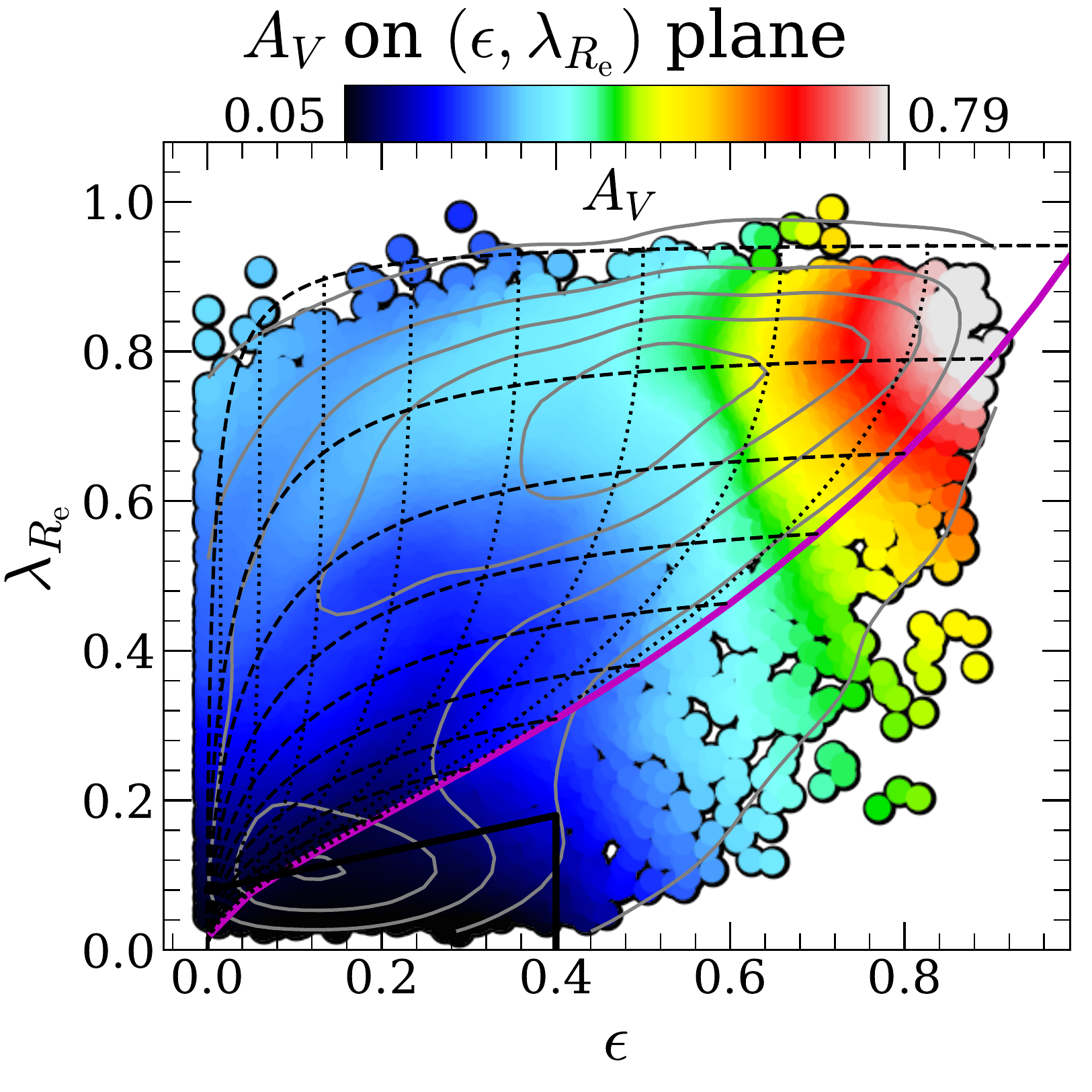}
\caption{Distribution of $A_V$ (the attenuation of the spectrum at $\lambda=5500\Angstrom$, i.e. $V-$band) on the $(\epsilon, \lambda_{R_{\rm e}})$ plane (see \autoref{sec:jam_props} for definitions of these two parameters). The black polygon is defined by $\lambda_{R_{\rm hsm}}=\epsilon/4+0.08$ and $\epsilon=0.4$, and encloses the region of slow-rotating ETGs \citep[eq.~19]{Cappellari(2016)}. The magenta curve shows the edge-on view for axisymmetric galaxies following a relation between intrinsic shape and anisotropy $\beta=0.7\epsilon_{\rm intr}$ \citep[fig.~9]{Cappellari2007}, with the black dotted lines showing its change with different inclinations ($\Delta i=10^{\circ}$). The black dashed lines show the theoretical distribution of galaxies versus inclinations, given different $\varepsilon_{\rm intr}$ values ($\Delta \varepsilon_{\rm intr}=0.1$). A kernel density estimation of the galaxy number density is indicated by the grey contours. This result comes from the {\sc ppxf} fitting with the {\sc fsps} stellar model.}
\label{fig:dust}
\end{figure}

To quantitatively evaluate the dust attenuation effect, we show, in \autoref{fig:dust}, the distribution of $A_V$ on the $(\epsilon, \lambda_{R_{\rm e}})$ plane (see \autoref{sec:jam_props} for definitions of these two parameters). The $(\epsilon, \lambda_{R_{\rm e}})$ plane is usually used to classify slow- and fast-rotators in early-type galaxies (e.g. \citealt{Emsellem_et_al.(2007),Emsellem_et_al.(2011),Fogarty_et_al.(2015)}; see \citealt{Cappellari(2016)} for a review). Here, however, we plot both early-type and late-type galaxies to investigate their dust attenuation variation on the $(\epsilon,\lambda_{R_{\rm e}})$ plane as a continuous sample. Before plotting, we make use of the {\sc python} implementation \citep{Cappellari_et_al.(2013b)} of the two-dimensional Locally Weighted Regression ({\sc loess}\footnote{Version 2.1.2, available from \url{https://pypi.org/project/loess/}}, \citealt{Cleveland_and_Devlin(1988)}) method to obtain a smoothed distribution of the color-coding values. A small smoothing parameter \texttt{frac=0.1} is adopted for all these kinds of figures (including Figs.~\ref{fig:sp2sigma}, \ref{fig:gammatot}, \ref{fig:sp_mass_size}, \ref{fig:sp_lambda_epsilon}, \ref{fig:slope_mass_size}, \ref{fig:slope_age2sigma}, \ref{fig:slope_lambda_epsilon}, \ref{fig:mass_light_lambda_epsilon}, and \ref{fig:t5090}) in this work. As can be seen, dust attenuation effect show obvious variation on the plane. Firstly, the slow-rotating galaxies (those within the region of black lines) appear to experience weaker dust attenuation effect (i.e. lower $A_V$ value) than the fast-rotating galaxies. At fixed $\epsilon$, $A_V$ increases (i.e. dust attenuation effect getting stronger) significantly from the low $\lambda_{R_{\rm e}}$ region (most of which are quenched ellipticals) to the high $\lambda_{R_{\rm e}}$ region (most of which are star-forming disk galaxies). Besides, galaxies show obvious increasing trend of $A_V$ (i.e. dust attenuation effect getting stronger) from the face-on view to the edge-on view of the galaxies (i.e. along the black dashed curves), especially for those which are intrinsically rotational-supported. All the findings above are consistent with our previous understanding of the dust content in different galaxies, confirming the robustness of {\sc ppxf} in dealing with dust attenuation effects.

We note here that the {\sc loess-}smoothed values will be biased to a narrower range, and as a result, we shall not focus on the exact values of the properties, but focus on the overall trend of these quantities. The original values of all the population-related properties (including the dust attenuation parameters used here, the global stellar population properties used in \autoref{sec:global_sp}, stellar population gradients used in \autoref{sec:sp_gradients}, star-formation history properties used in \autoref{sec:sfh}), the stellar population profiles, and the spatially resolved stellar population maps for all the $\sim 10000$ galaxies are available from the website of MaNGA DynPop (\url{https://manga-dynpop.github.io}). The readers are referred to \autoref{table:catalog} for detailed explanations of the quantities. A brief comparison between this work and the existing stellar population catalogues for the MaNGA full sample, namely the {\sc pipe3d} catalogue \citep{pipe3d} and the {\sc firefly} catalogue \citep{firefly} is given in Appendix~\ref{apx:comparison}.

\subsubsection{Global and spatially resolved stellar population properties}
\label{sec:global_and_maps}
\begin{figure}
\centering
\includegraphics[width=1\columnwidth]{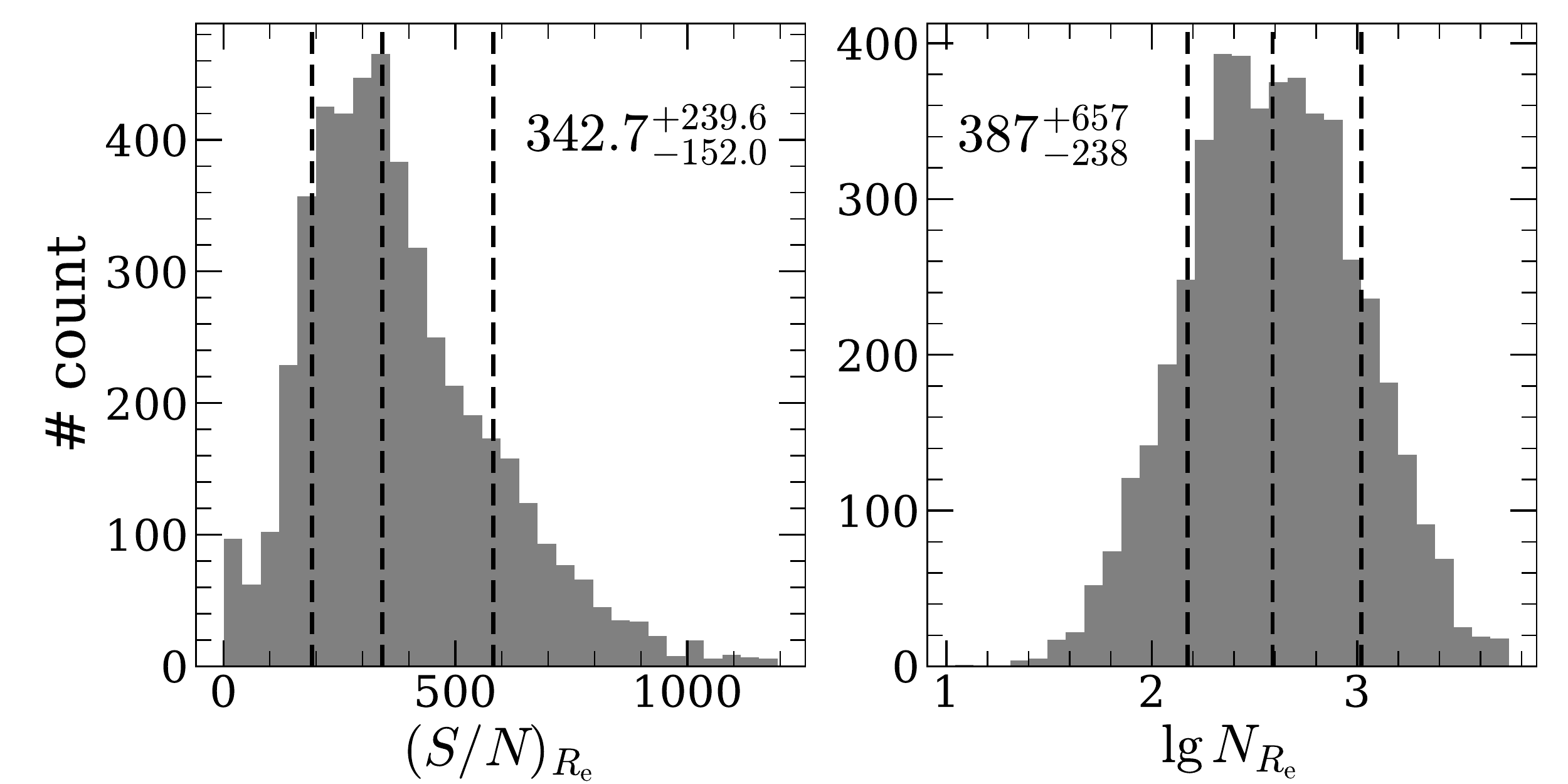}
\caption{Distributions of signal-to-noise ratio of stacked spectra within half-light isophotes (left) and the number of spaxels within half-light isophotes (right). In each panel, the three dashed lines represent $16\%$, $50\%$, and $84\%$ percentiles from left to right, with the numbers indicating the corresponding values.}
\label{fig:snr}
\end{figure}

Spatially resolved spectra are typically binned to a higher signal-to-noise ratio ($S/N$) in order to get more accurate estimates of stellar population properties. In this work, we bin the spatially resolved spectra for each galaxy in two ways for different purposes:
\begin{enumerate}
    \item We take the luminosity-weighted stellar population within the elliptical half-light isophote as the global stellar population properties of galaxies. To do so, we first bin, for each galaxy, all the spectra within the elliptical half-light isophote (the aperture size, ellipticity, and position angle of the elliptical half-light isophote can be obtained from \citetalias{Zhu_et_al.(2023a)}) together and perform {\sc ppxf} fitting on the derived spectra. In \autoref{fig:snr}, we show the distributions of the signal-to-noise ratios of the stacked spectra within half-light isophote and the number of stacked spaxels within this aperture. The $S/N$ here is calculated as the ratio between the median values of flux and noise of the stacked spectra within the wavelength range from $4730\,\Angstrom$ to $4780\,\Angstrom$, which does not include obvious emission and absorption lines. As can be seen, the elliptical half-light isophote typically contains a sufficiently large number of spaxels (with the median value being 387) and thus reaches a sufficiently high signal-to-noise ratio (with the median value being $S/N\sim 342$), with which the global stellar population properties can be accurately estimated.

    \item To get the population property maps and further the population profiles/gradients, we first bin the spatially resolved spectra to $S/N=30$ using the two-dimensional adaptive spatial binning method \citep{Cappellari_et_al.(2003)} with the {\sc VorBin}\footnote{Version 3.1.5, available from \url{https://pypi.org/project/vorbin/}} software and perform {\sc ppxf} to binned spectrum of each bin. We explain the calculation of population profiles/gradients in \autoref{sec:sp_gradients}.
\end{enumerate}

For both the stacked spectra within half-light isophote and the binned, spatially resolved spectra, we apply the {\sc ppxf} fitting with the settings and assumptions described above to obtain the best-fitted template weights. With the best-fitted template weights from {\sc ppxf}, luminosity age and metallicity of the spectrum can be calculated as: 
\begin{equation}
    \label{eq:lwsp}
    x = \frac{\sum_{k} w_k L_k x_k}{\sum_k w_k L_k},
\end{equation}
where $x$ can be $\lg\rm Age$ or $[Z/H]$; $w_k$ and $L_k$ are the best-fitted weight and SDSS $r$-band luminosity of the $k-$th template, respectively. Stellar mass-to-light ratio is defined as:
\begin{equation}
    \label{eq:ml}
    M_{\ast}/L = \frac{\sum_k  w_k M_{\ast,k}}{\sum_k w_k L_k},
\end{equation}
where $M_{\ast,k}$ is the stellar mass (including the mass of living stars and stellar remnants, but excluding the gas lost during stellar evolution) of $k-$th templates. The summation goes over all the templates used in the fitting. The calculation of stellar mass-to-light ratio is automatically done by the \texttt{\sc mass\_to\_light} routine of the {\sc ppxf} software, which is independent of the dust attenuation effect. Throughout this paper, we only investigate luminosity-weighted stellar population properties, while the mass-weighted quantities will also be provided in the full catalogue (see \autoref{table:catalog}).

\begin{figure*}
\centering
\includegraphics[width=0.68\columnwidth]{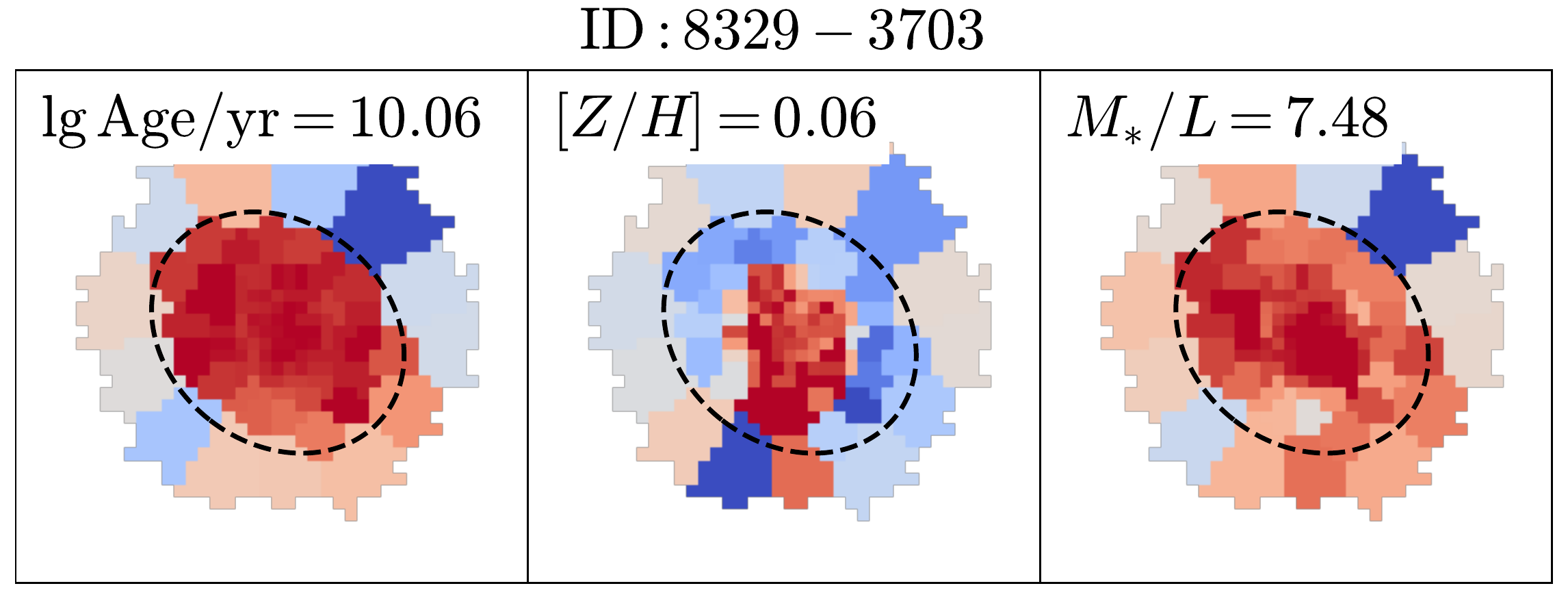}
\includegraphics[width=0.68\columnwidth]{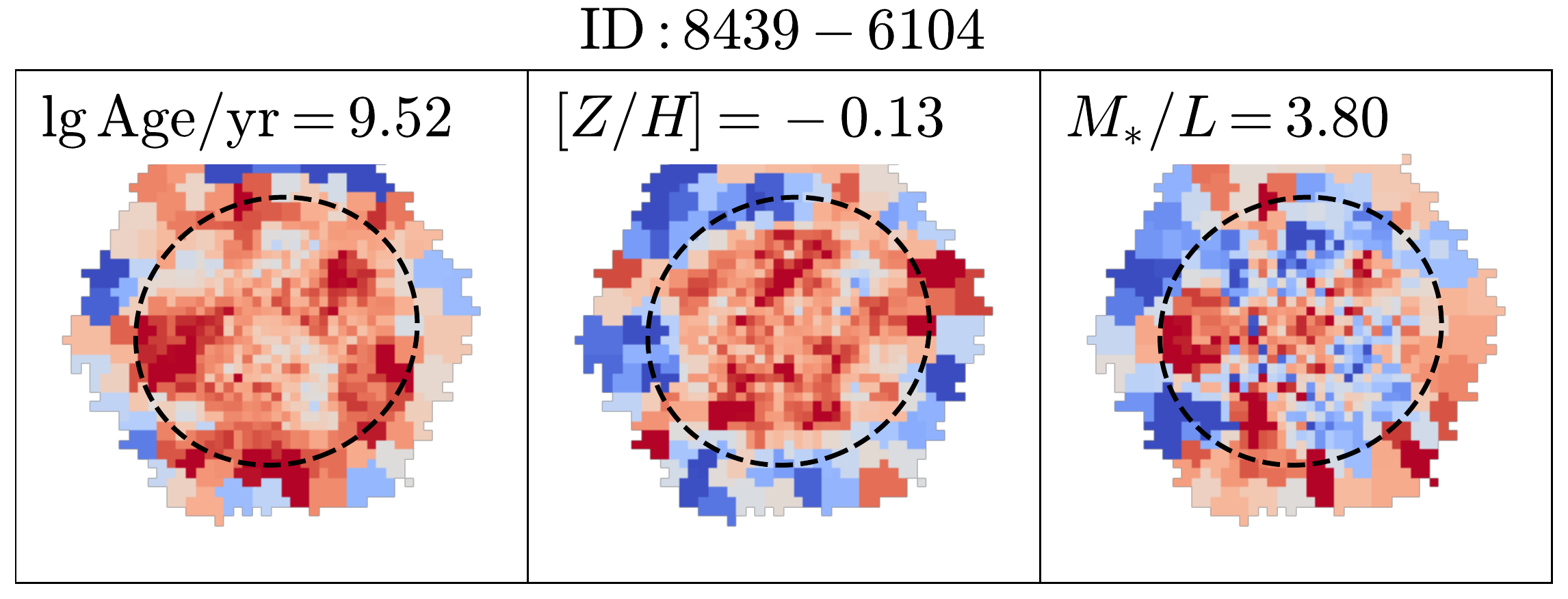}
\includegraphics[width=0.68\columnwidth]{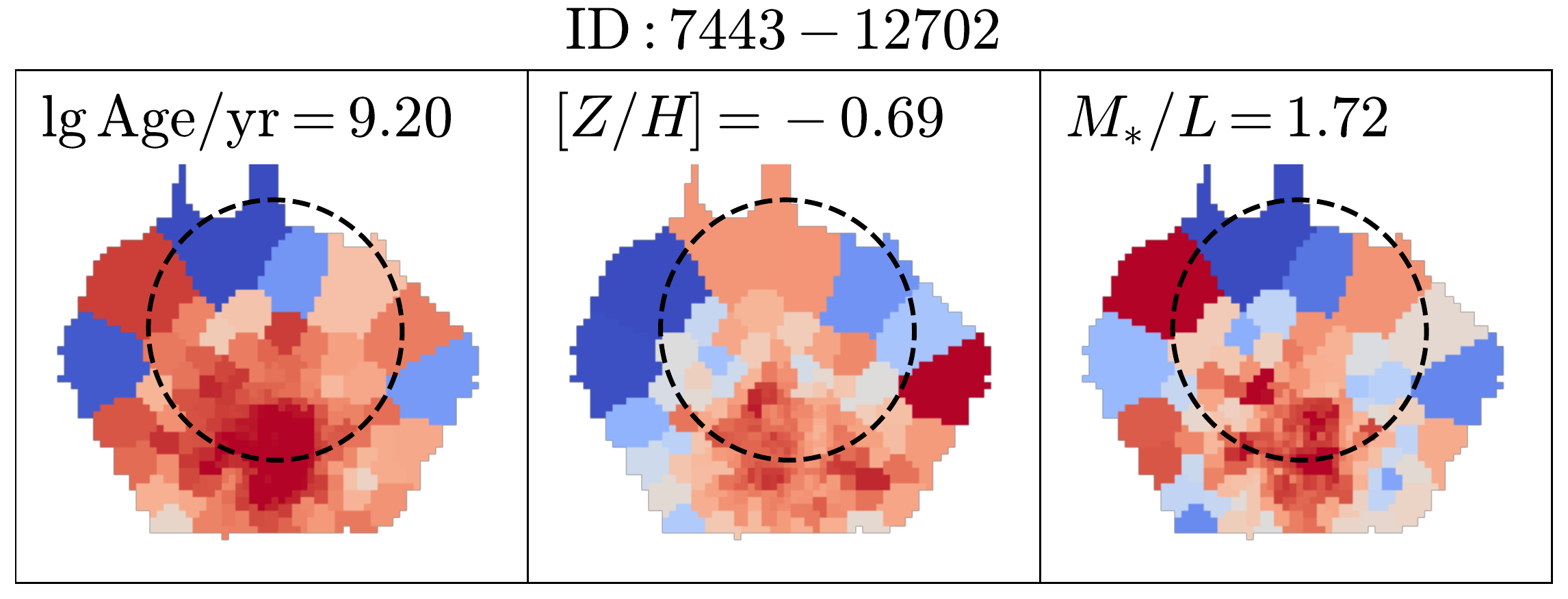}

\includegraphics[width=0.68\columnwidth]{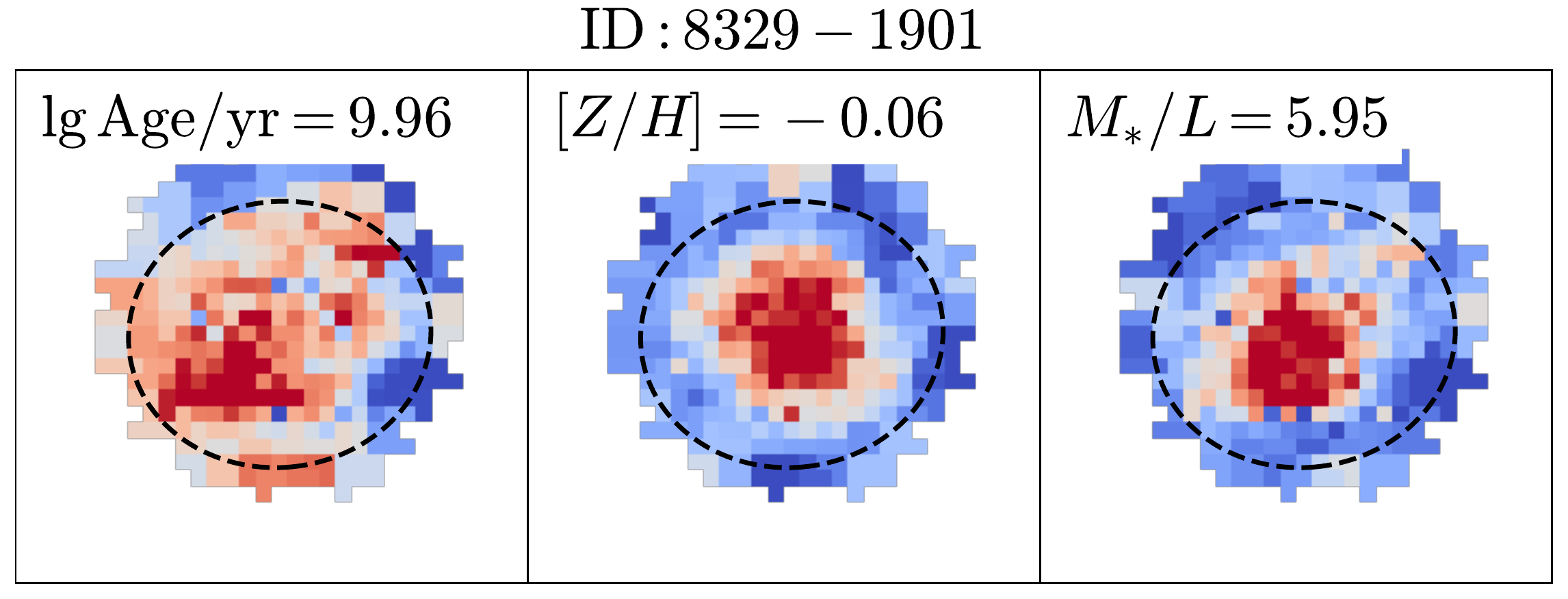}
\includegraphics[width=0.68\columnwidth]{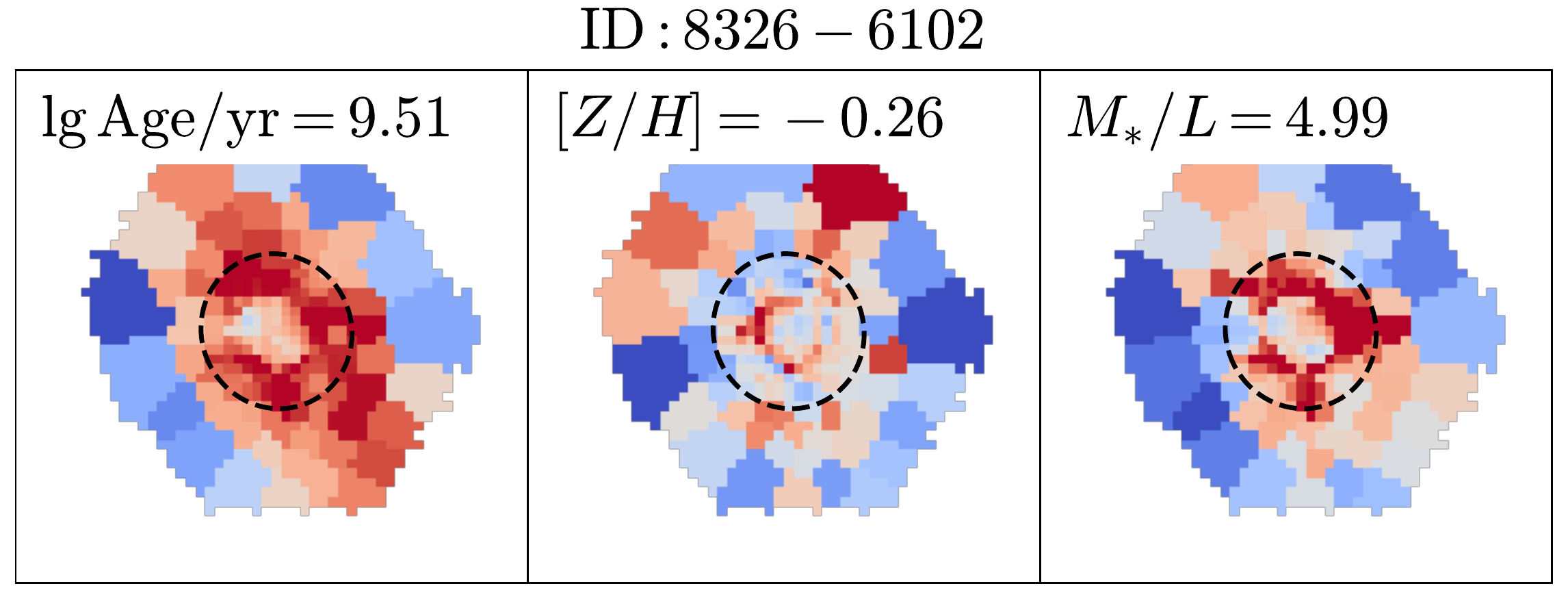}
\includegraphics[width=0.68\columnwidth]{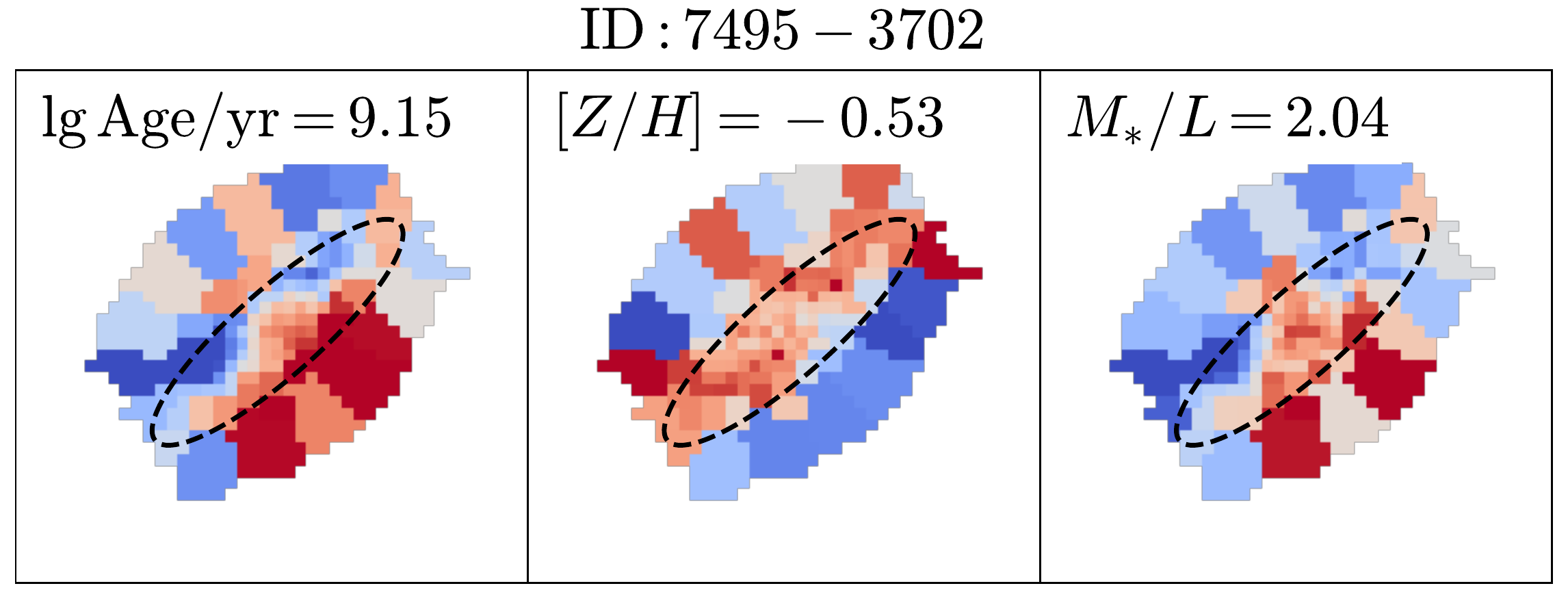}

\includegraphics[width=0.68\columnwidth]{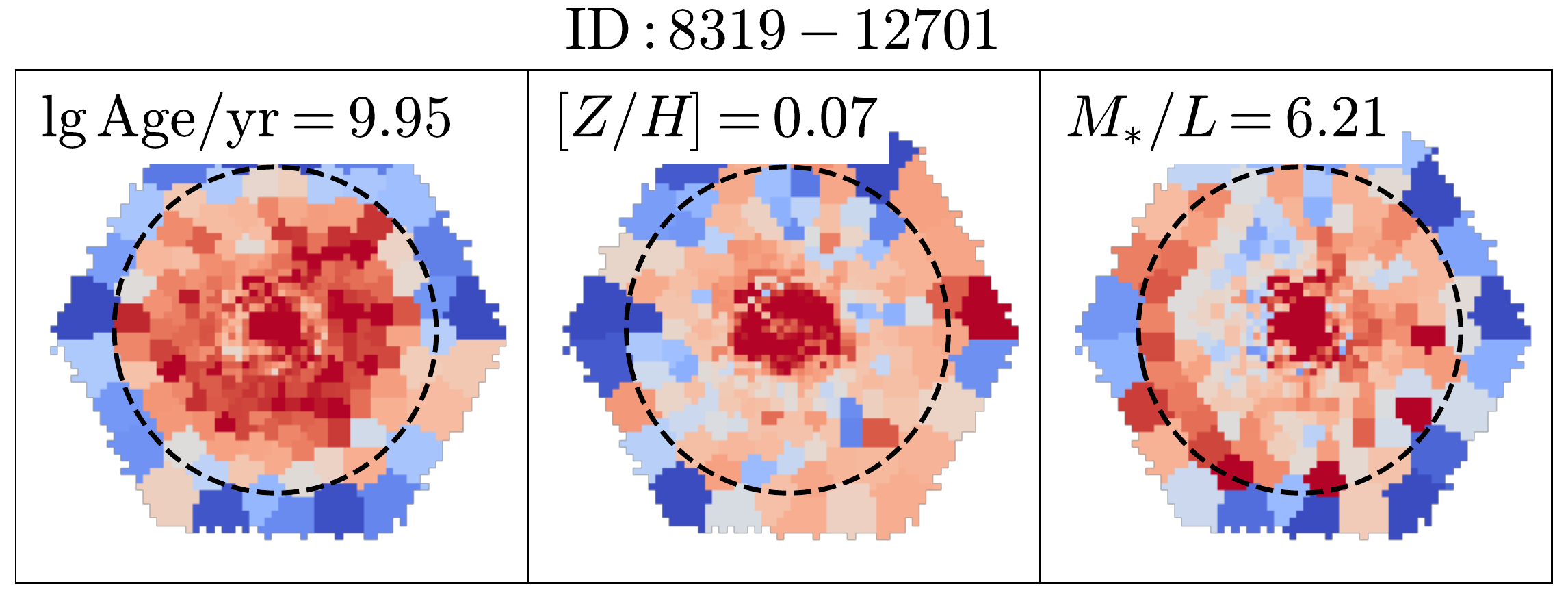}
\includegraphics[width=0.68\columnwidth]{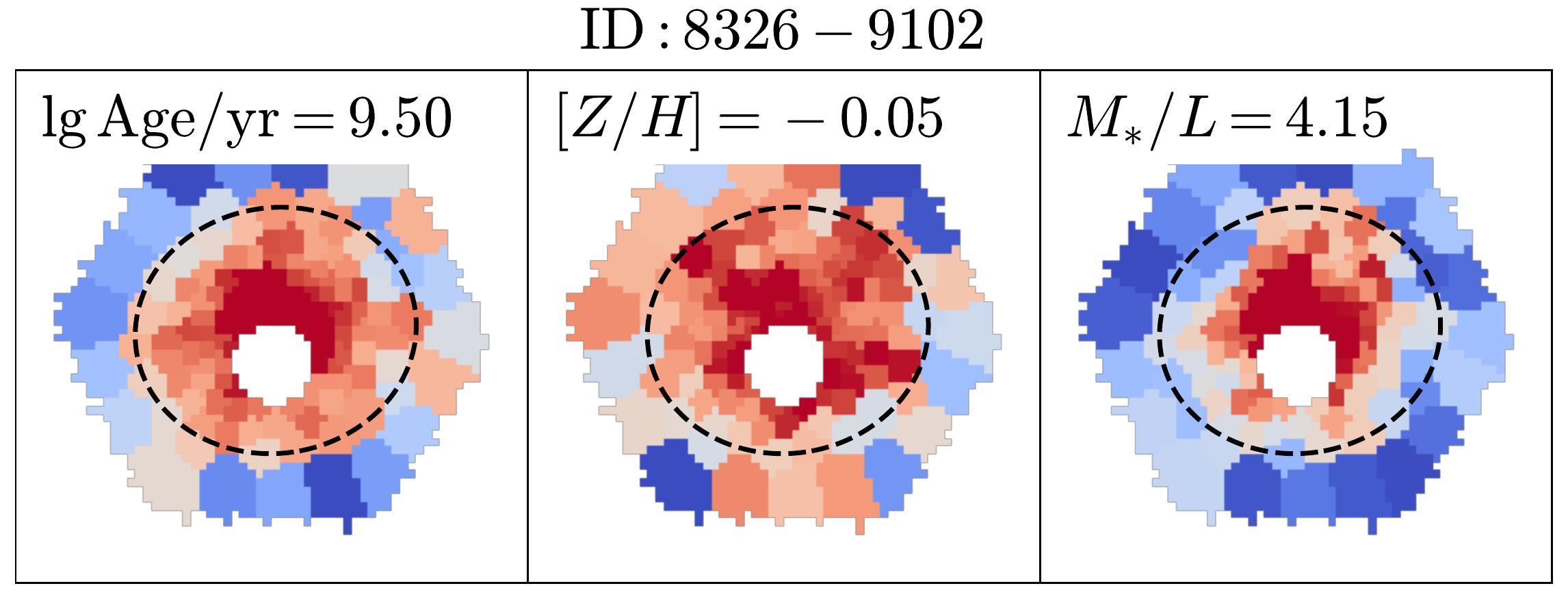}
\includegraphics[width=0.68\columnwidth]{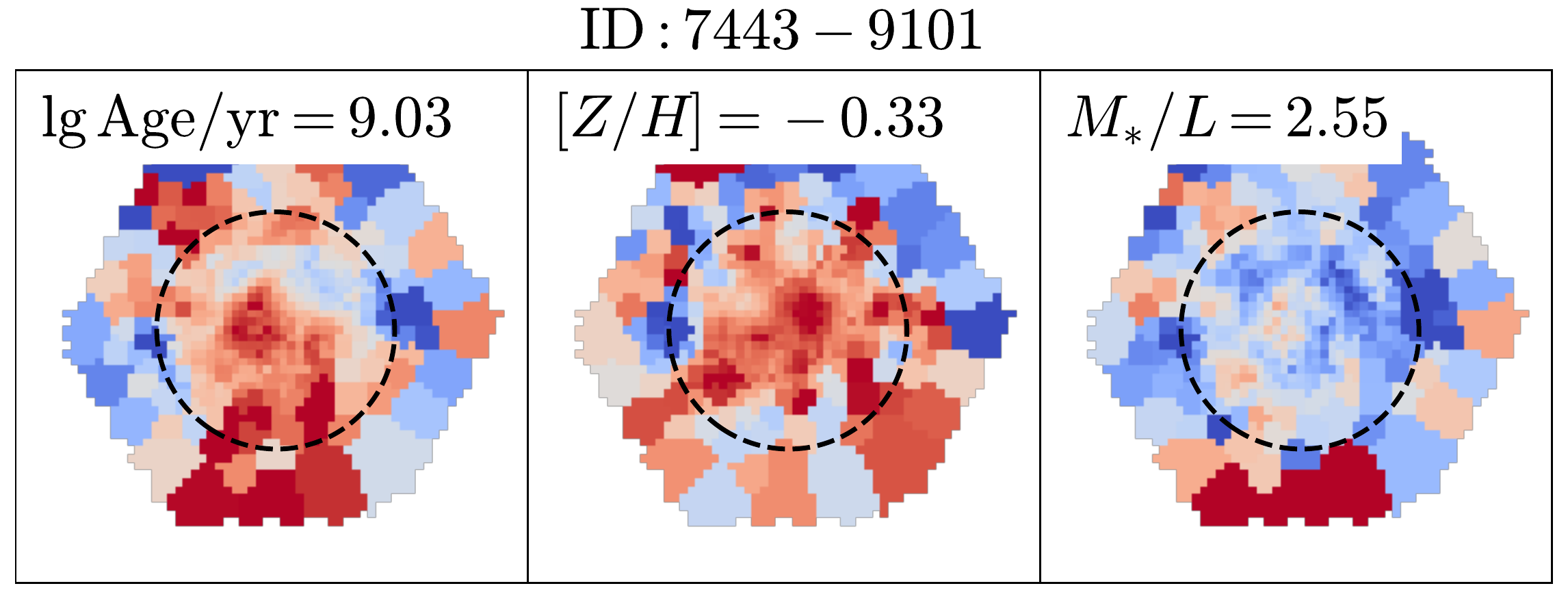}

\includegraphics[width=0.68\columnwidth]{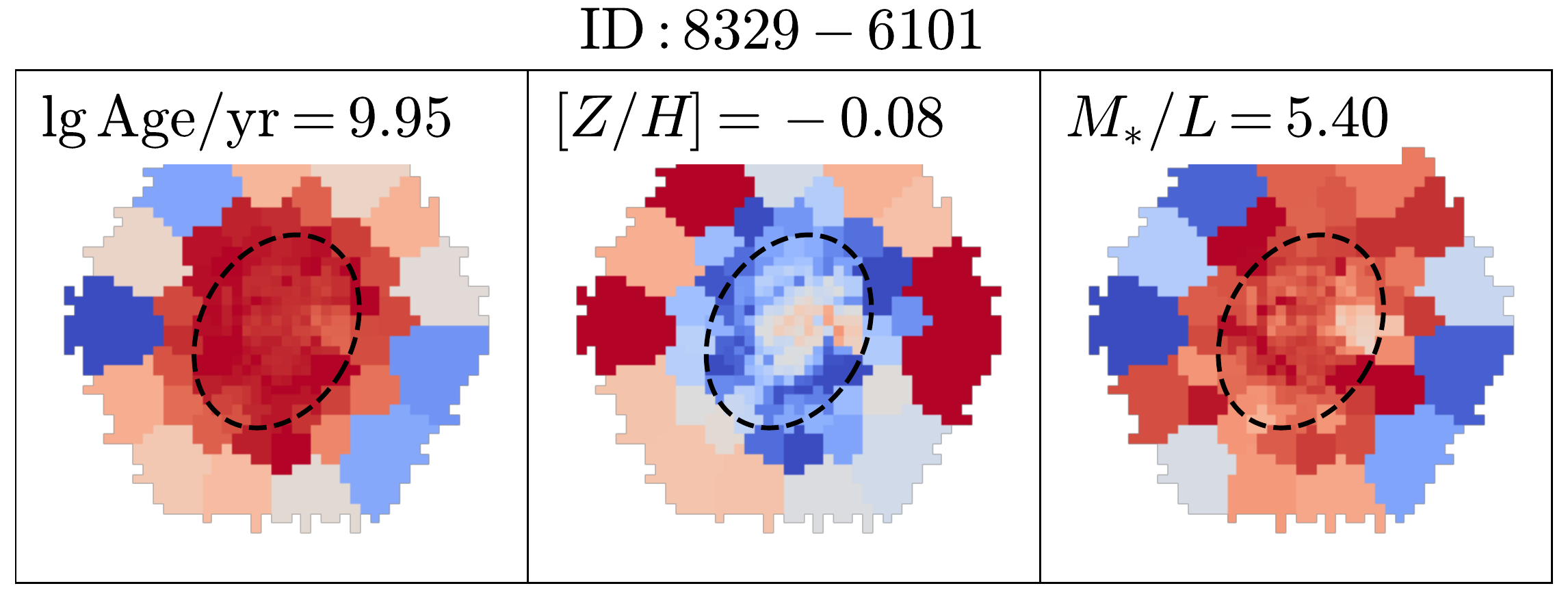}
\includegraphics[width=0.68\columnwidth]{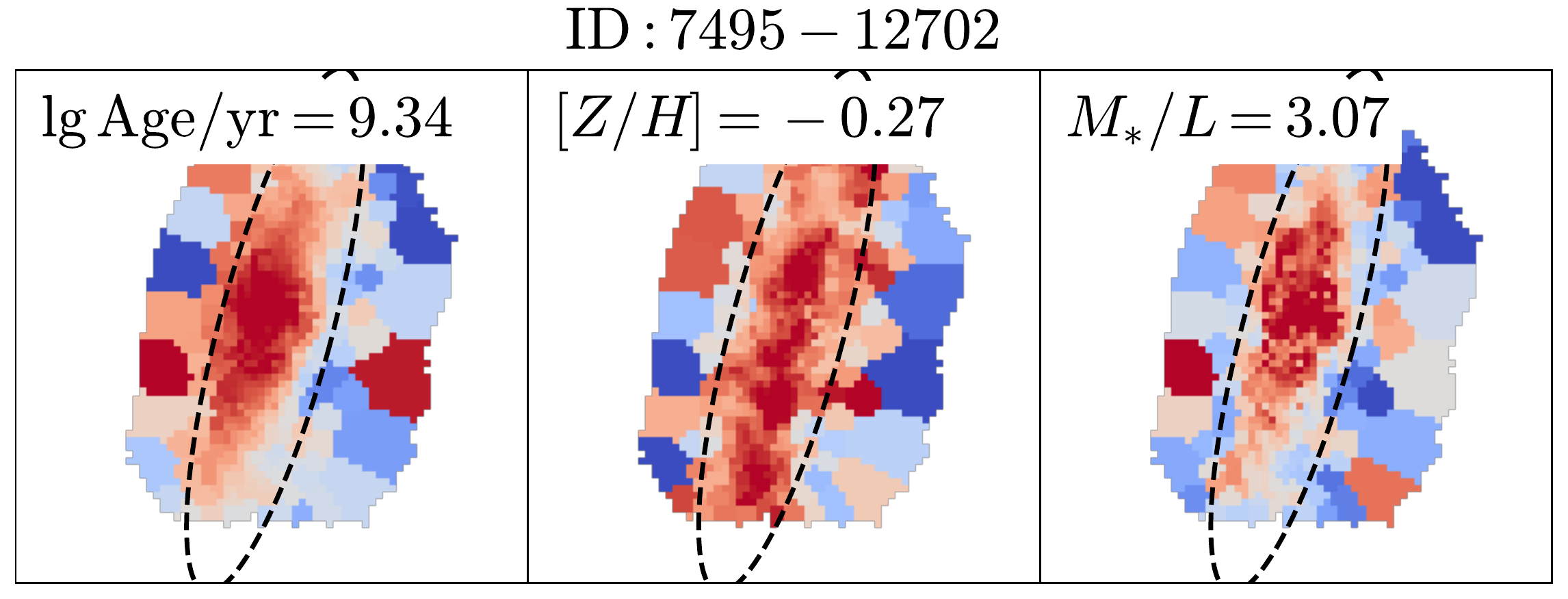}
\includegraphics[width=0.68\columnwidth]{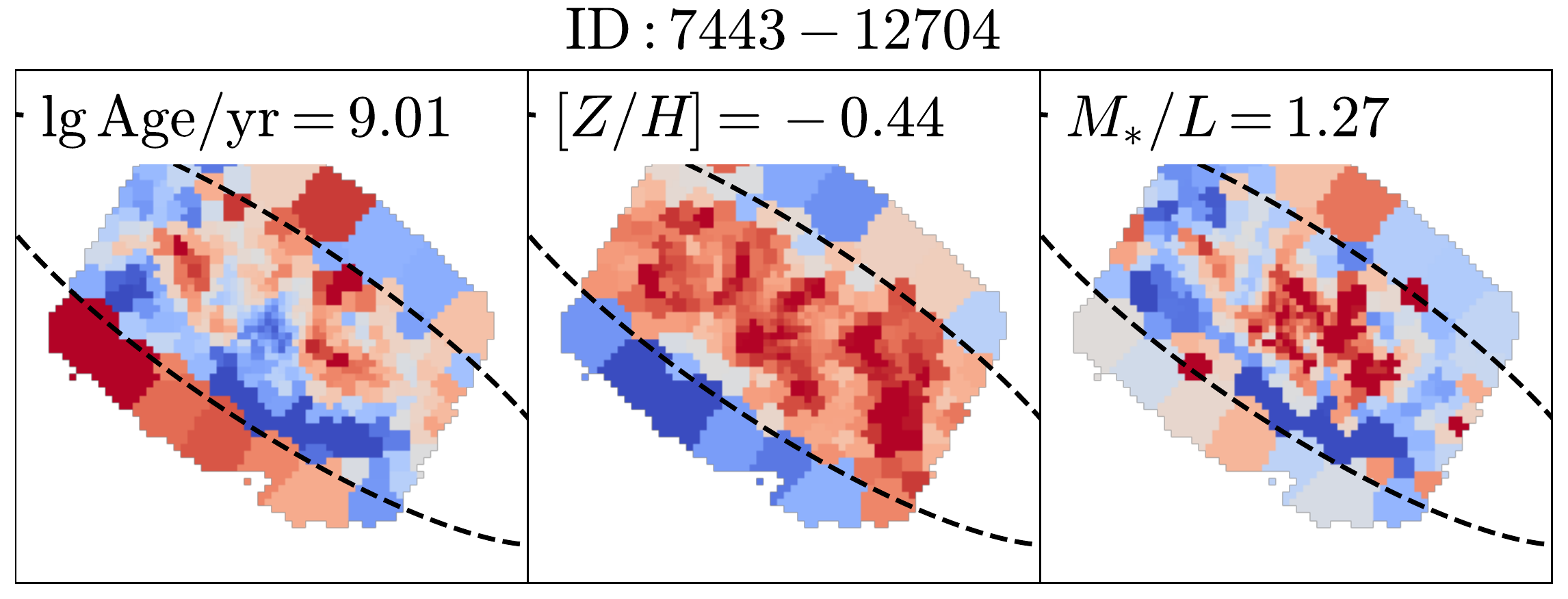}

\includegraphics[width=0.68\columnwidth]{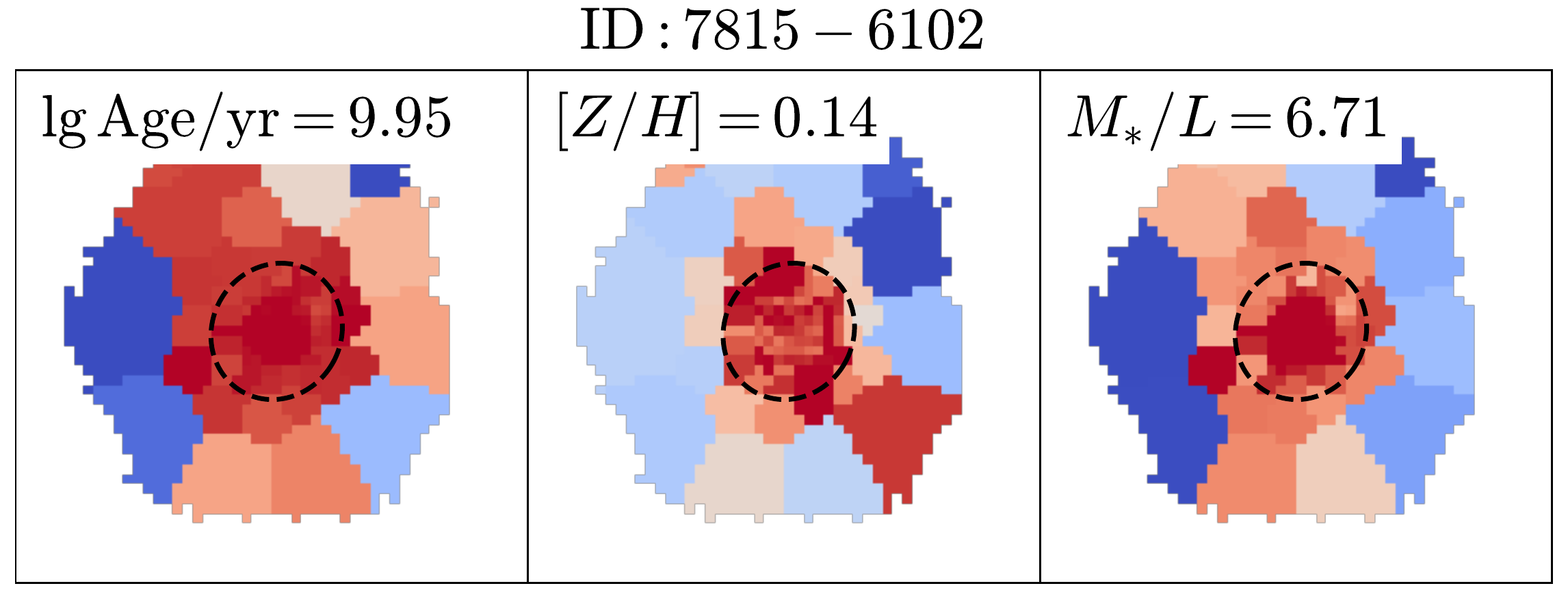}
\includegraphics[width=0.68\columnwidth]{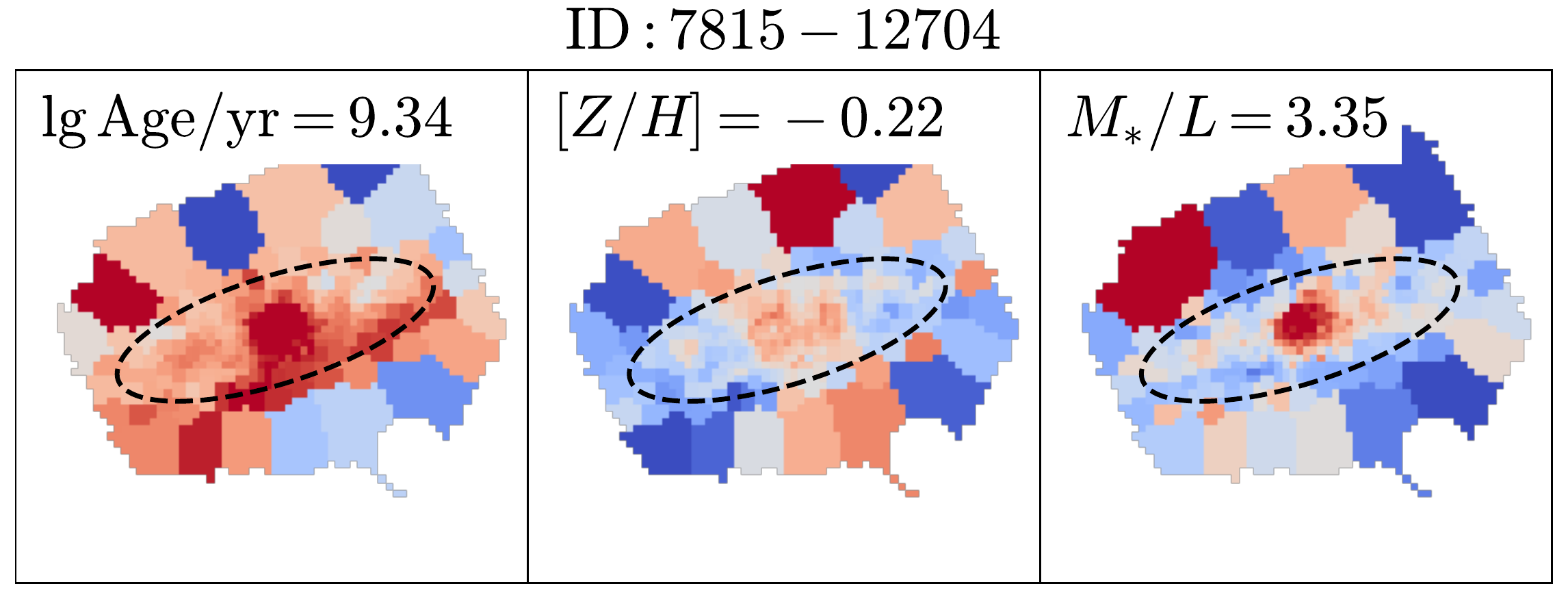}
\includegraphics[width=0.68\columnwidth]{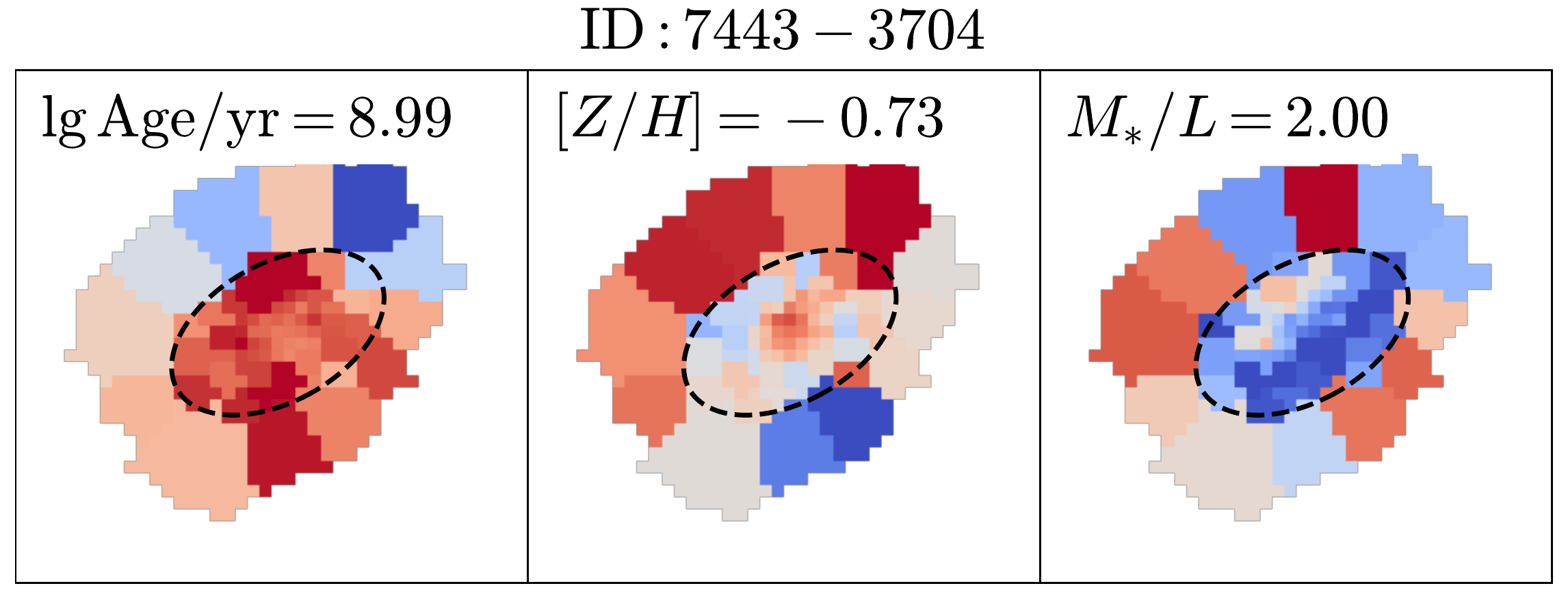}

\includegraphics[width=0.68\columnwidth]{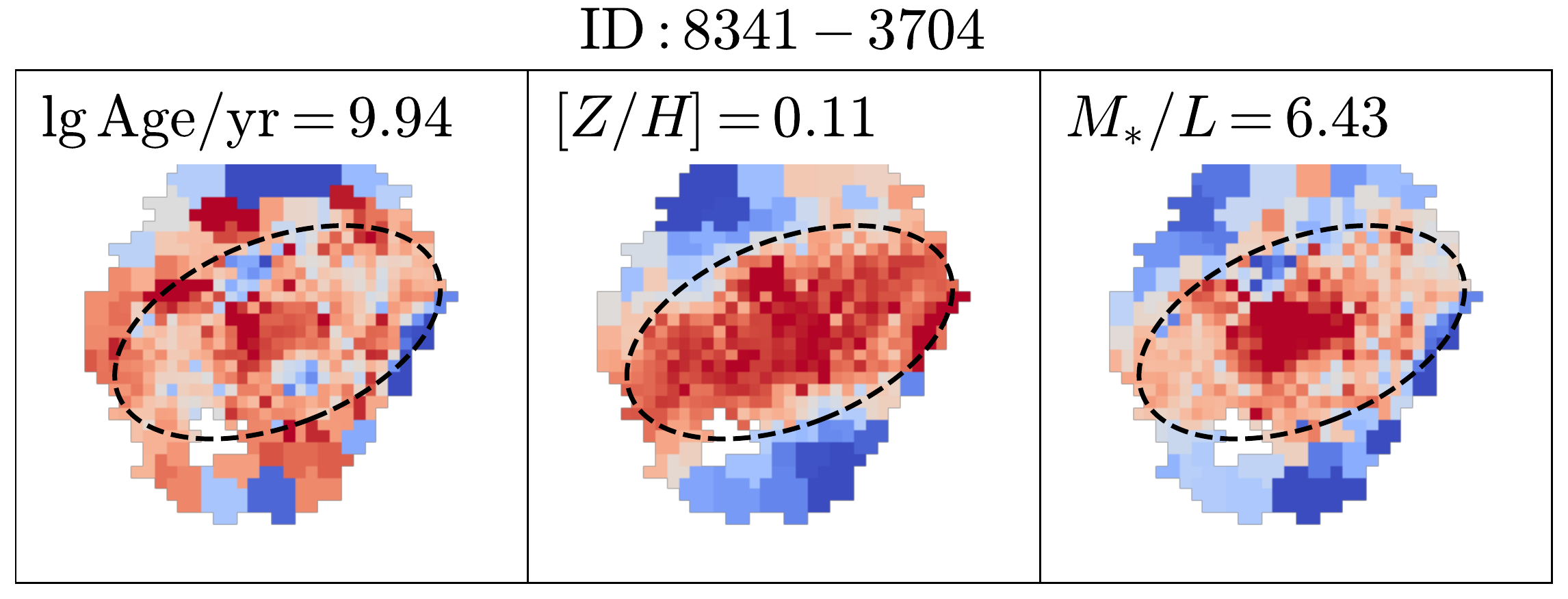}
\includegraphics[width=0.68\columnwidth]{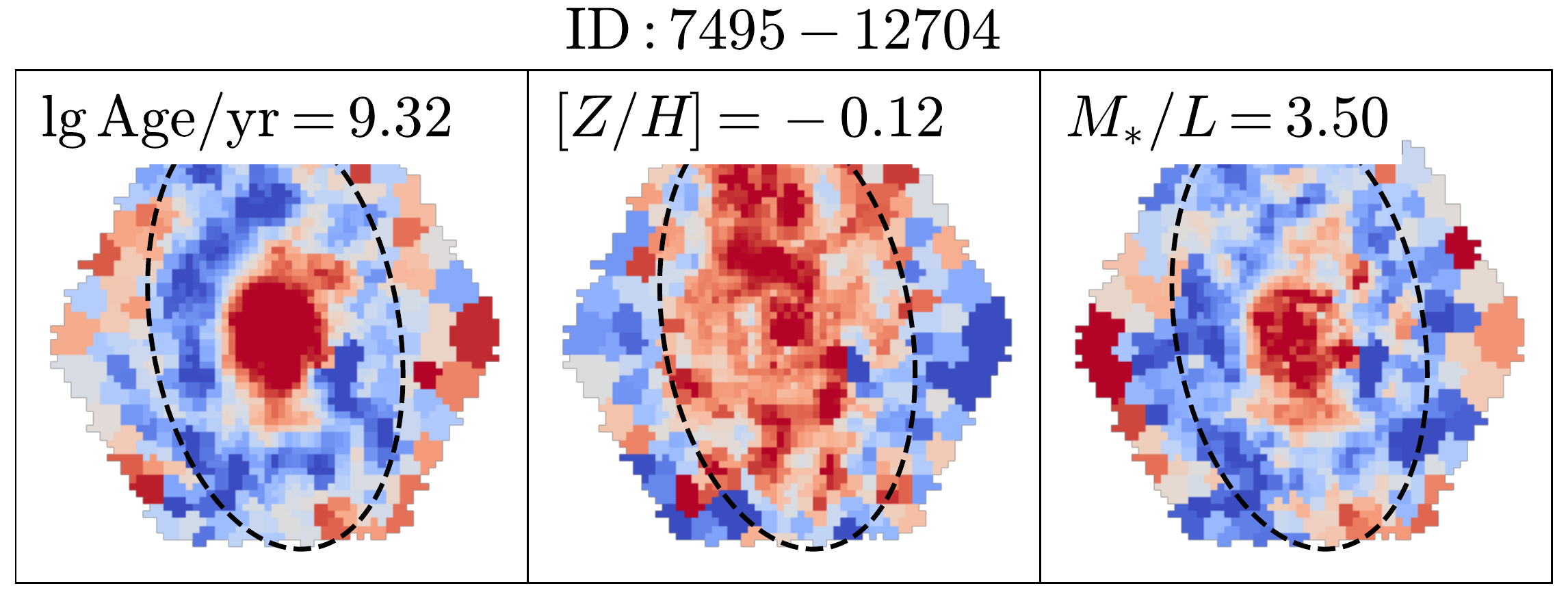}
\includegraphics[width=0.68\columnwidth]{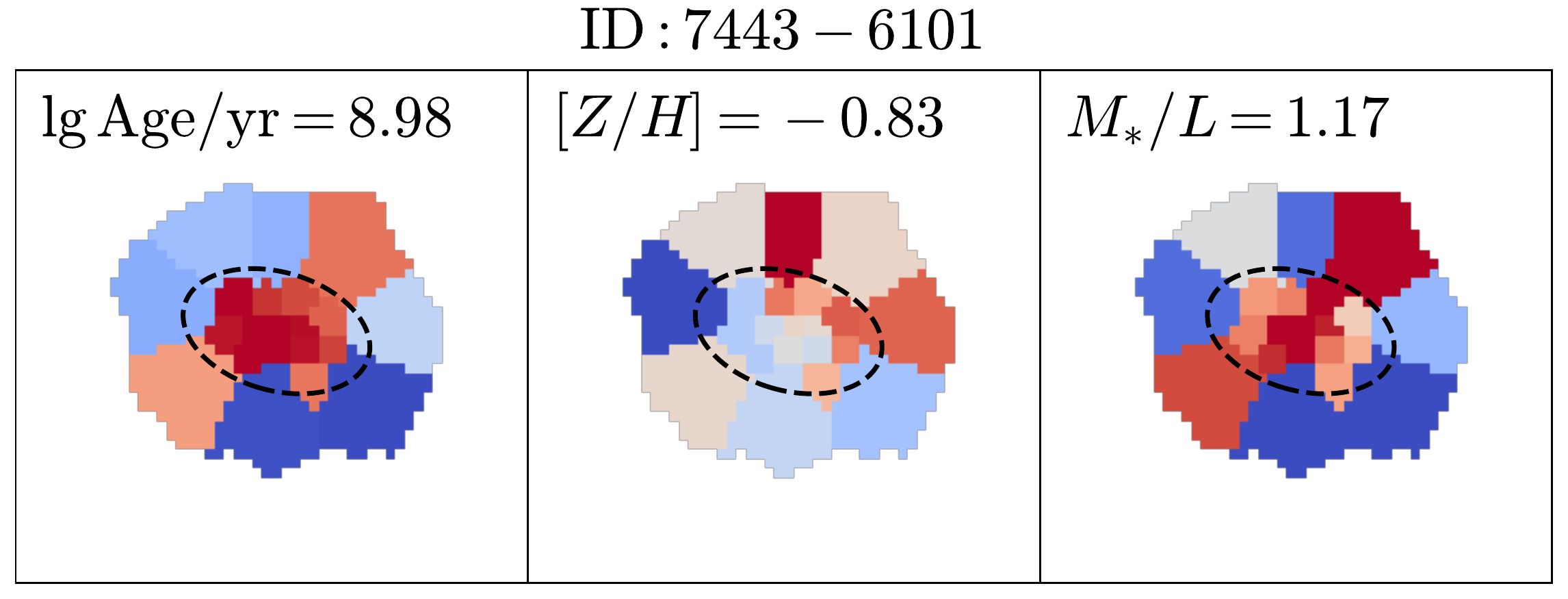}

\includegraphics[width=0.68\columnwidth]{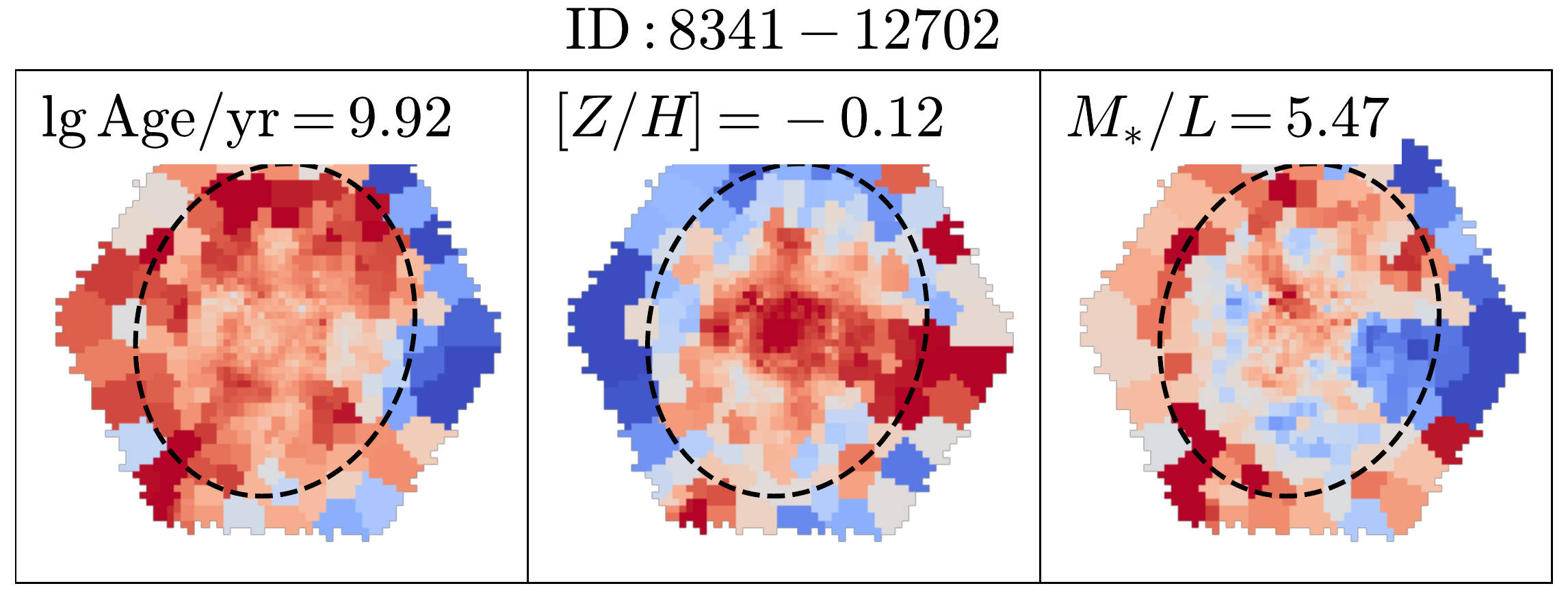}
\includegraphics[width=0.68\columnwidth]{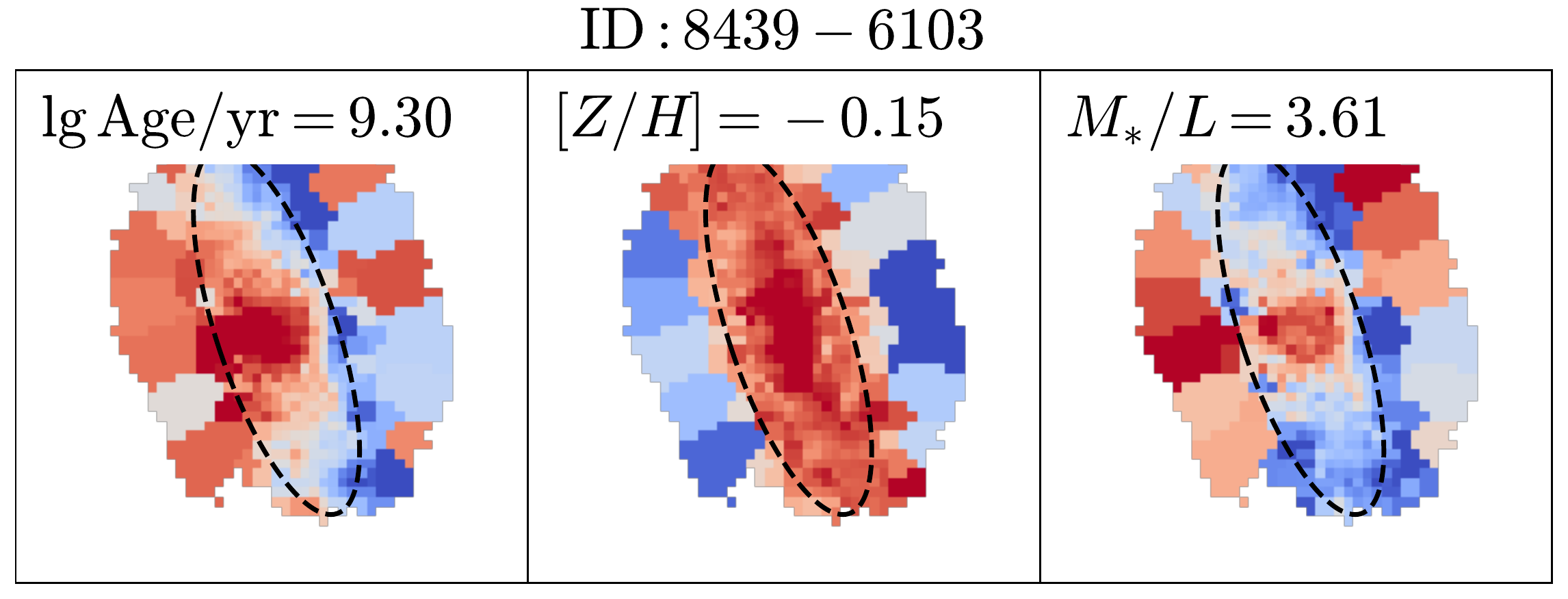}
\includegraphics[width=0.68\columnwidth]{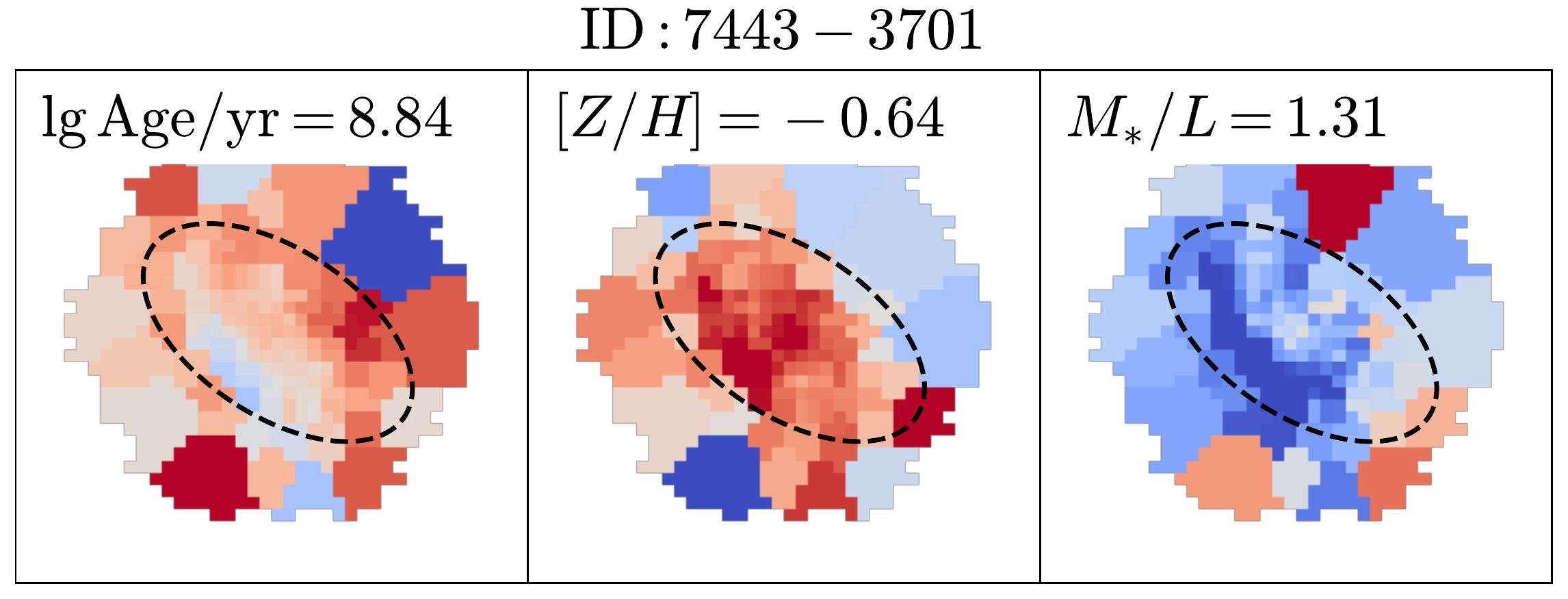}

\includegraphics[width=0.68\columnwidth]{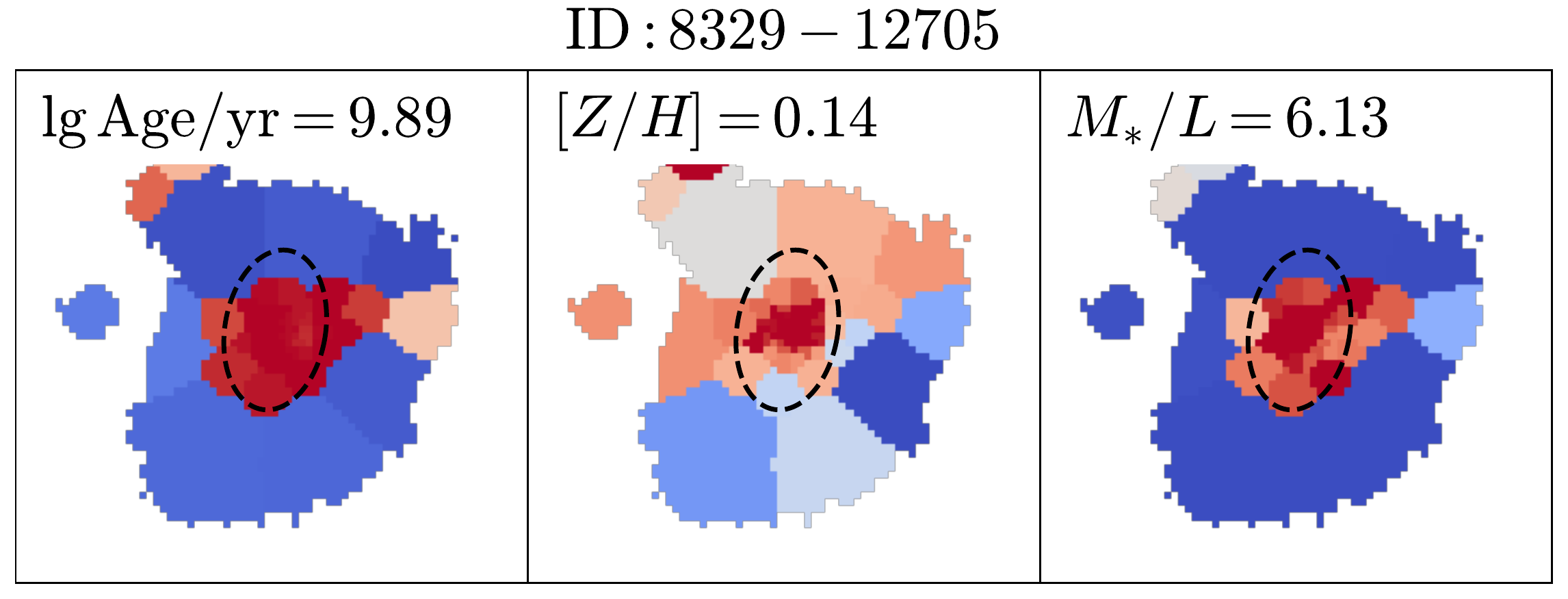}
\includegraphics[width=0.68\columnwidth]{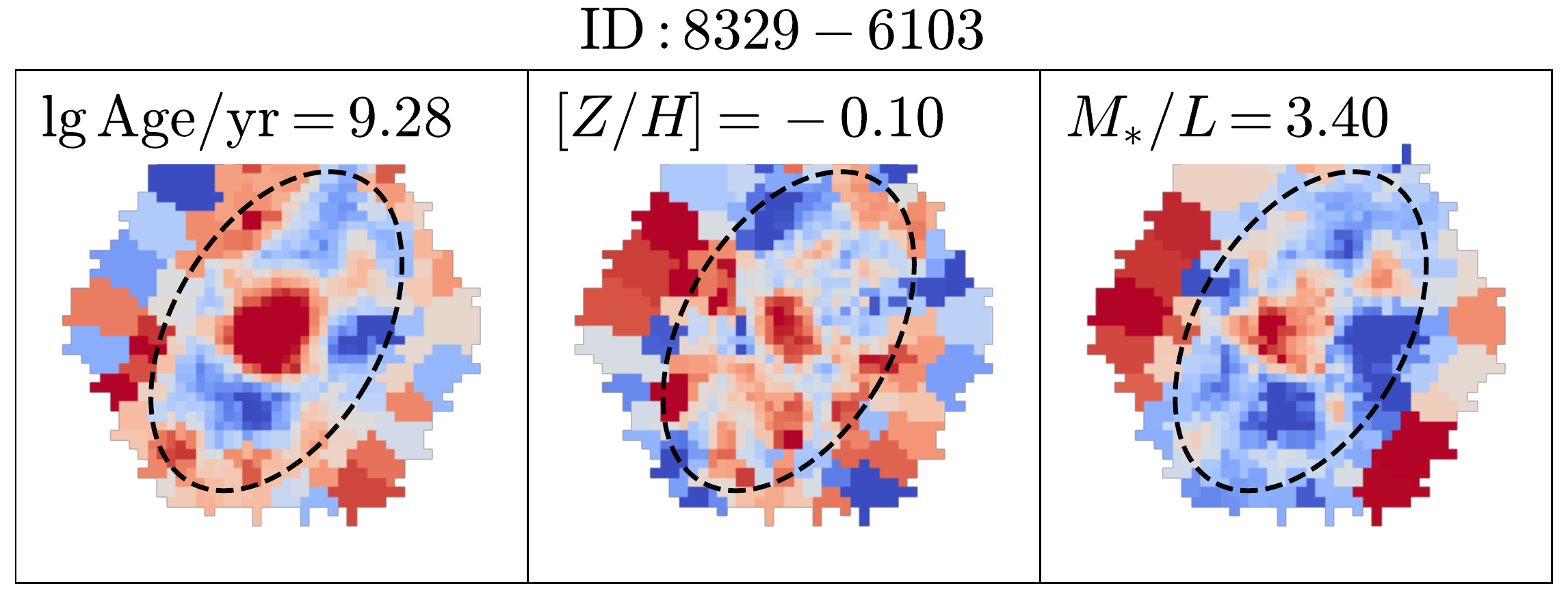}
\includegraphics[width=0.68\columnwidth]{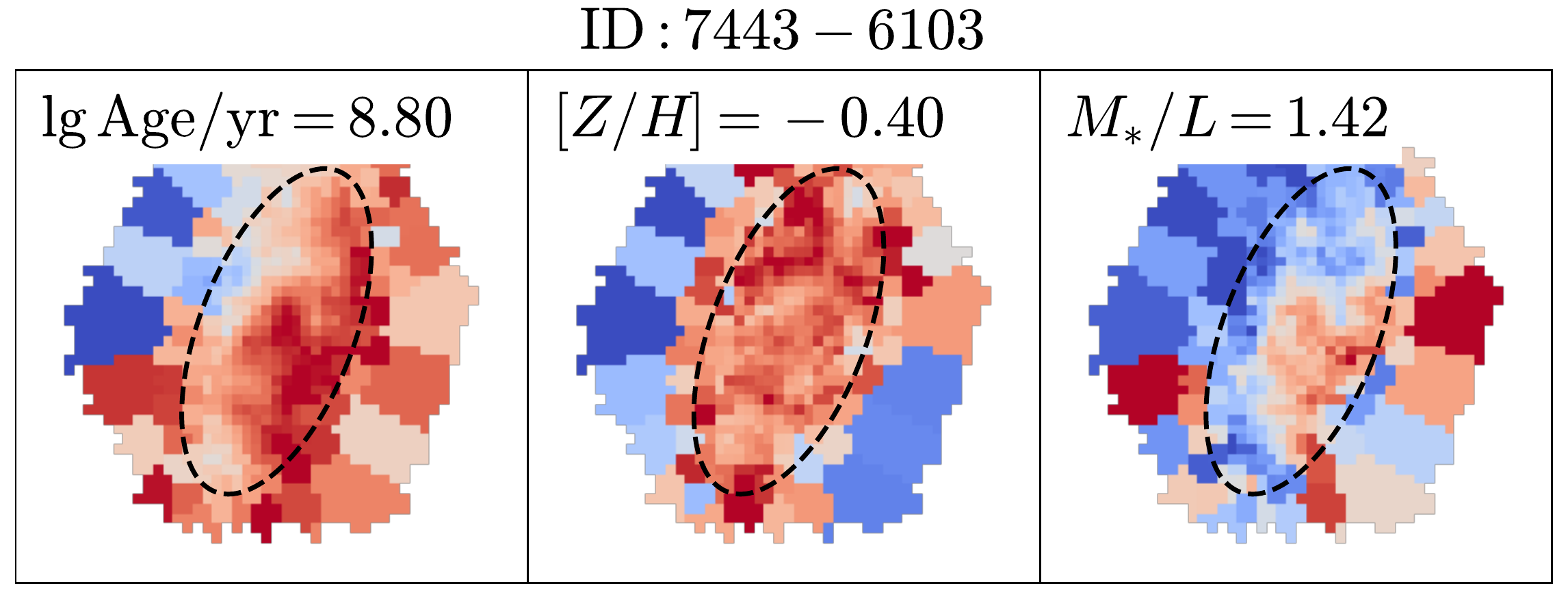}

\includegraphics[width=0.68\columnwidth]{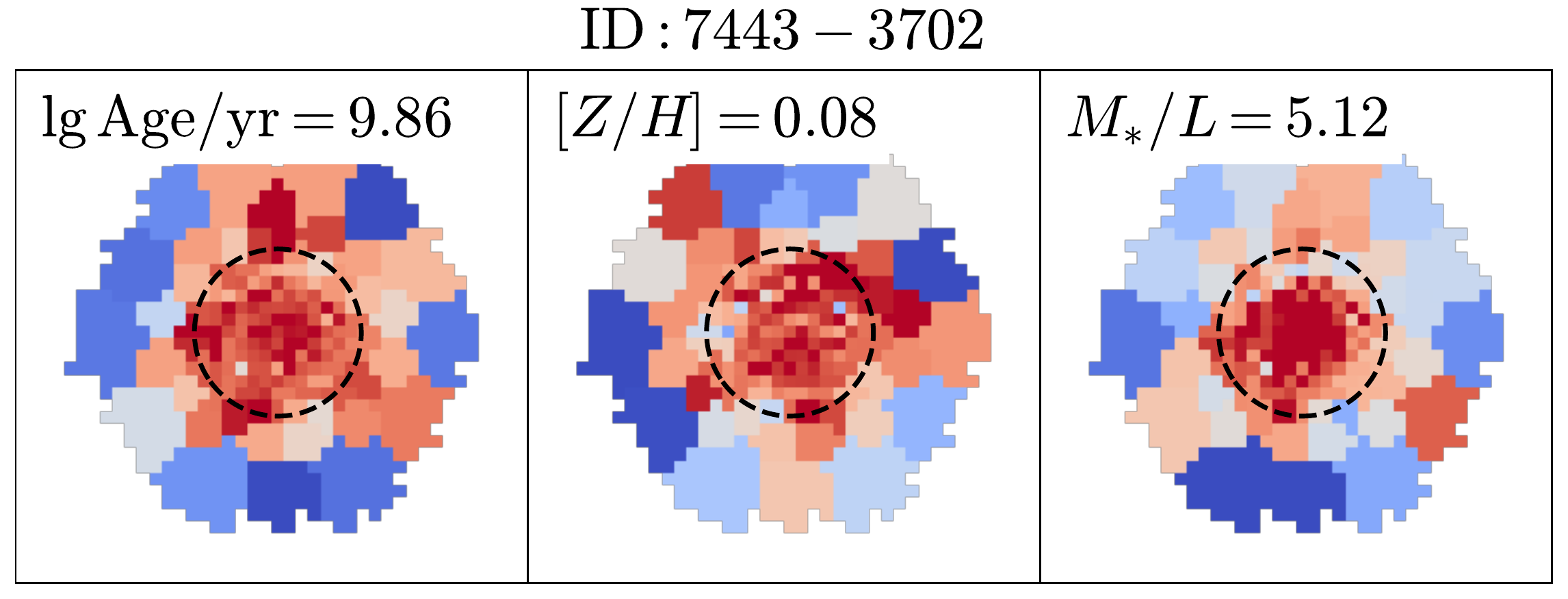}
\includegraphics[width=0.68\columnwidth]{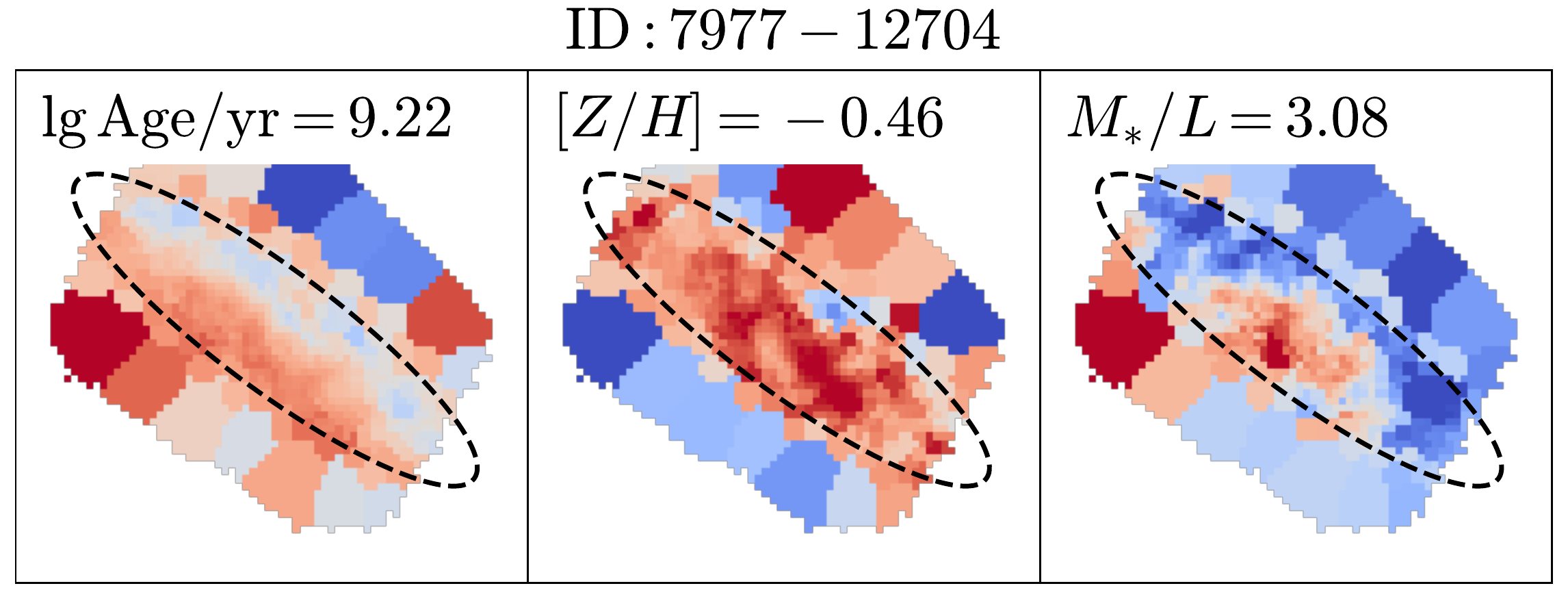}
\includegraphics[width=0.68\columnwidth]{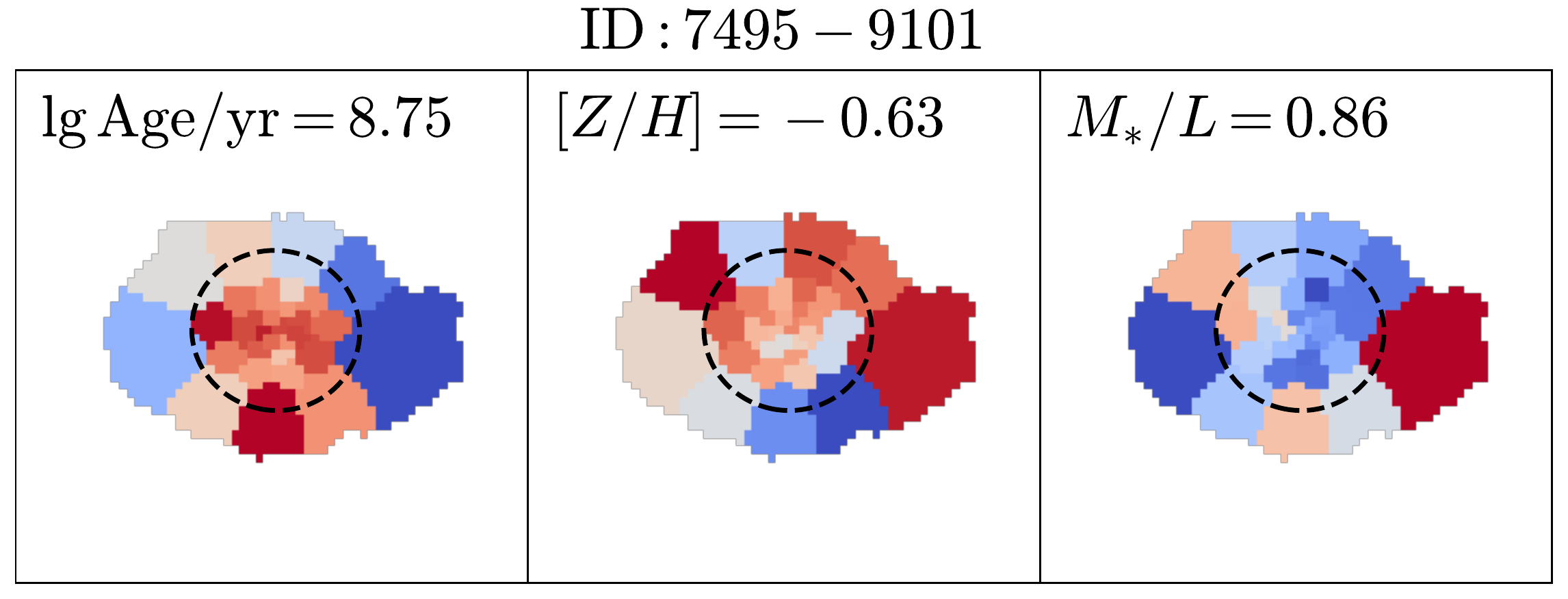}

\includegraphics[width=0.68\columnwidth]{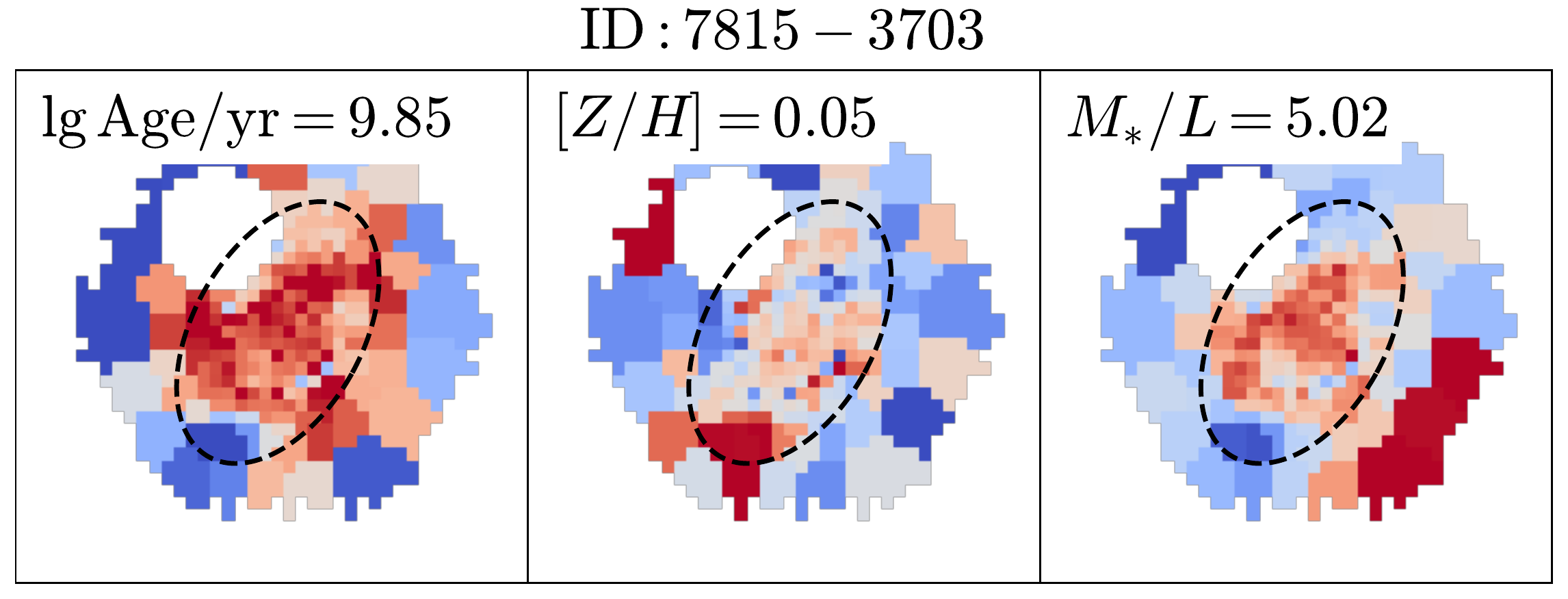}
\includegraphics[width=0.68\columnwidth]{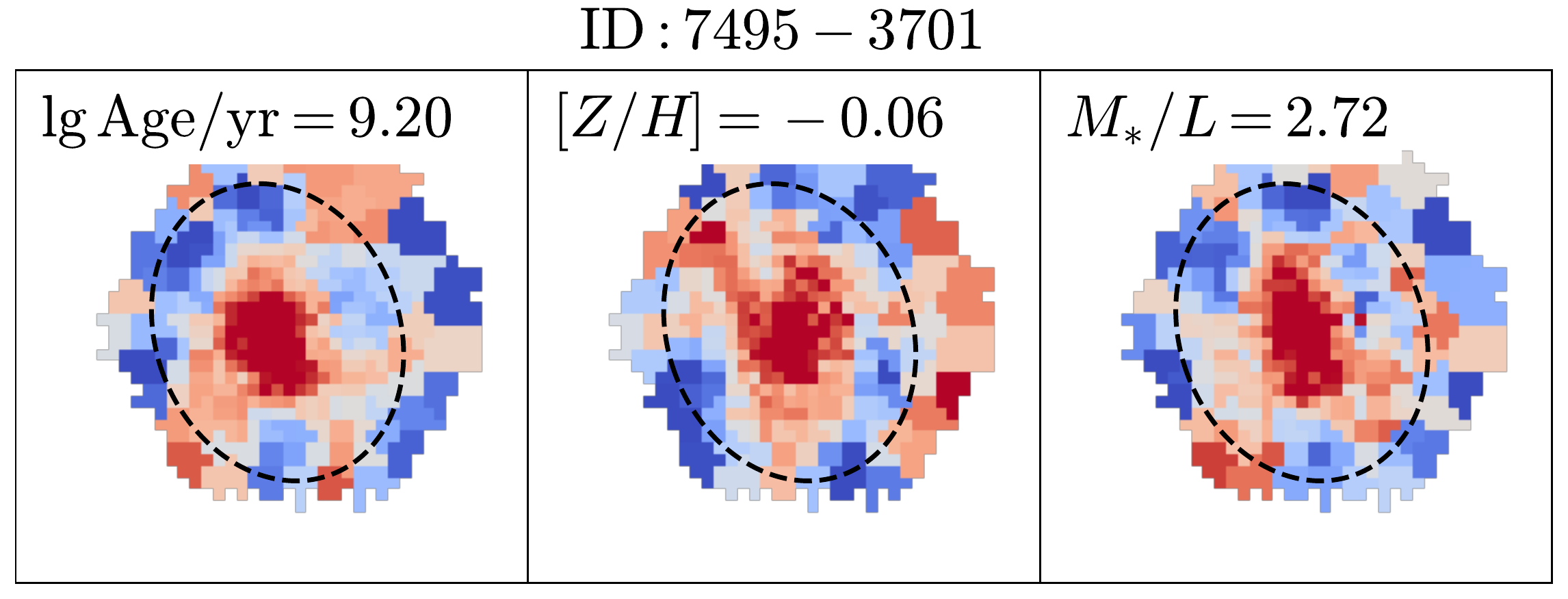}
\includegraphics[width=0.68\columnwidth]{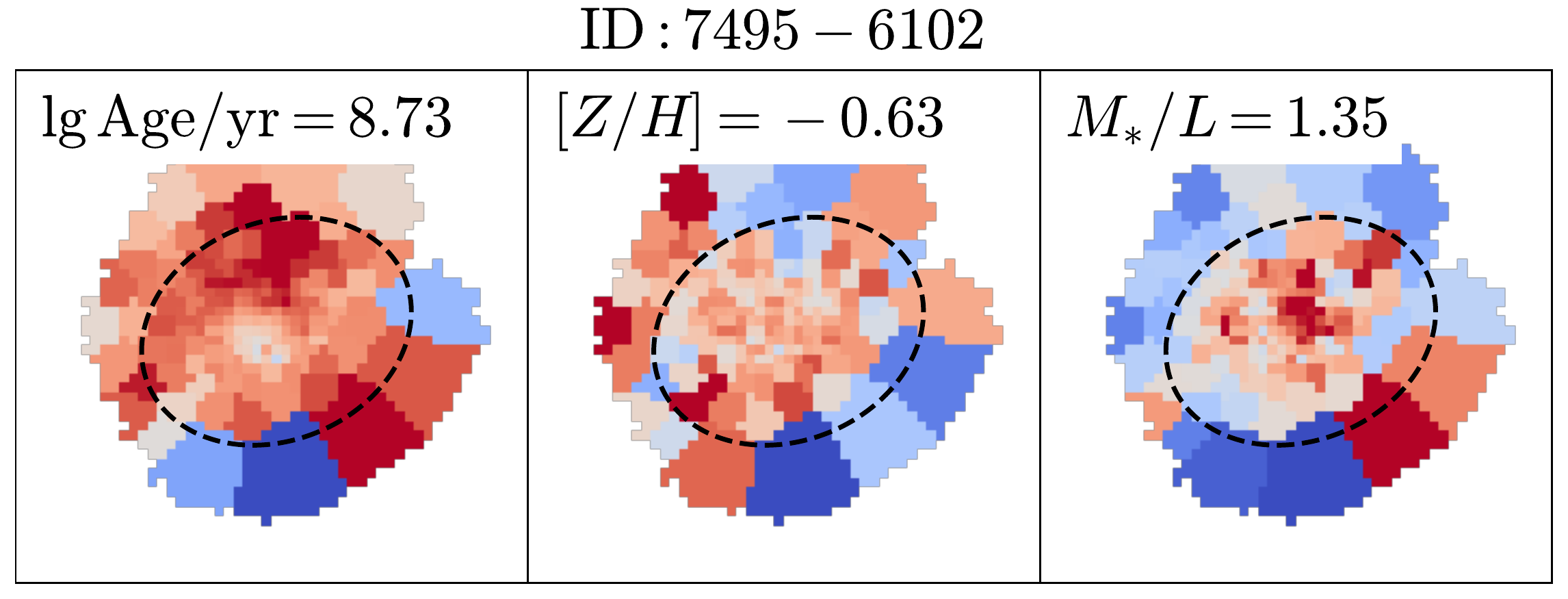}

\caption{Examples of stellar population maps for 10 old galaxies (the left column), 10 galaxies of intermediate age (the middle column), and 10 star-forming galaxies (the right column). For each galaxy, we show their maps of luminosity-weighted age, metallicity, and stellar mass-to-light ratio from left to right, with their global stellar population property values shown in each panel. Redder colour corresponds to older age, higher metallicity, and larger stellar mass-to-light ratio. The black dashed ellipse in each panel represents the elliptical half-light isophote.}
\label{fig:maps}
\end{figure*}

In \autoref{fig:maps}, we present the maps of stellar population properties (luminosity-weighted age, metallicity, and stellar mass-to-light ratio) of 30 examples (10 old galaxies, 10 galaxies with intermediate age, and 10 young galaxies). We note here that in principle the global stellar population properties can also be obtained by calculating the luminosity-weighted values of these properties within the half-light isophote on the spatially resolved population maps. Although there are discrepancies between the global stellar population properties within the half-light isophote derived from these two different ways, we confirm that our main results do not change with the definition of global stellar population properties.

\section{Global stellar population properties}
\label{sec:global_sp}

Stellar population properties have already been found to strongly correlate with structural and dynamical properties of galaxies, such as galaxy mass or velocity dispersion (e.g. \citealt{McDermid_et_al.(2015),Scott_et_al.(2017),Li_et_al.(2018)}; see \citealt[][sec.~4.3]{Cappellari(2016)} for a review). In this section, we will present the correlations of galaxy (global) stellar population properties, including global age ($\lg\,\mathrm{Age}$), metallicity ($[Z/H]$), and stellar mass-to-light ratio ($M_{\ast}/L$), against structural and dynamical properties of galaxies for the MaNGA sample. 

\subsection{One-dimensional relation of stellar populations}
\label{sec:1d_sp_relation}
\begin{figure*}
\centering
\includegraphics[width=2\columnwidth]{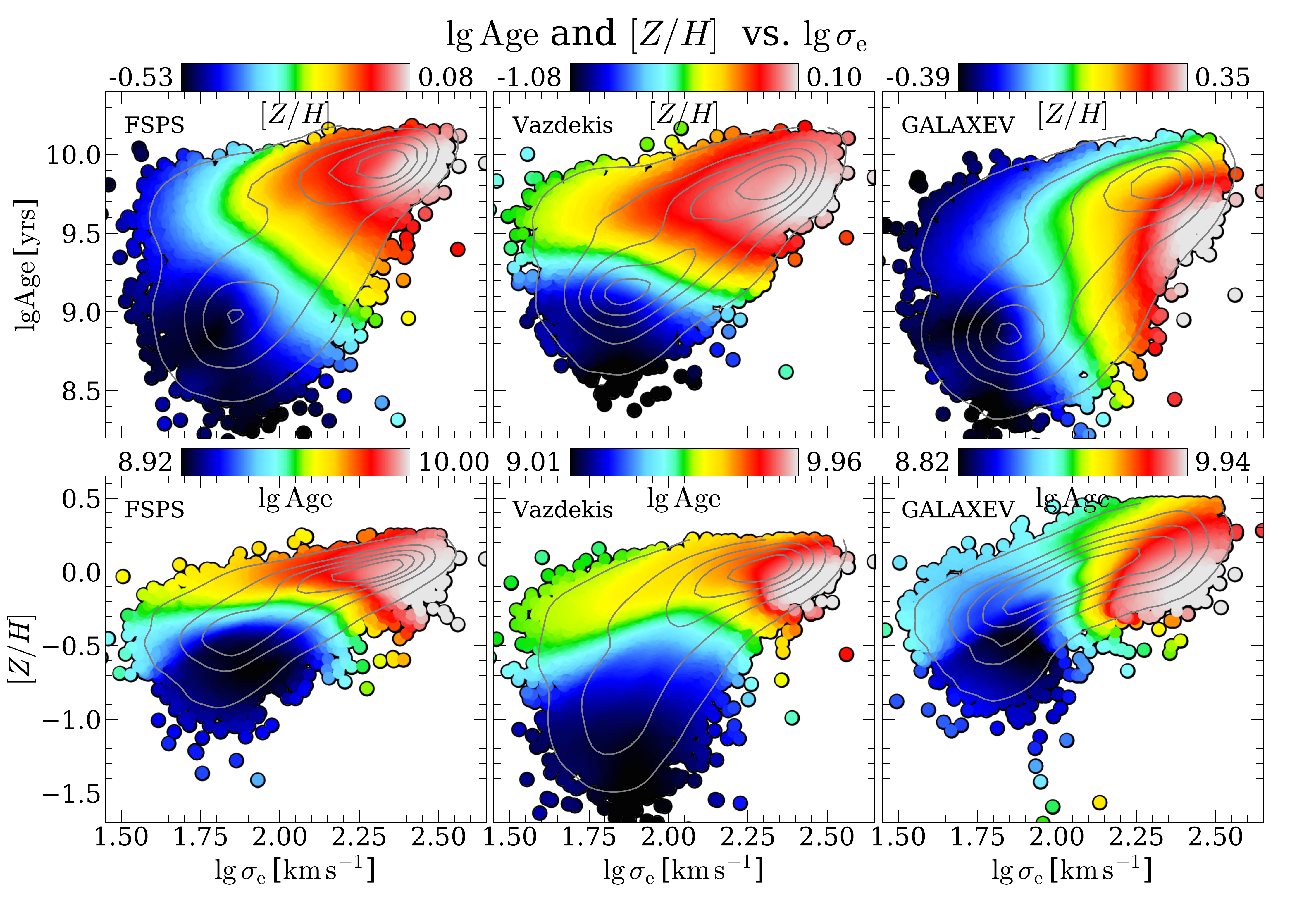}
\caption{Top: the distribution of global metallicity $[Z/H]$ on the $(\sigma_{\rm e},\mathrm{Age})$ plane for the {\sc fsps}, Vazdekis, and {\sc galaxev} models from left to right. Bottom: the distribution of age $\lg\,\mathrm{Age}$ on the $(\sigma_{\rm e},[Z/H])$ plane for the three stellar libraries. In each panel, a kernel density estimation of the galaxy number density is indicated by the grey contours.}
\label{fig:sp2sigma}
\end{figure*}

In \autoref{fig:sp2sigma}, we first present the one-dimensional correlation between galaxy stellar population properties (i.e. age and metallicity) and the velocity dispersion for all the three SSP models (i.e. the {\sc fsps}, the Vazdekis, and the {\sc galaxev} models; see \autoref{sec:libraries} for details of the three models). As shown in the figure, both age and metallicity show increasing trend with velocity dispersion for all the three SSP models, consistent with many previous studies (e.g. \citealt{Thomas2010,Kuntschner2010,McDermid_et_al.(2015),Scott_et_al.(2017),Li_et_al.(2018)}). On the $(\sigma_{\rm e},\mathrm{Age})$ plane (top panels of \autoref{fig:sp2sigma}), galaxies show clear bimodal distribution for all the three SSP models, classifying the galaxies into the well-known ``red sequence'' (old and massive), ``blue cloud'' (young and low-mass), and ``green valley'' (intermediate age and intermediate mass) groups (e.g. \citealt{Schawinski2014}). It is present in all the three SSP models that the red sequence galaxies have relatively shallower $\sigma_{\rm e}-\mathrm{Age}$ relation, while blue cloud galaxies have a steeper one. This is consistent with the findings in \citet{Li_et_al.(2018)}, in which early-type galaxies are seen to have smaller slope of the relation between velocity dispersion and galaxy age than spiral galaxies (see fig.~5 therein). All these confirms the robustness of our stellar population synthesis results. 

Importantly, we find that metallicity shows obvious variation on the $(\sigma_{\rm e},\mathrm{Age})$ plane but it is not only driven by $\sigma_{\rm e}$. At fixed $\sigma_{\rm e}$, metallicity still increases with increasing age, especially for galaxies with $\lg\,(\sigma_{\rm e}/\mathrm{km\,s^{-1}}) \lesssim 2.3$. This strong decrease in metallicity with decreasing ages at fixed $\sigma_{\rm e}$ was also seen in a study at redshift $z\approx0.8$ by \citet[fig.~6]{Cappellari2020}, which included both photometry and spectroscopy. The generality of this result, with completely different data and samples, confirms its robustness. The {\sc fsps} and Vazdekis models show similar patterns of metallicity on the $(\sigma_{\rm e},\mathrm{Age})$ plane, with the {\sc fsps} model showing more galaxies with younger age than the Vazdekis model. This is because that the age of the youngest stellar template in {\sc fsps} is $\lg\,\mathrm{Age/yr}=6$, significantly younger than that of the Vazdekis model (with the youngest age being $\lg\,\mathrm{Age/yr}=7.8$). Besides, the Vazdekis model seems to provide too low metallicities due to the Age--Metallicity degeneracy \citep{Worthey(1994)}: the fit tries to compensate for the lack of sufficiently young models by lowering the metallicity, which produces, to first order, a similar effect as lowering the age. The {\sc galaxev} model, however, seems to be problematic and does not show similar metallicity pattern on the $(\sigma_{\rm e},\mathrm{Age})$ plane as the {\sc fsps} and Vazdekis models do: galaxies with intermediate age ($\lg\,\mathrm{Age/yr}\sim 9.5$) with {\sc galaxev} model appear to have the highest metallicity, while for the {\sc fsps} and Vazdekis models, galaxies in the red sequence have the highest metallicity. This inconsistency is also seen in the bottom panels of \autoref{fig:sp2sigma}, where we show the correlation between $[Z/H]$ and $\lg\,\sigma_{\rm e}$, colour-coded by galaxy age. As can be seen, the {\sc fsps} and Vazdekis models show similar age distribution on the $(\sigma_{\rm e},[Z/H])$ plane, with the Vazdekis model having more galaxies with lower metallicity, due to the lower metallicity boundary of the Vazdekis model (see \autoref{fig:libs}). Besides, galaxy age show stronger correlation with metallicity than with $\sigma_{\rm e}$ as we can see that the age varies roughly along the direction of metallicity on the $(\sigma_{\rm e},[Z/H])$ plane. The {\sc galaxev} model again shows difference with the {\sc fsps} and Vazdekis models with metallicity being more strongly correlated with $\sigma_{\rm e}$. The reason for the significantly different behaviour of the \textsc{galaxev} model would require further studies, but highlights the importance on not relying on a single set of modelling assumptions. Considering the problematic results of the {\sc galaxev} model and the lack of youngest templates of the Vazdekis model, we take the {\sc fsps} model as our default stellar library in the following sections. We note here that to fully understand the main driving factor of the stellar population variations as shown in \autoref{fig:sp2sigma}, as well as the difference between the stellar libraries, a principal component analysis (PCA) or a partial correlation analysis may be helpful, which is, however, beyond the scope of this paper. We encourage the interested readers to perform those analyses if necessary.

\begin{figure}
\centering
\includegraphics[width=1\columnwidth]{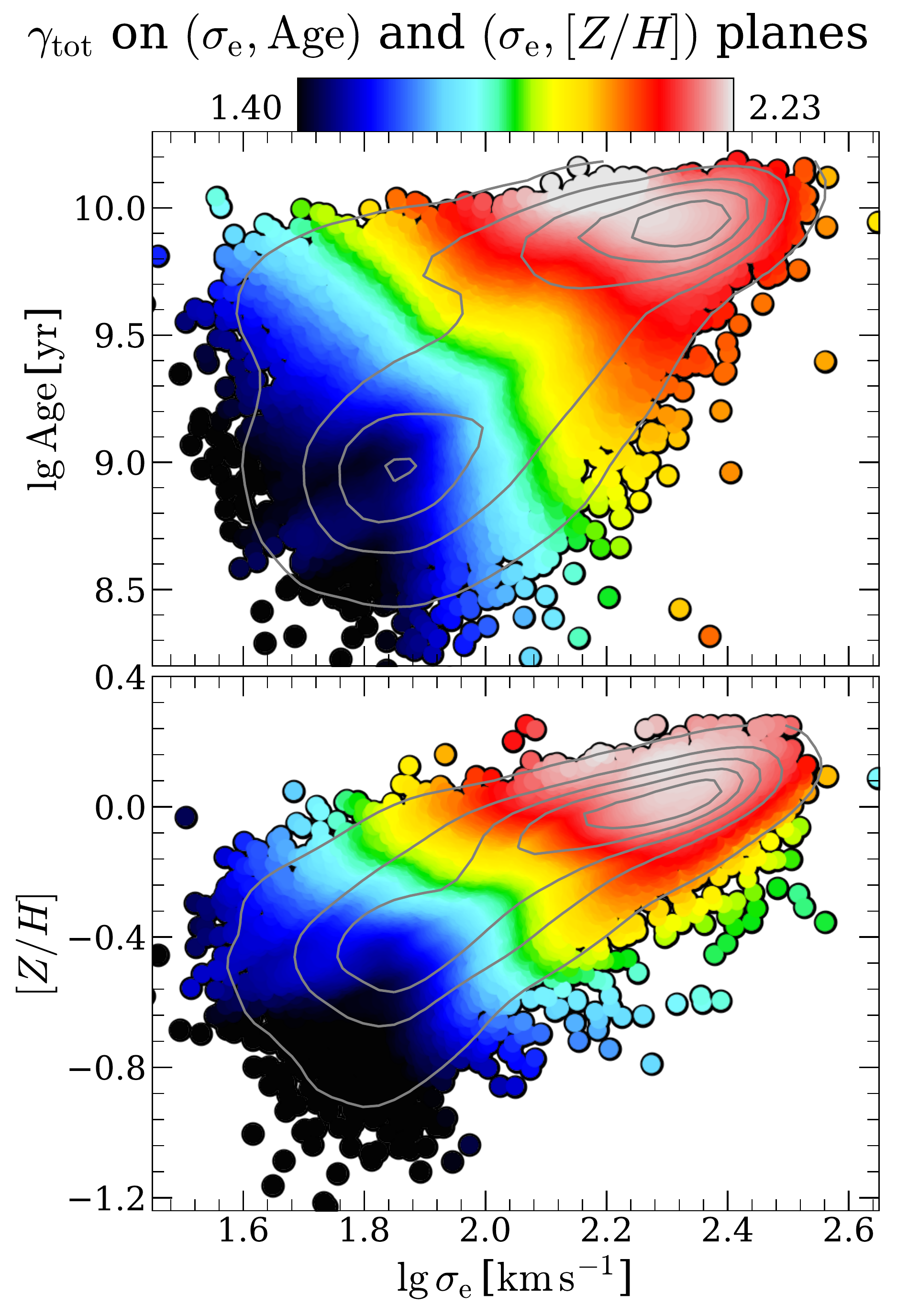}
\caption{Distribution of the dynamically-determined total density slope on $(\sigma_{\rm e},\mathrm{Age})$ (top) and $(\sigma_{\rm e},[Z/H])$ (bottom) planes. Age and metallicity used here are derived with {\sc fsps} models (see \autoref{sec:libraries} for details of stellar models). In each panel, a kernel density estimation of the galaxy number density is indicated by the grey contours.}
\label{fig:gammatot}
\end{figure}

In \autoref{fig:gammatot}, we present the distribution of total density slope of galaxies (see \autoref{sec:jam_props} for definition) on both the $(\sigma_{\rm e},\mathrm{Age})$ and the $(\sigma_{\rm e},[Z/H])$ planes. As can be seen, despite some subtle discrepancies, the total density slope shows an overall similar distribution as galaxy age (metallicity) on the two planes: galaxies with higher $\sigma_{\rm e}$, older age, and higher metallicity typically have larger total density slopes (i.e. steeper total mass profiles). The similar pattern on the two planes indicates a strong correlation between galaxy mass density slopes and stellar population properties, consistent with \citet[][fig.~22]{Cappellari(2016)}, where the total density slope of galaxies shows similar trend as stellar population properties on the mass-size plane. Interestingly, the total density slope is still seen to vary with age and metallicity at fixed $\sigma_{\rm e}$, indicating that the correlation between stellar population and density slope is not solely driven by $\sigma_{\rm e}$, consistent with the findings in \citet{Lu_et_al.(2020)}.

\subsection{Stellar population on two-dimensional planes}
\label{sec:2d_sp_relation}
Apart from the one-dimensional correlation between galaxy stellar population properties and galaxy velocity dispersion, we also show the distributions of stellar population properties on two-dimensional dynamical/structural phase spaces. 

\begin{figure*}
\centering
\includegraphics[width=2\columnwidth]{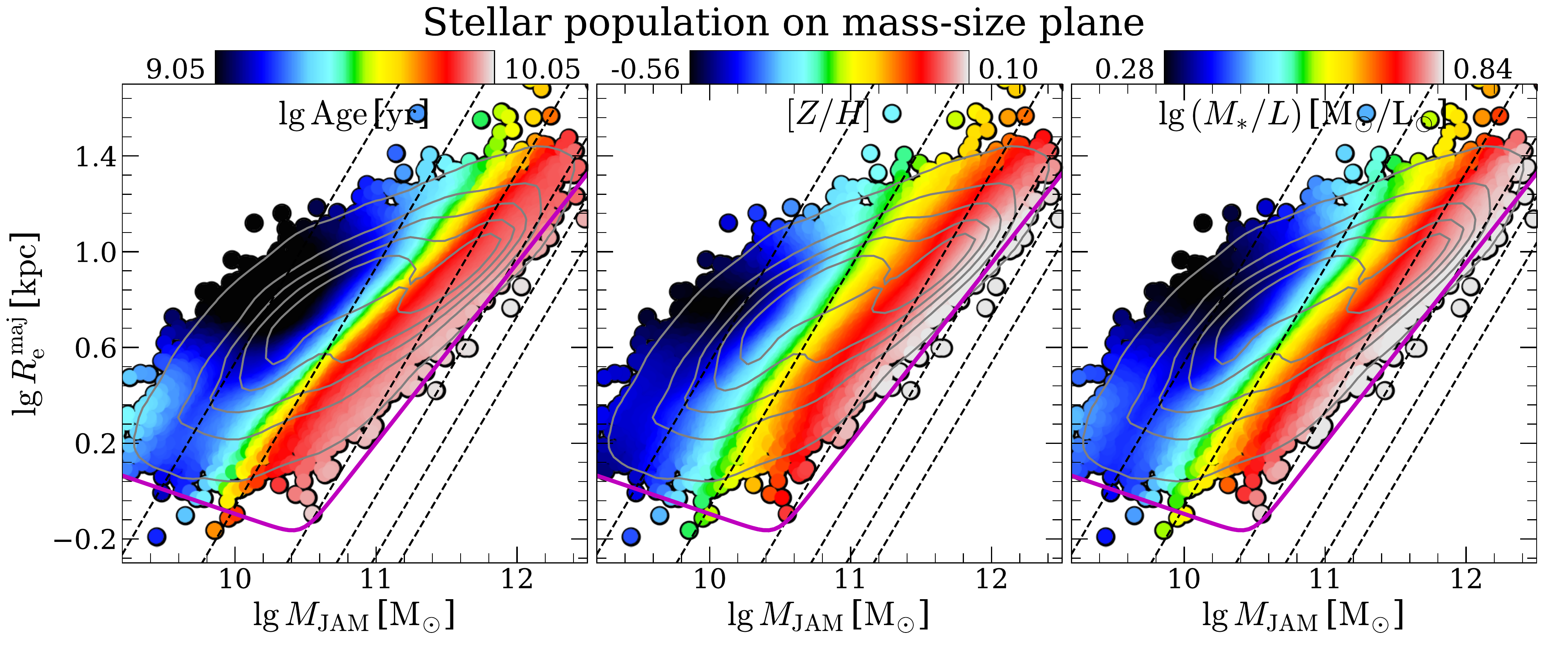}
\caption{Distributions of age (left), metallicity (right), and stellar mass-to-light ratio (right) on the mass-size plane ($\lg\,R_{\rm e}^{\rm maj}$ versus $\lg\,M_{\rm JAM}$, see \autoref{sec:jam_props} for definitions of the two parameters). In each panel, the black dashed lines indicate the lines of constant $\sigma_{\rm e}$: 50, 100, 200, 300, 400, and 500 $\mathrm{km\,s^{-1}}$. The magenta curve is the zone of exclusion defined in \citet{Cappellari_et_al.(2013b)}. The galaxy number density is indicated by the grey contours.}
\label{fig:sp_mass_size}
\end{figure*}

In \autoref{fig:sp_mass_size}, we show the distributions of galaxy age, metallicity, and stellar mass-to-light ratio on the mass-size plane, $(M_{\rm JAM},R_{\rm e}^{\rm maj})$ (see \autoref{sec:jam_props} for definitions of the two parameters). As shown in the figure, age, metallicity, and stellar mass-to-light ratio of galaxies all show systematic variations on the mass-size plane, ``roughly'' along the direction of the virial estimate of the velocity dispersion $\sigma_{\rm e}$, which accurately follows lines of $R_{\rm e}^{\rm maj}\propto M_{\rm JAM}$ \citep{Cappellari_et_al.(2013a)}. This is consistent with the positive correlation between stellar population properties and $\sigma_{\rm e}$ seen from \autoref{fig:sp2sigma}. 

In previous studies, the stellar velocity dispersion $\sigma$ was found to be a better predictor of the galaxies $M/L$, which traces their stellar population, than stellar mass \citep{Cappellari2006}. Later, it was found to better predict galaxy colours \citep{Franx2008} and stellar population from single fibres \citep{Graves2009}. These early results were significantly strengthened by numerous studies using dynamical masses from integral field spectroscopy (IFS), all confirming that stellar population is better predicted by $\sigma$ than $M_{\ast}$. This was pointed out with ATLAS$^{\rm 3D}$ \citep{McDermid_et_al.(2015),Cappellari(2016)}, SAMI \citep{ Scott_et_al.(2017),Barone2018,Barone2020} and MaNGA \citep{Li_et_al.(2018)}, and even at significant redshift $z\approx0.8$ \citep{Cappellari(2023)}. 

Here we confirm again the general result, but we also find that the variation of stellar population properties are not strictly parallel with the constant-$\sigma_{\rm e}$ lines. At fixed $\sigma_{\rm e}$, more massive galaxies appear to be slightly younger, more metal-poor, and have lower stellar mass-to-light ratio than the less massive galaxies. This is consistent with the findings of \citet{Lu_et_al.(2020)}, where the correlations between stellar population properties and other dynamical properties (i.e. the slope of galaxy velocity dispersion) at fixed $\sigma_{\rm e}$ are firmly observed. It means that mass density gradients (indicated by the velocity dispersion gradients) contain information of the stellar population which is not fully accounted for by $\sigma_{\rm e}$. In addition, metallicity appear to have a better parallel correlation between constant-metallicity lines and constant-$\sigma_{\rm e}$ lines than age, also consistent with the smaller scatter of $\sigma_{\rm e}-[Z/H]$ relation than that of the $\sigma_{\rm e}-\mathrm{Age}$ relation in \autoref{fig:sp2sigma}. It is worth pointing out, however, that the detailed location of galaxies on the mass-size diagram depends on how these quantities are measured, and one should be careful with interpreting subtle differences.

\begin{figure*}
\centering
\includegraphics[width=2\columnwidth]{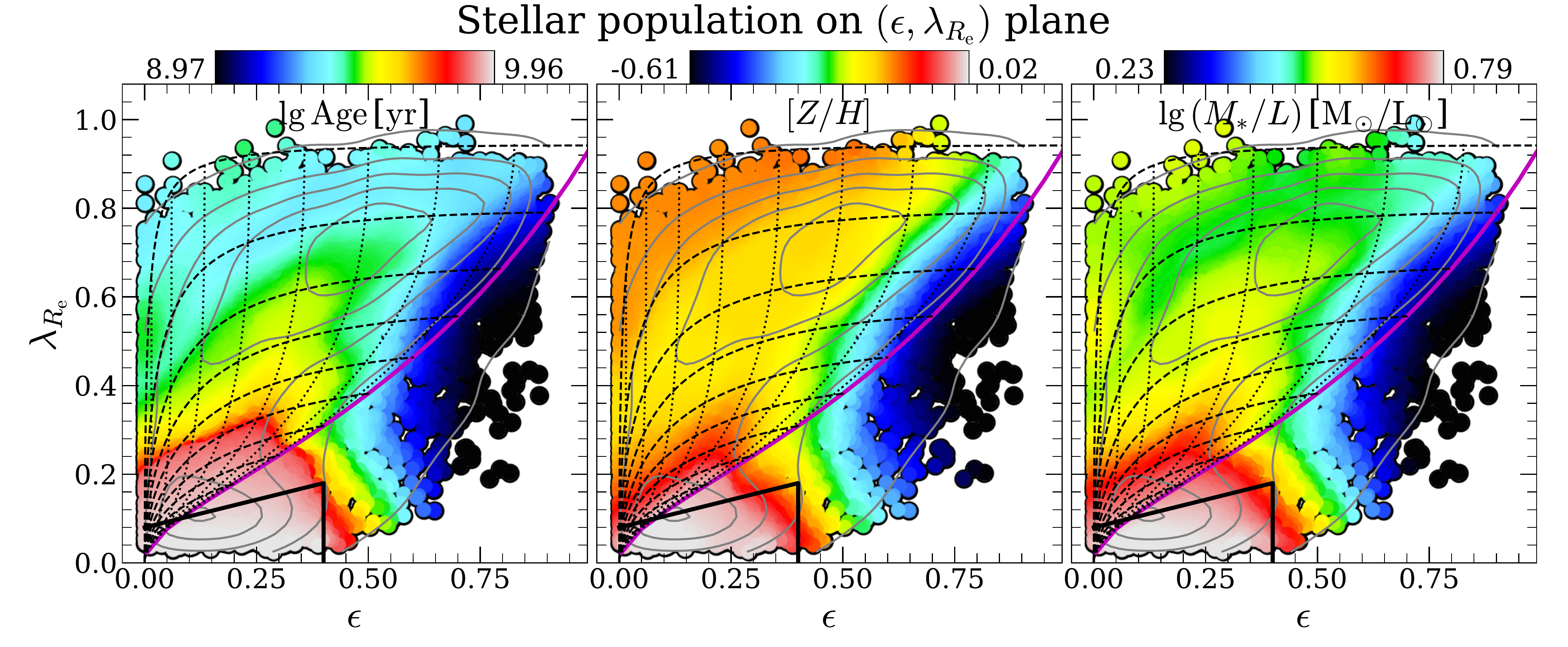}
\caption{Distributions of age (left), metallicity (right), and stellar mass-to-light ratio (right) on the $(\epsilon,\lambda_{R_{\rm e}})$ plane (see \autoref{sec:jam_props} for definitions of the two parameters). The symbols are the same as \autoref{fig:dust}.}
\label{fig:sp_lambda_epsilon}
\end{figure*}

In \autoref{fig:sp_lambda_epsilon}, we show the distribution of stellar population properties on the $(\epsilon,\lambda_{R_{\rm e}})$ plane. As can be seen, slow-rotating galaxies (within the region of black lines) tend to be the oldest among all the galaxies and galaxy age show obvious decreasing trend with increasing $\lambda_{R_{\rm e}}$ (i.e. become younger), consistent with \citet{van_de_Sande_et_al.(2018)}, in which the age distribution of 843 galaxies from the SAMI Galaxy Survey on the $(\epsilon,(V/\sigma)_{\rm e})$ plane is studied\footnote{$(V/\sigma)_{\rm e}-\epsilon$ relation is physically equivalent to $\lambda_{R_{\rm e}}-\epsilon$ relation. $(V/\sigma)_{\rm e}$ and $\lambda_{R_{\rm e}}$ can be converted to each other using the empirical calibration of \citet{Emsellem_et_al.(2007),Emsellem_et_al.(2011)}.}. That is because rotation-dominated galaxies are more likely to be disk galaxies on the star-formation main sequence, as shown by \citet[fig.~2]{Wang2020} for MaNGA galaxies (see also \citealt[][fig.~27]{Fischer_et_al.(2019)}).

We also investigate the distributions of stellar metallicity and $M_{\ast}/L$ of galaxies on the $(\epsilon,\lambda_{R_{\rm e}})$ plane (the middle and right panels of \autoref{fig:sp_lambda_epsilon}), where we also see their systematic variations on the plane, which has never been seen before. We find that faster-rotating galaxies tend to be more metal-poor and have lower $M_{\ast}/L$. This is consistent with the trends we see on the $(M_{\rm JAM},R_{\rm e}^{\rm maj})$ plane in \autoref{fig:sp_mass_size}, given that fast-rotating galaxies are those with smaller bulges and lower $\sigma_{\rm e}$. To first order, the contours of constant ages or metallicity or $M_{\ast}/L$ follow the lines at which the galaxies of given flattening project for different inclination (the black dashed curves). This makes sense as one does not expect inclination to change the properties of the stellar population. However, the diagram also reveals some significant variations along the lines of constant intrinsic flattening. This is likely an effect due to dust and field coverage: for an edge-on view of galaxies, the observed population is dominated by the disk, which is typically more star-forming and thus is younger, more metal-poor, and has lower $M_{\ast}/L$. The effect of inclination on the stellar population is generally ignored, because it is difficult to correct for it, but we show here how one can reveal its importance with the $(\lambda_{R_{\rm e}},\epsilon)$ diagram.

\section{Population property gradients}
\label{sec:sp_gradients}

\begin{figure*}
\centering
\includegraphics[width=0.68\columnwidth]{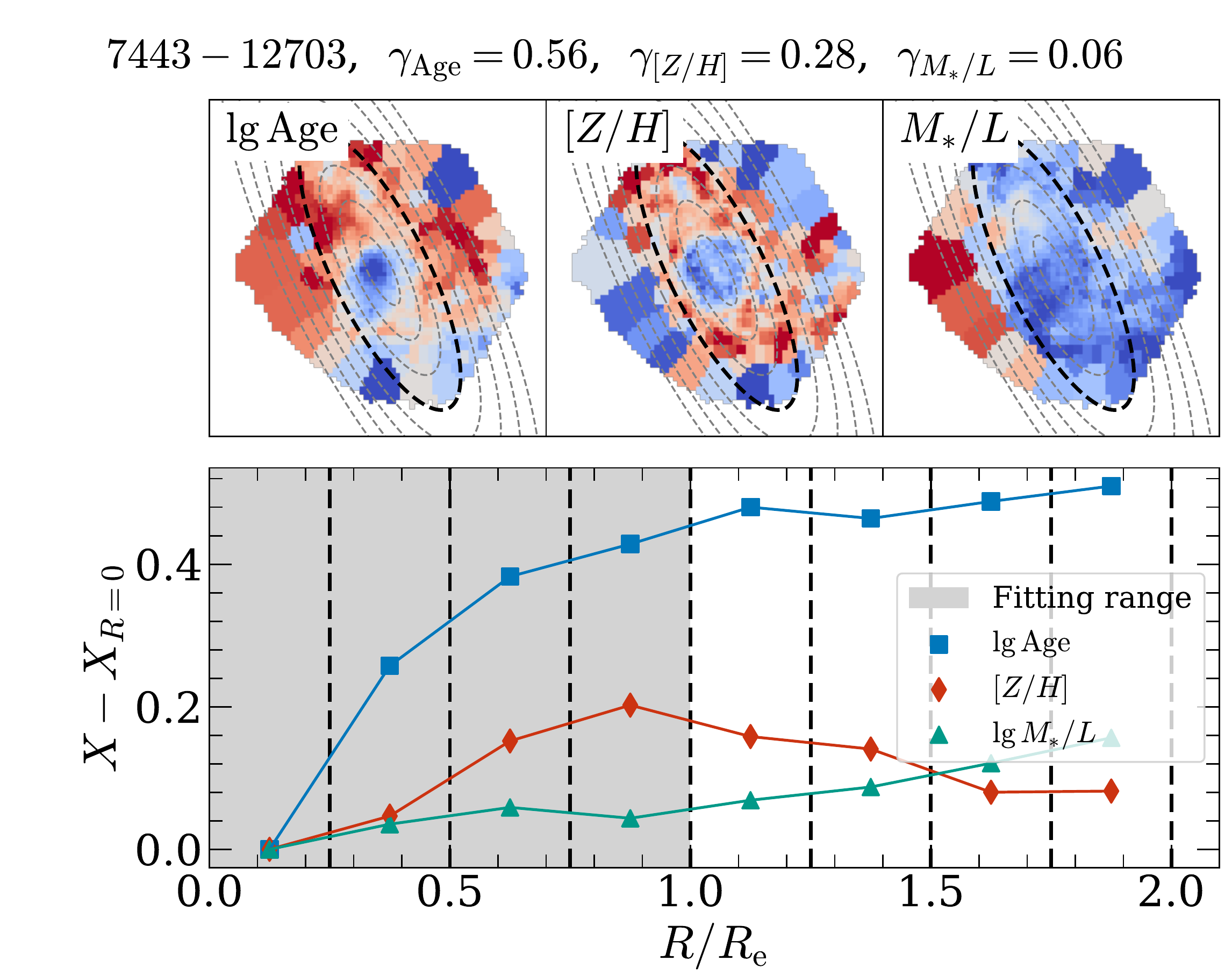}
\includegraphics[width=0.68\columnwidth]{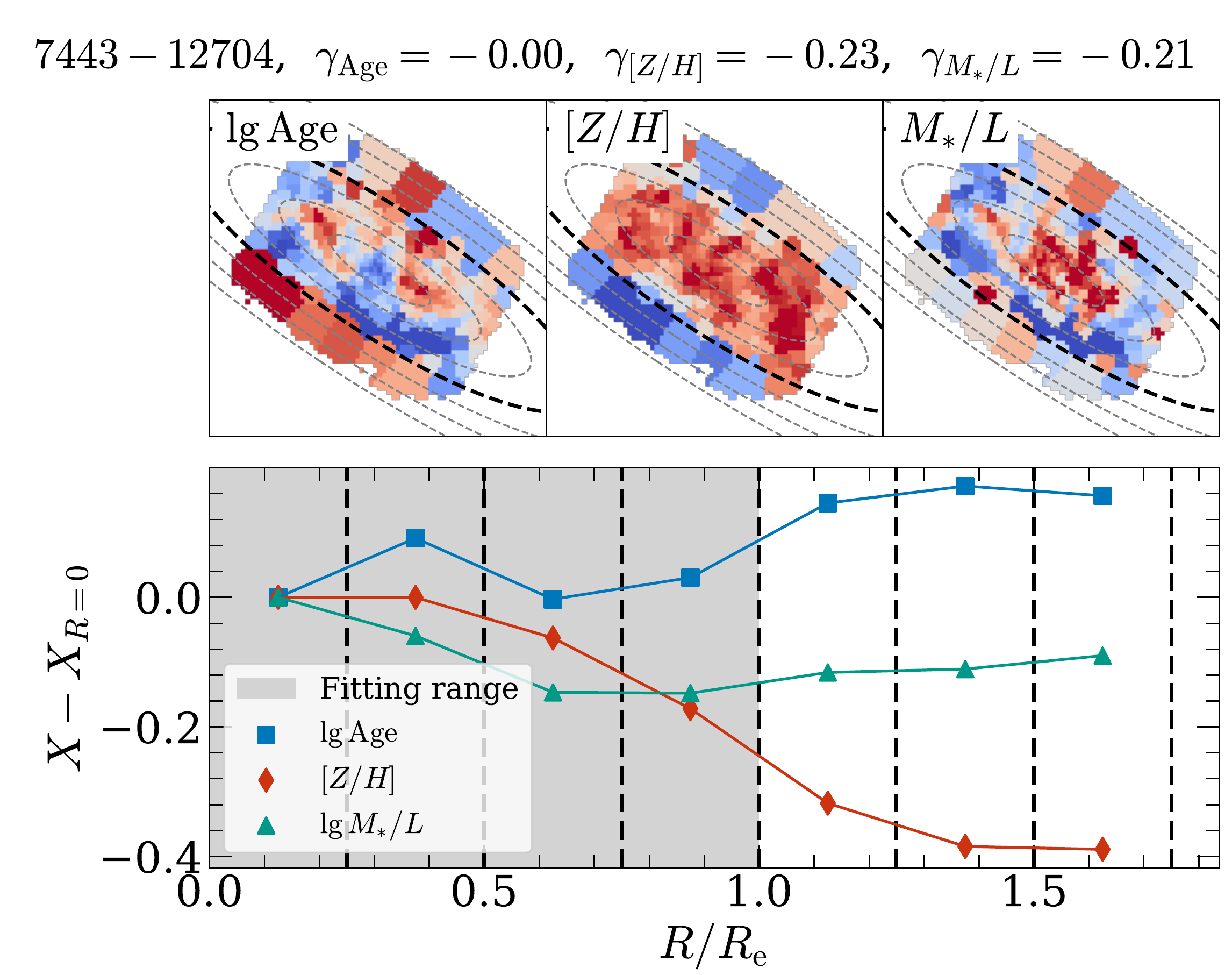}
\includegraphics[width=0.68\columnwidth]{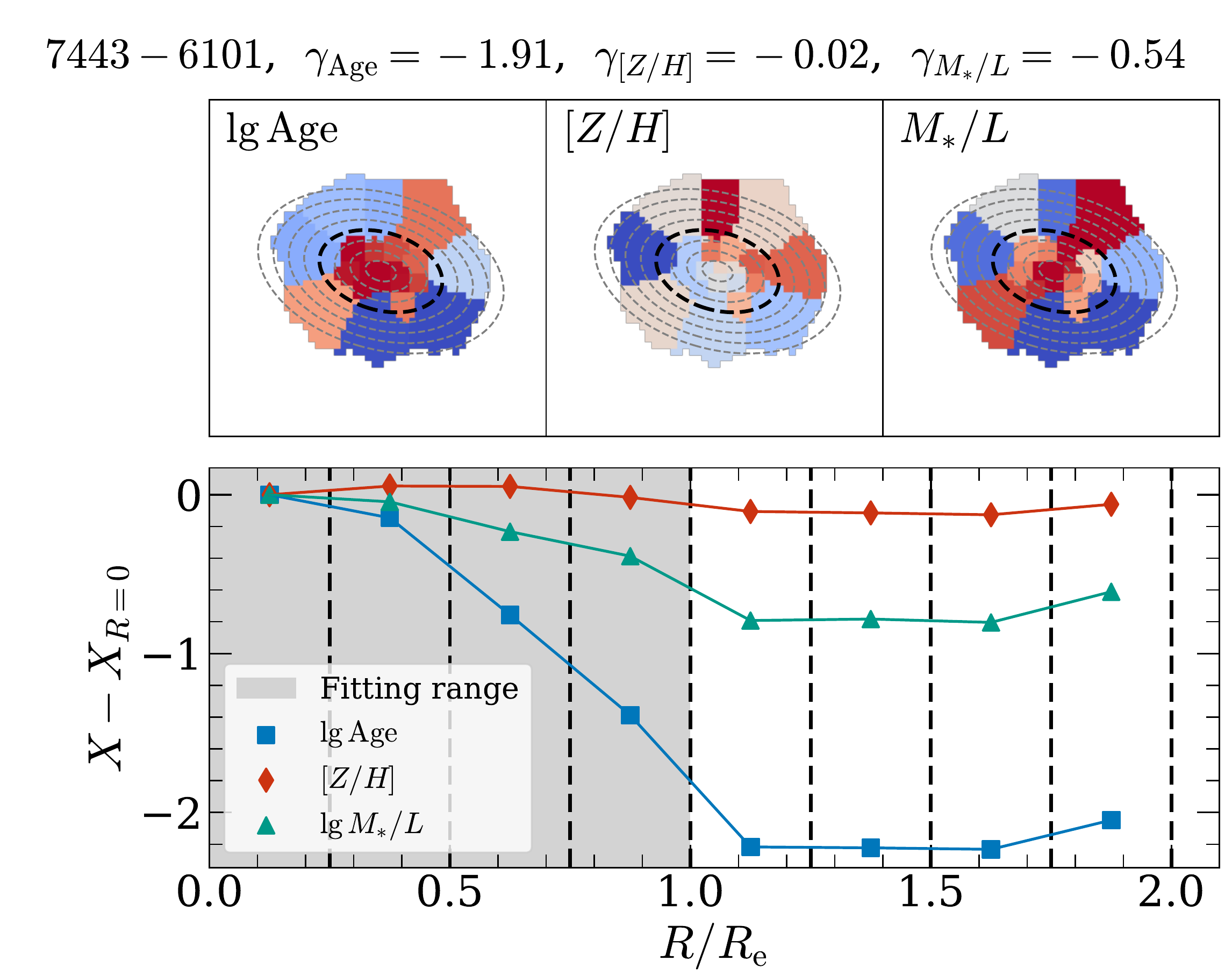}

\includegraphics[width=0.68\columnwidth]{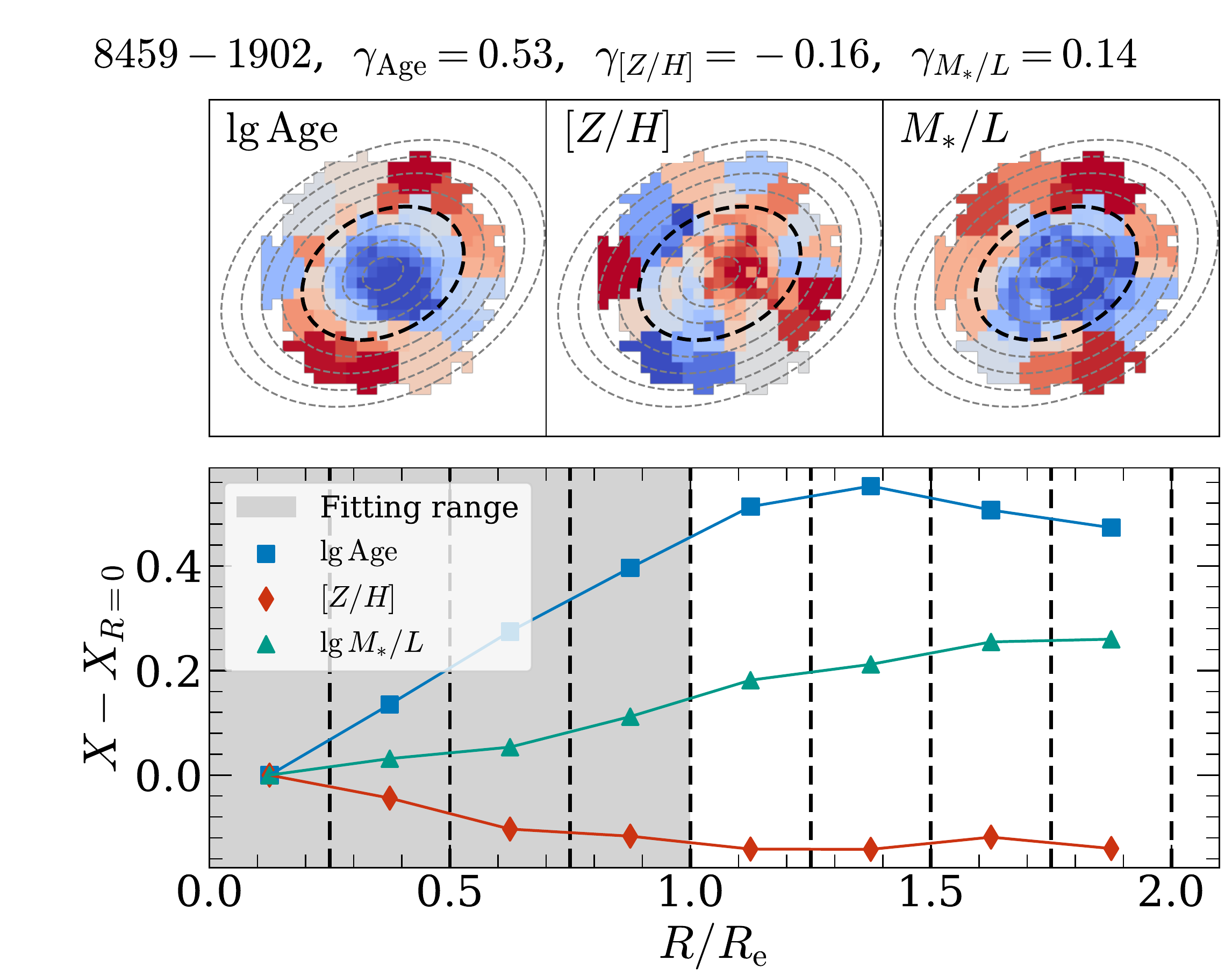}
\includegraphics[width=0.68\columnwidth]{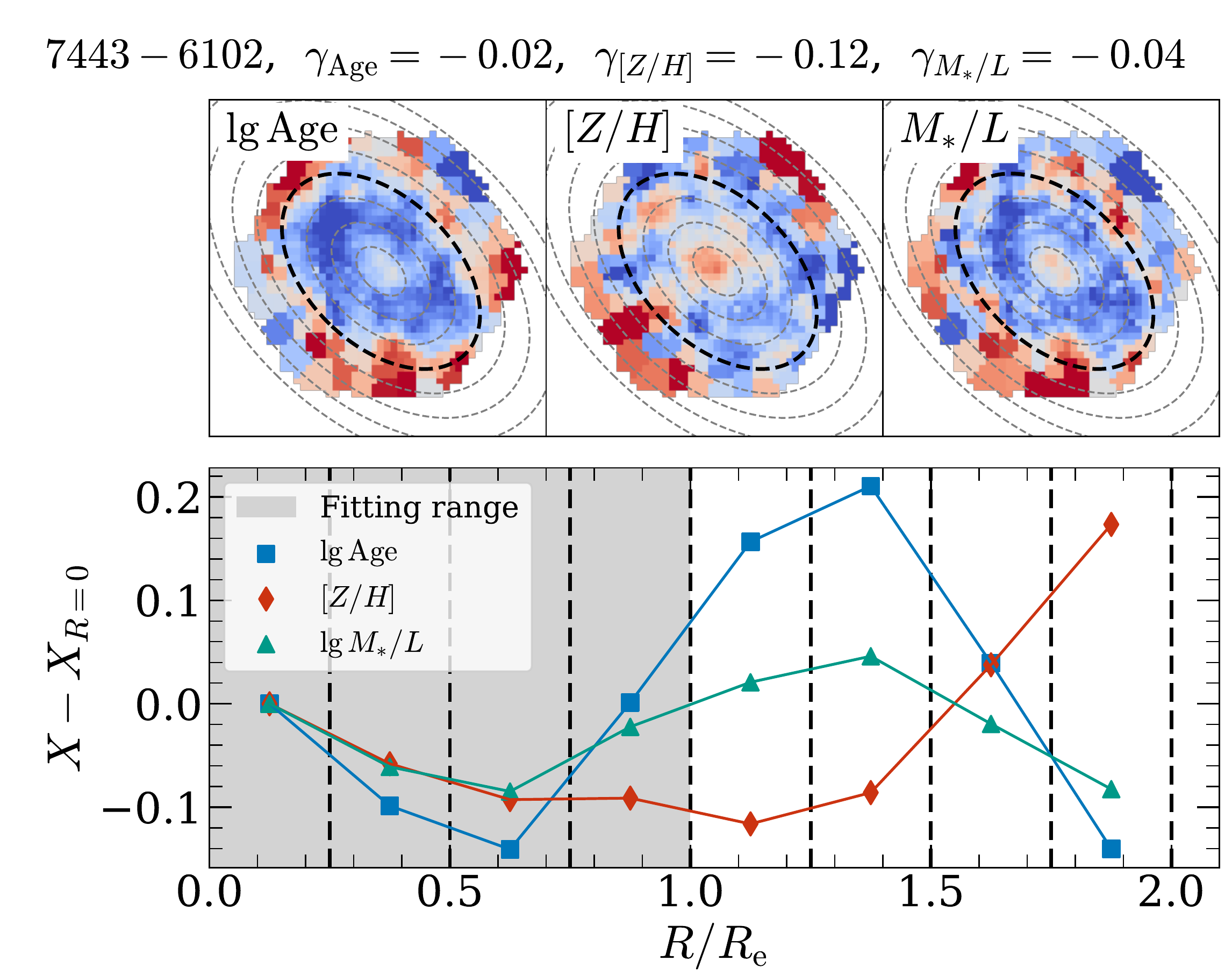}
\includegraphics[width=0.68\columnwidth]{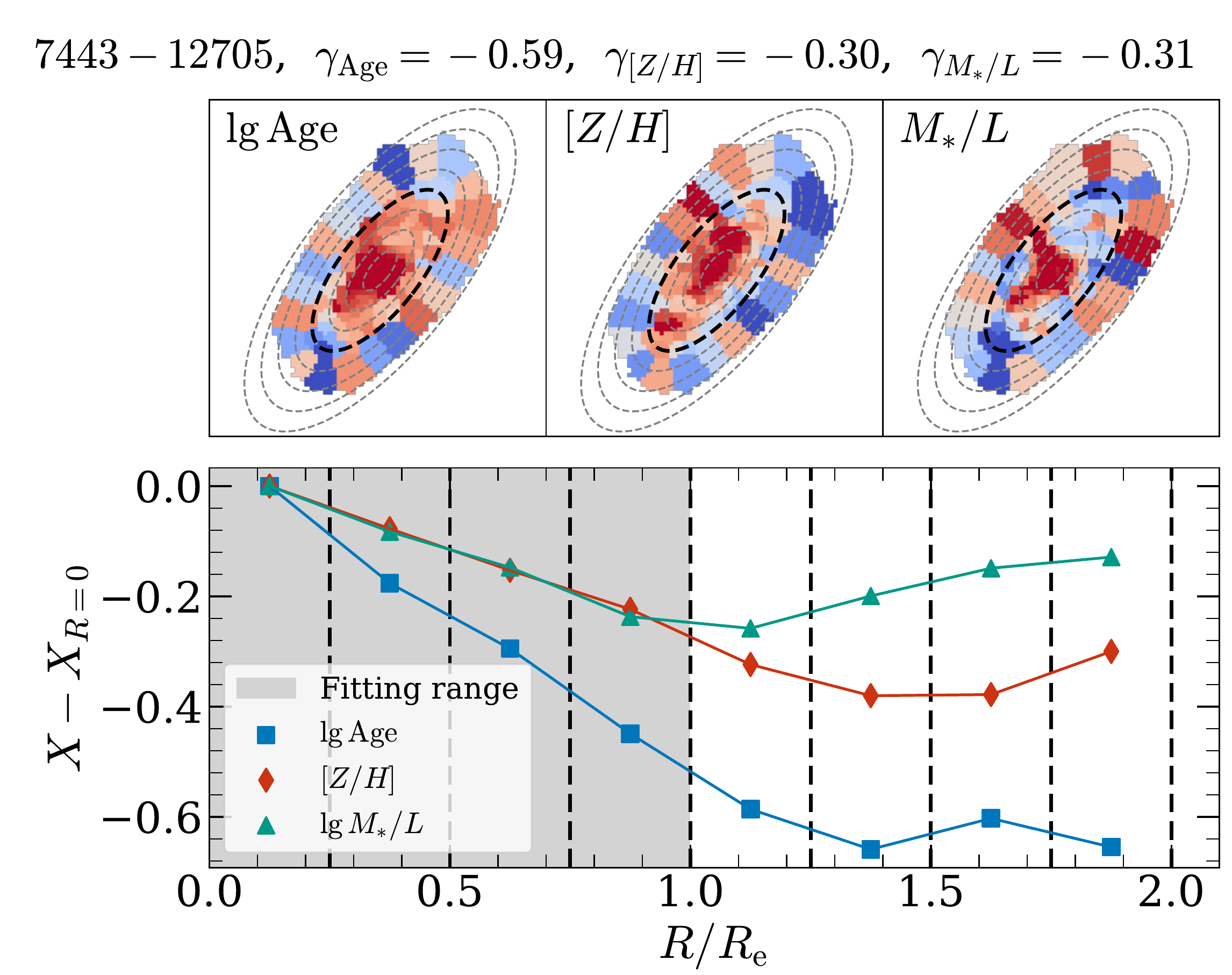}

\includegraphics[width=0.68\columnwidth]{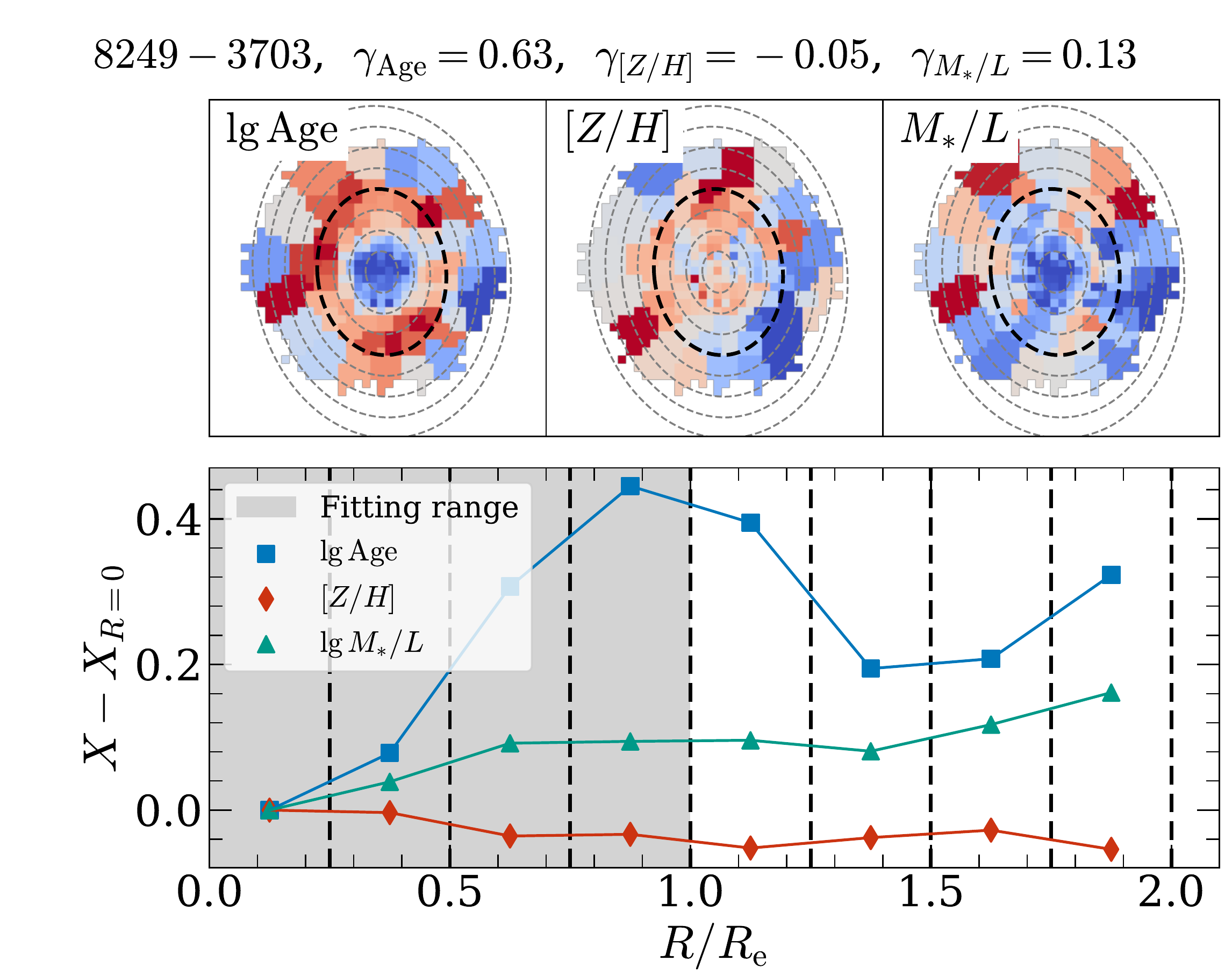}
\includegraphics[width=0.68\columnwidth]{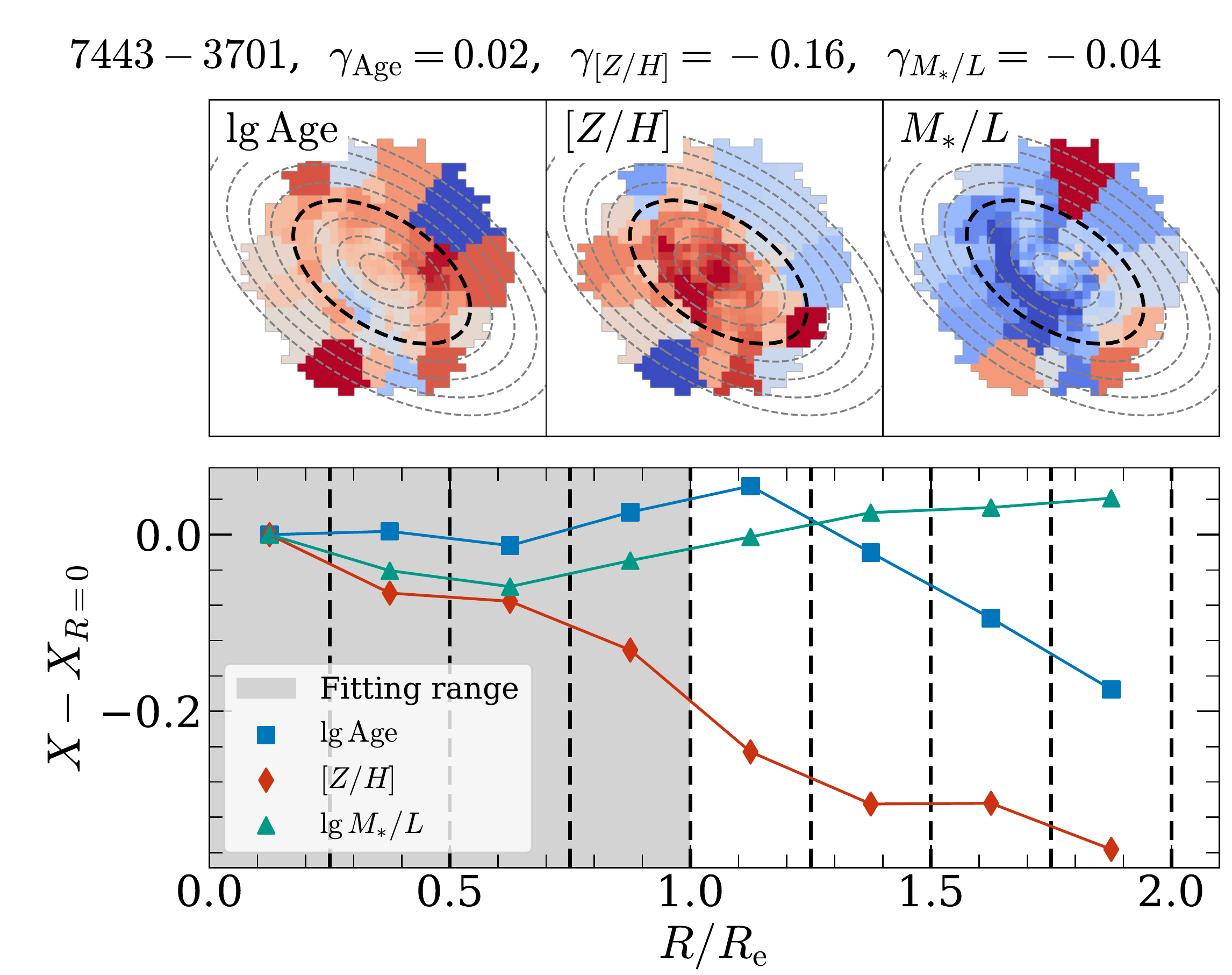}
\includegraphics[width=0.68\columnwidth]{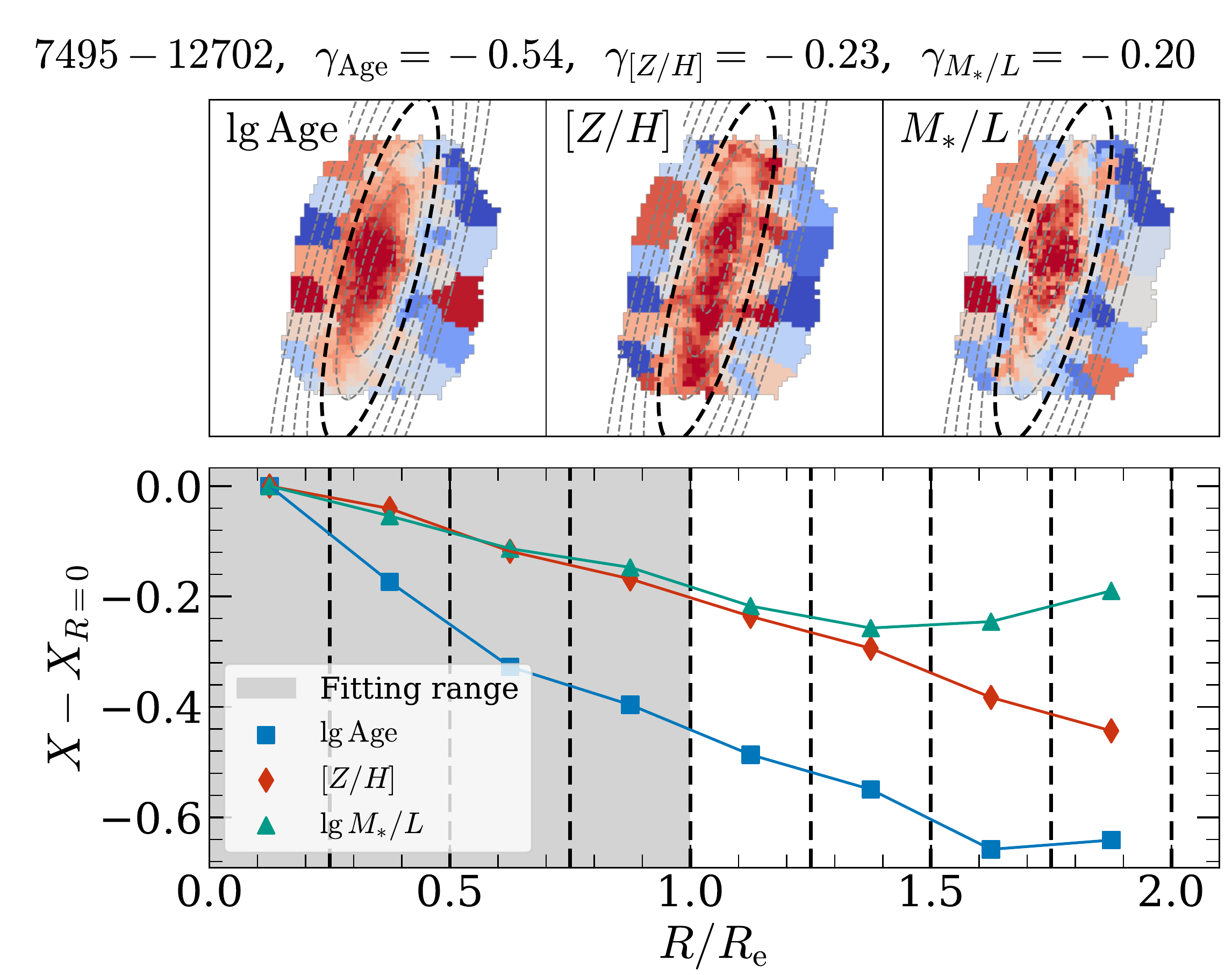}

\includegraphics[width=0.68\columnwidth]{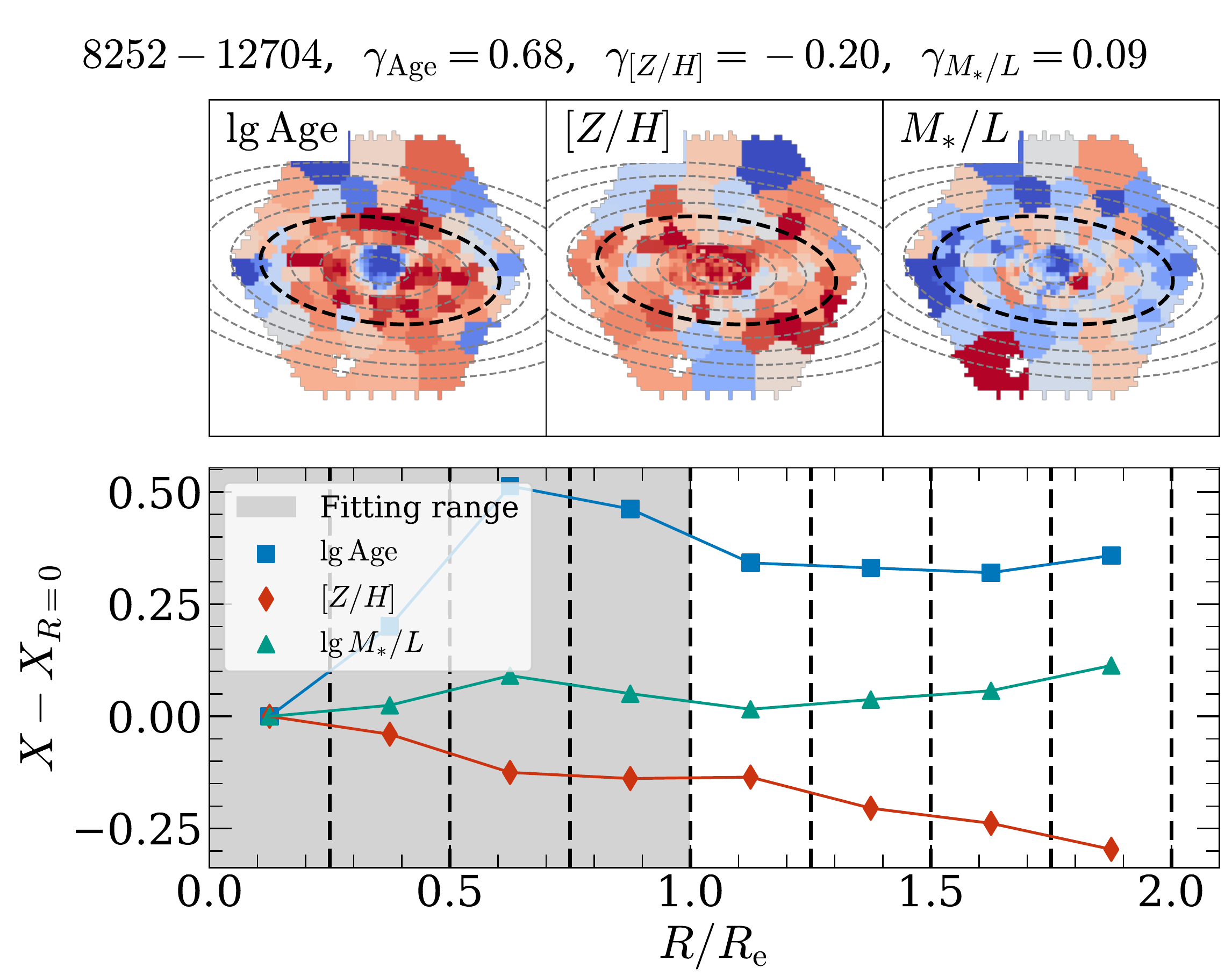}
\includegraphics[width=0.68\columnwidth]{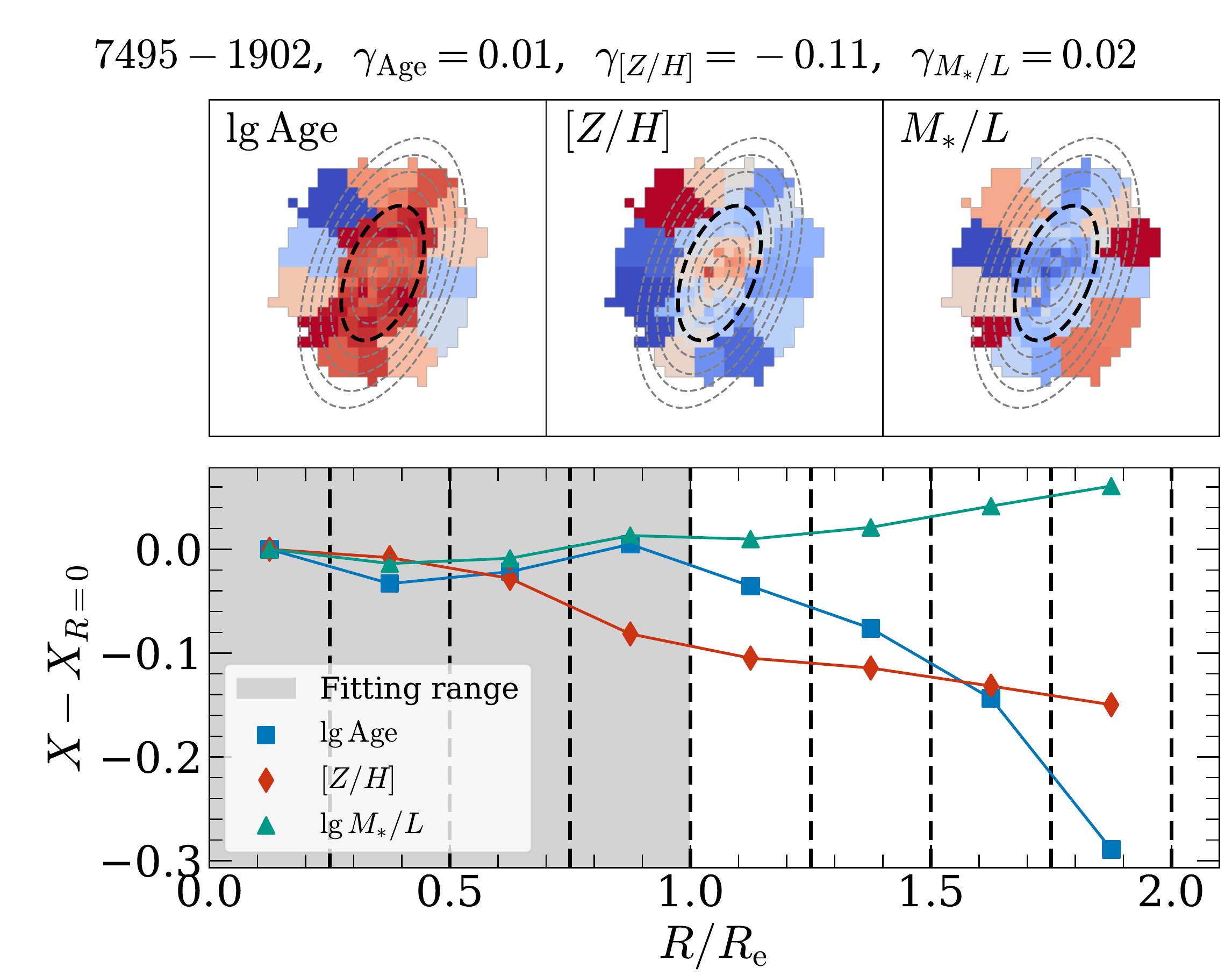}
\includegraphics[width=0.68\columnwidth]{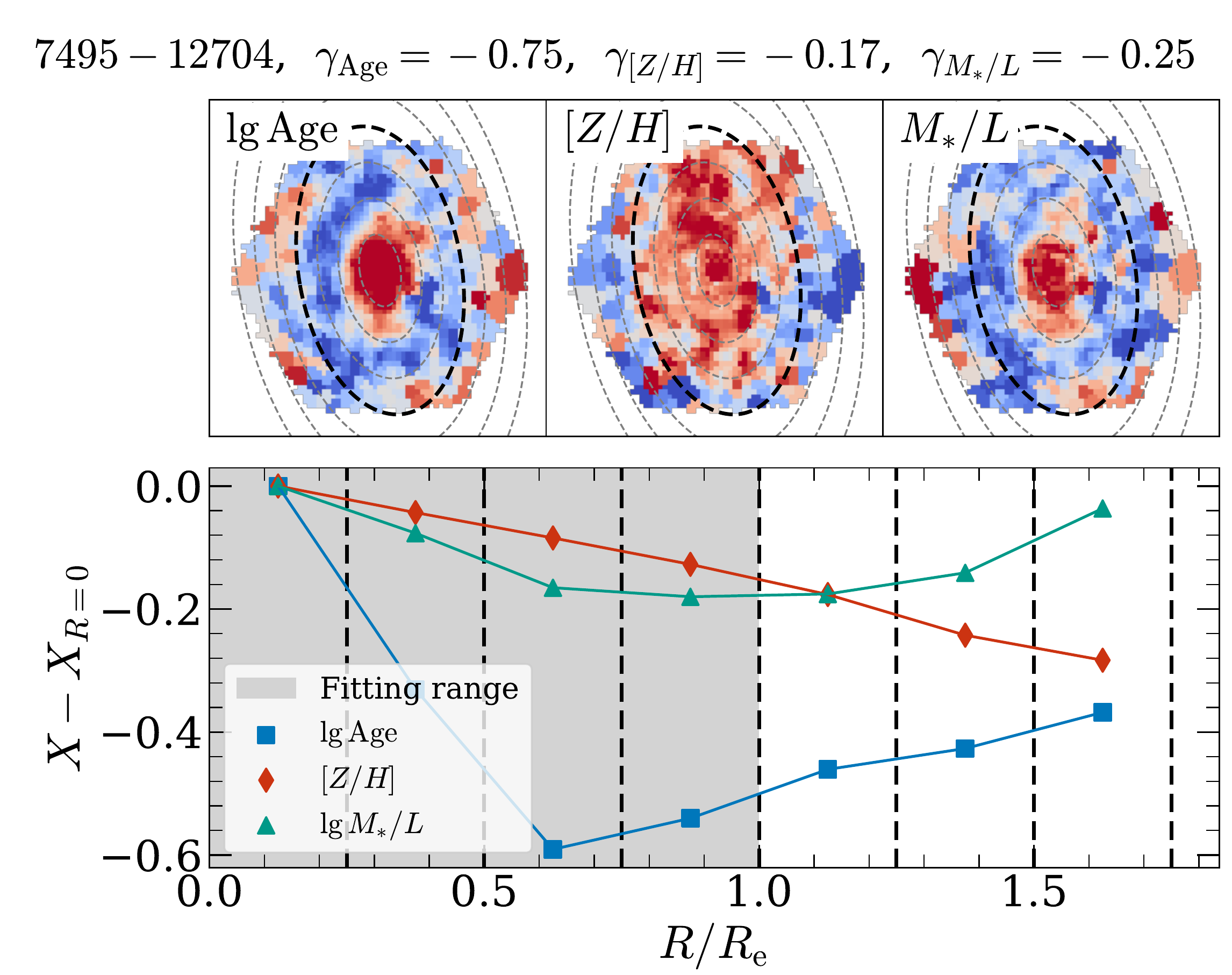}

\caption{Examples of stellar population profiles for 4 galaxies with significant positive age gradient ($\gamma_{\rm Age}>0.5$, i.e. increasing age profile within half-light isophote; the left column), 4 with age gradient close to 0 ($|\gamma_{\rm Age}|<0.05$, i.e. flat age profile within half-light isophote; the middle column), and 4 with significant negative age gradient ($\gamma_{\rm Age}<-0.5$, i.e. decreasing age profile within half-light isophote; the right column). For each galaxy, we show their age, metallicity, and stellar mass-to-light ratio maps (from left to right) in the top of each sub-figure, with thin dashed grey ellipses indicating the elliptical annuli within which the stellar population profiles are calculated, and the thick dashed black ellipse indicating the elliptical half-light isophote, within which the population gradients are calculated (see \autoref{sec:sp_gradients} for details). In the bottom panel of each sub-figure, the age, metallicity, and stellar mass-to-light ratio profiles are shown with blue, red, and green curves respectively. Before plotting, the three profiles are shifted to have the same value in the innermost bin ($0<R<0.25R_{\rm e}$), so that the three profiles of different magnitudes can be plotted together. $X$ in Y-axis of the panel can be $\lg\,\mathrm{Age}$, $[Z/H]$, and $\lg\,M_{\ast}/L$. The black dashed lines in the bottom panel of each sub-figure indicate the boundaries of each radial bin. The grey shaded region indicates $r<R_{\rm e}$, within which the gradients of stellar population properties are calculated by performing a linear fitting to the profiles. Galaxy ID and the gradients of age, metallicity, and stellar mass-to-light ratio of each galaxy are shown on the top of each sub-figure.}
\label{fig:age_profile_calculation}
\end{figure*}

\begin{figure}
\centering
\includegraphics[width=1\columnwidth]{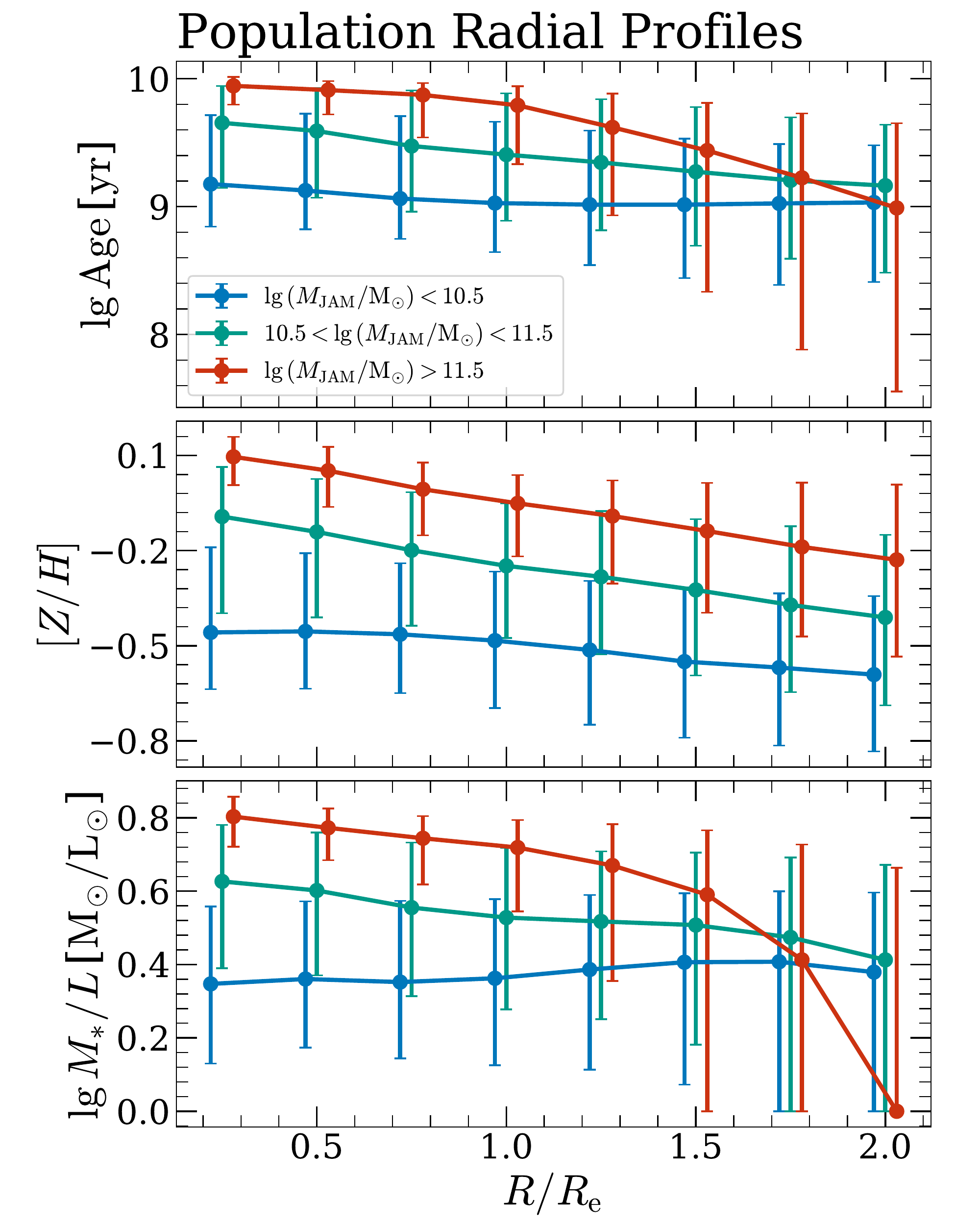}
\caption{Radial profiles of age (top), metallicity (middle), and stellar mass-to-light ratio (bottom) of galaxies. In each panel, the median values of population radial profiles for galaxies with different masses are indicated by curves of different colours, with the error bars indicating the range from the 16th to the 84th percentiles ($\pm 1\sigma$).}
\label{fig:sp_profile}
\end{figure}

\begin{figure*}
\centering
\includegraphics[width=2\columnwidth]{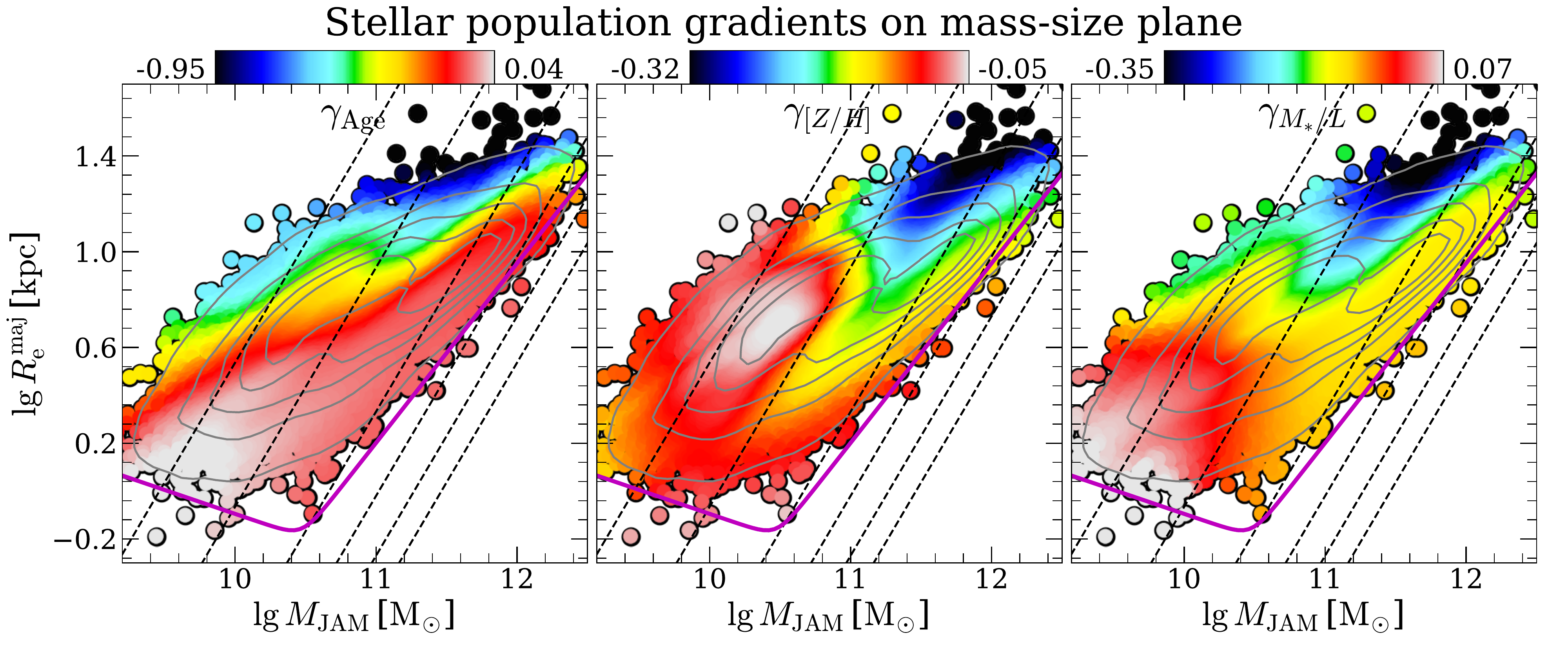}
\caption{Distribution of age gradient ($\gamma_{\rm Age}$, left), metallicity gradient ($\gamma_{[Z/H]}$, middle), and stellar mass-to-light gradient ($\gamma_{M_{\ast}/L}$, right) on the mass-size plane ($\lg\,R_{\rm e}^{\rm maj}$ versus $\lg\,M_{\rm JAM}$, see \autoref{sec:jam_props} for definitions of the two parameters). The symbols are the same as \autoref{fig:sp_mass_size}.}
\label{fig:slope_mass_size}
\end{figure*}

\begin{figure*}
\centering
\includegraphics[width=2\columnwidth]{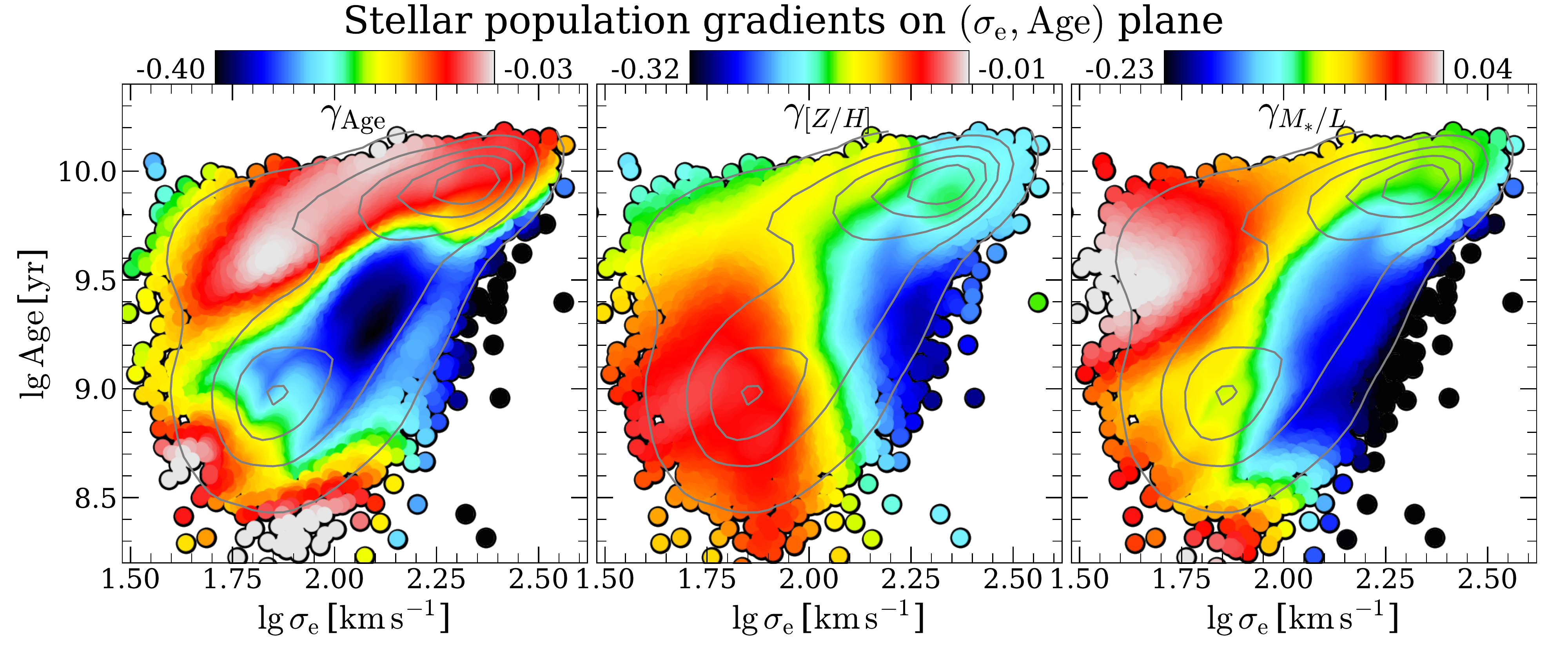}
\caption{Distribution of age gradient ($\gamma_{\rm Age}$, left), metallicity gradient ($\gamma_{[Z/H]}$, middle), and stellar mass-to-light gradient ($\gamma_{M_{\ast}/L}$, right) on the $(\sigma_{\rm e},\mathrm{Age})$ plane. The galaxy number density is indicated by the grey contours.}
\label{fig:slope_age2sigma}
\end{figure*}

\begin{figure*}
\centering
\includegraphics[width=2\columnwidth]{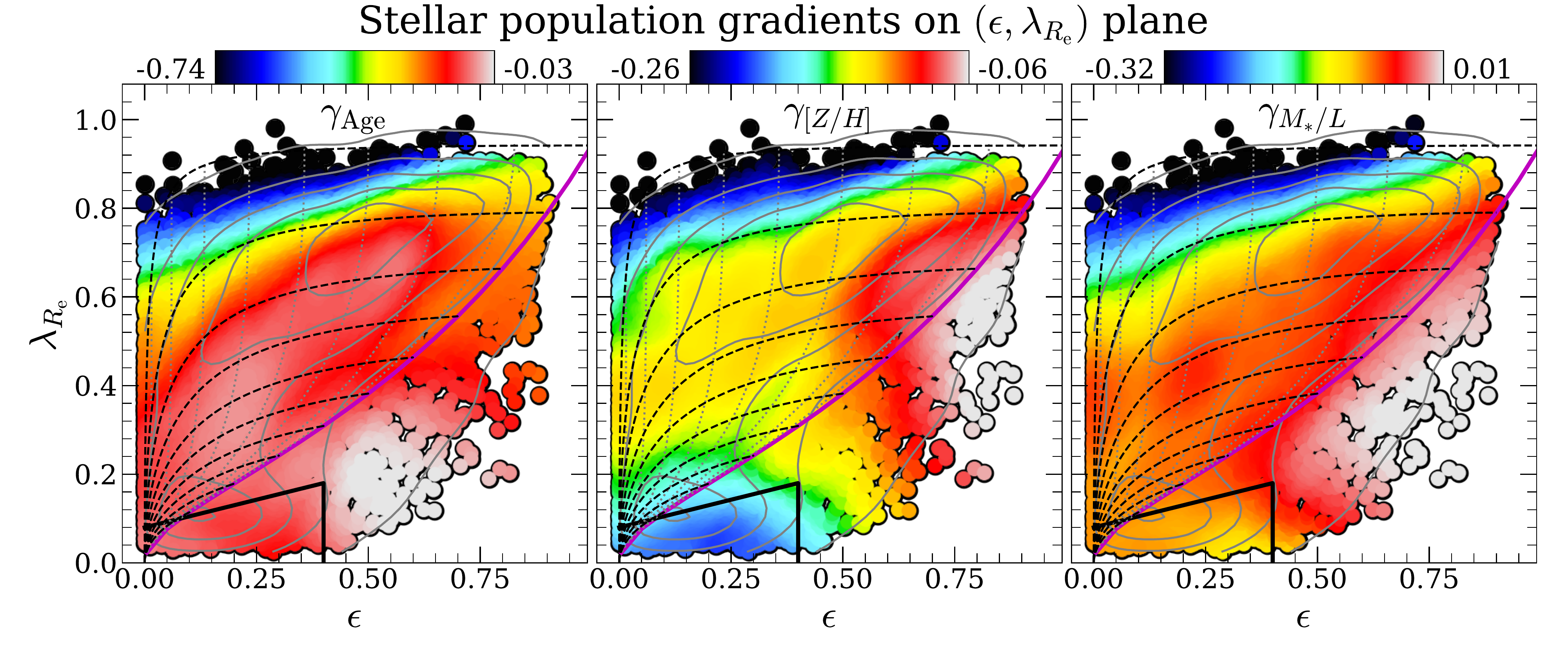}
\caption{Distribution of age gradient ($\gamma_{\rm Age}$, left), metallicity gradient ($\gamma_{[Z/H]}$, middle), and stellar mass-to-light gradient ($\gamma_{M_{\ast}/L}$, right) on the $(\epsilon,\lambda_{R_{\rm e}})$ plane (see \autoref{sec:jam_props} for definitions of the two parameters). The symbols are the same as \autoref{fig:dust}.}
\label{fig:slope_lambda_epsilon}
\end{figure*}

\begin{figure*}
\centering
\includegraphics[width=1.5\columnwidth]{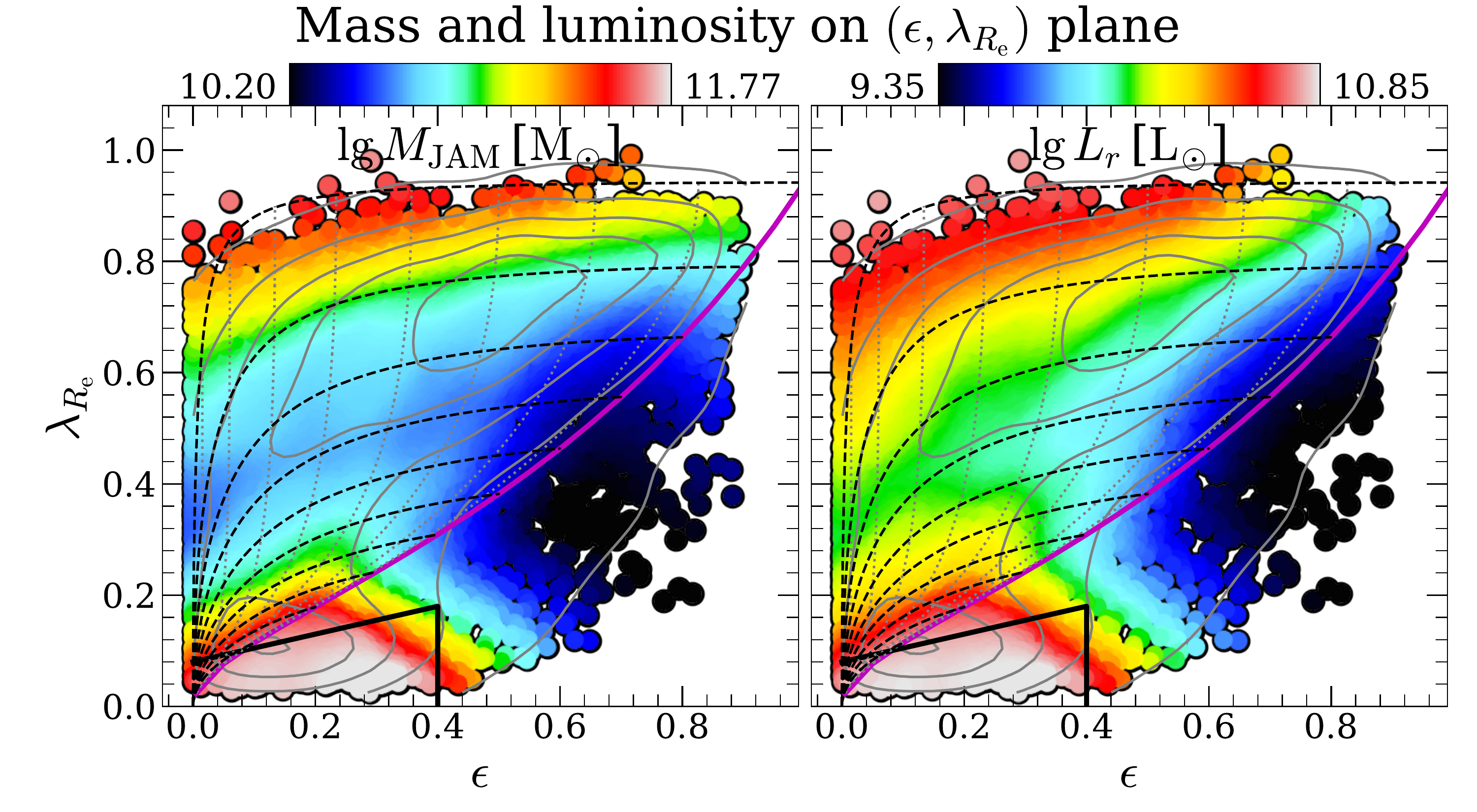}
\caption{Distributions of galaxy mass ($\lg\,M_{\rm JAM}$, left) and $r-$band luminosity ($\lg\,L_{r}$, right) on the $(\epsilon,\lambda_{R_{\rm e}})$ plane (see \autoref{sec:jam_props} for definitions of the two parameters). The symbols are the same as \autoref{fig:dust}.}
\label{fig:mass_light_lambda_epsilon}
\end{figure*}

Apart from the scaling relation of global stellar population properties, we also study the scaling relations for stellar population gradients in this section. As mentioned in \autoref{sec:sp}, the population profiles and gradients are derived from the spatially resolved stellar population maps, which are first Voronoi binned to $S/N\sim 30$ before {\sc ppxf} fitting. Pixels that are associated to the same Voronoi bin share the same age, metallicity, and stellar mass-to-light ratio. To calculate the stellar population profiles and gradients, we first divide the galaxies into different elliptical annuli with the radial step of each annulus being $0.25R_{\rm e}^{\rm maj}$ along the major axis. The position angle and ellipticity of each annulus are set to be the same as the elliptical half-light isophote of the galaxy. For each annulus, we take the average luminosity-weighted stellar population properties as the value at this radius and then perform a linear fit to the radius-population relation. For age and stellar mass-to-light ratio, we take their values in logarithmic space into account (i.e. we fit the $\lg\,\mathrm{Age}-R/R_{\rm e}$ and $\lg\,M_{\ast}/L-R/R_{\rm e}$ relations), while for metallicity, we fit the $[Z/H]-R/R_{\rm e}$ correlation. We note here that due to the different observation ranges for different galaxies in MaNGA, the elliptical annuli are divided out the $2R_{\rm e}^{\rm maj}$ along the major axis if applicable, while the stellar population gradients are only calculated within the elliptical half-light isophotes. For simplicity, we adopt the expression of ``stellar population property at $nR_{\rm e}$'', instead of ``stellar population within the elliptical annulus with semi-major axis being $nR_{\rm e}^{\rm maj}$''. 

To better demonstrate this process, we present \autoref{fig:age_profile_calculation}, where we show 12 examples of stellar population profile/gradient calculation, among which 4 galaxies have strong increasing age profiles within half-light isophote (i.e. significantly positive age gradient), 4 have nearly flat age profiles within half-light isophote (i.e. age gradient close to 0), and 4 have strong decreasing age profiles within half-light isophote (i.e. significantly negative age gradient). The thin grey dashed ellipses in each map are the boundaries of the radial bins and the thick black dashed ellipse is the elliptical half-light isophote, within which the stellar population gradients are calculated.

Interestingly, galaxies can have distinct profiles of different stellar population properties. For example, Galaxy 8459-1902 and Galaxy 8249-3703 (the middle two galaxies in the left column of \autoref{fig:age_profile_calculation}) have significantly increasing age profiles from the galaxy center towards the outskirts of the galaxies, while their metallicity profiles appear to be flat or even mildly decreasing. On the contrary, Galaxy 7443-12704 and Galaxy 7443-3701 (the first and third galaxies in the middle column of \autoref{fig:age_profile_calculation}) have flat age profiles, but show significantly decreasing metallicity profiles from inner part to the outer part of the galaxies. In addition, the steepness of the stellar population profiles can change dramatically at different radii. For example, the luminosity-weighted age of Galaxy 7443-12703 (the first galaxy in the left column of \autoref{fig:age_profile_calculation}) increases rapidly at small radii, but is nearly constant at $R>R_{\rm e}$. Galaxy 7443-6102 (the second galaxy in the middle column of \autoref{fig:age_profile_calculation}) has an even more complicated age profile. This results in the fact that the stellar population gradients can be sensitive to the radial range they are estimated. Thus, the readers should be careful when comparing the results from different studies.

In \autoref{fig:sp_profile}, we present the radial profiles of age, metallicity, and stellar mass-to-light ratio for galaxies in different mass bins ($\lg\,M_{\rm JAM}<10.5$, $10.5<\lg\,M_{\rm JAM}<11.5$, and $\lg\,M_{\rm JAM}>11.5$). As shown in the figure, galaxies with the lowest mass ($\lg\,M_{\rm JAM}<10.5$) tend to have the youngest center and a nearly flat age profile. It means that the galaxies with the lowest mass have active/recent star-formation at all radii. The intermediate-mass galaxies ($10.5<\lg\,M_{\rm JAM}<11.5$) appear to have a gradually decreasing age profile from galaxy center to the outskirts. Interestingly, the most massive galaxies ($\lg\,M_{\rm JAM}>11.5$) appear to have the oldest center with the age profile being nearly flat within $R\lesssim R_{\rm e}$, while show obviously decreasing age profiles towards larger radii at $R\gtrsim R_{\rm e}$ with a large scatter. This again raises the importance of clarifying the radial range of gradient calculation. For metallicity, nearly all the galaxies in different mass bins show similar, decreasing metallicity profiles from the central region to the outer part of the galaxies, with the most massive galaxies ($\lg\,M_{\rm JAM}>11.5$) having the highest metallicity at all radii. This is consistent with the findings of previous studies (e.g. \citealt{Kuntschner2010,Goddard2017,Zheng_et_al.(2017),Li_et_al.(2018),Santucci_et_al.(2020)}), where galaxies are found more likely to have negative metallicity gradients (higher metallicity at galaxy center). Besides, galaxies with the lowest mass have a mildly increasing $M_{\ast}/L$ profile, with the outer part of the galaxies having slightly higher $M_{\ast}/L$. It means that on average, star-formation is slightly more active at the outskirts of these galaxies. The intermediate-mass galaxies again show a gradually decreasing $M_{\ast}/L$ profile and the most massive galaxies again have gradually-decreasing $M_{\ast}/L$ profile at small radii ($R\lesssim 1.5R_{\rm e}$), which becomes steeper at larger radii ($R\gtrsim 1.5R_{\rm e}$) with large scatters. This result shows remarkable agreement with the study of \citet[][fig.~4]{Ge_et_al.(2021)}, where $M_{\ast}/L$ profile is also seen to be mass-dependent, and the steepness of $M_{\ast}/L$ profile varies with radius for galaxies in the most massive bin, confirming the robustness of our work.

The young and low-$M_{\ast}/L$ outer region of massive galaxies and their large scatters at large radii ($R\sim 2R_{\rm e}$) indicate that galaxies in this mass bin (i.e. $\lg\,M_{\rm JAM}>11.5$) does not only consist of totally quenched galaxies but also contains massive disk galaxies. These disk galaxies typically have large, old, and quenched bulges at the galaxy center, and gas-rich disks at the outskirts, which are still actively forming stars, making the outer part younger and to have lower stellar mass-to-light ratio (e.g. \citealt{Lu_et_al.(2022a)} in simulations and \citealt{Zhang_et_al.(2019)} in observations). 

This interpretation can be confirmed by \autoref{fig:slope_mass_size}, where we show the distributions of the gradients of age, metallicity, and stellar mass-to-light ratio ($\gamma_{\rm Age}$, $\gamma_{[Z/H]}$, and $\gamma_{M_{\ast}/L}$) on the mass-size plane. Unlike global stellar population properties (see \autoref{fig:sp_mass_size}), stellar population gradients show more complicated distributions on the mass-size plane. At the low mass end ($\lg\,M_{\rm JAM}\sim 9.6-9.8$), galaxies with smaller sizes are more likely to have positive $\gamma_{\rm Age}$ (younger centers and older outer parts), compared to those with the same mass but larger sizes. This is consistent with \citet{Lu_et_al.(2021)}, where part of low-mass disk galaxies have experienced more retrograde (with respect to their stellar spin) mergers and thus have counter-rotating gas disks and a high fraction of counter-rotating stars (with respect to the bulk of stellar rotation), making the galaxies to have smaller sizes and more centrally-concentrated star-formation, i.e. younger centre than outer part. Besides, galaxies with ($\lg\,M_{\rm JAM}\sim 11-12$) and $\lg\,R_{\rm e}^{\rm maj}\gtrsim 1.2$ have the lowest $\gamma_{\rm Age}$ (the steepest decreasing age profile from center to the outskirts), consistent with the cartoon by \citet[fig.~23]{Cappellari(2016)}, in which we can see that galaxies in this region are mainly massive disk galaxies with large, old, and quenched bulges at galaxy center and star-forming disks at the outskirts of the galaxies. According to \citet[fig.~23]{Cappellari(2016)}, massive galaxies contain both quenched elliptical galaxies, as well as the massive disk galaxies, consistent with our guess for the origin of the young outskirts and large scatters in age and stellar mass-to-light ratio of massive galaxies ($\lg\,M_{\rm JAM}>11.5$) in \autoref{fig:sp_profile}. The gradients of metallicity and stellar mass-to-light ratio show similar distributions as the age gradient on the mass-size plane, with a subtle difference in metallicity gradient, where the galaxies with the flattest metallicity profiles locate in the region with intermediate mass and intermediate size ($\lg\,R_{\rm e}^{\rm maj}\sim 0.6-1$ and $\lg\,M_{\rm JAM}\sim 10-11$). All the detailed features of the variation of galaxy gradients in this diagram agree remarkably well with the independent study by \citet[fig.7]{Li_et_al.(2018)}, confirming the robustness of the trends.

The mass-size plane does not capture the main features of population gradients in galaxies. In fact, we find that there are major systematic trends in the gradients at fixed $\sigma_{\rm e}$, namely at fixed location on the mass-size plane, as a function of galaxy age. In \autoref{fig:slope_age2sigma}, we present the stellar population gradients on the $(\sigma_{\rm e},\mathrm{Age})$ plane, where $\lg\,\mathrm{Age}$ is the global luminosity-weighted age of galaxies (see \autoref{sec:global_and_maps} for the definition of global population properties). As can be seen, older galaxies with $\lg\,\sigma_{\rm e}\gtrsim 2.0$ tend to have less negative age gradients (closer to 0, indicating a flat age profile). This indicates that old galaxies are almost equally old at different radii. It may be caused by both a physical origin, where old and quenched galaxies have longer time to radially mix their stellar content, and a technical origin, where age differences are difficult to measure at the age older than $8\,\rm Gyr$. Interestingly, galaxies in the green valley, namely between the two peaks in the galaxy number density, appear to have the lowest $\gamma_{\rm Age}$ (the most negative, indicating the steepest decreasing age profile from center to the outer part of galaxies). This shows that that green valley galaxies are typically massive disk galaxies, which have quenched centers (bulges) and maintain star-forming disks at the outskirts. These results are consistent with \citet{Lah_et_al.(2023)}, which studied the differences of stellar population of bulges and disks in MaNGA and found that early-type galaxies tend to have bulges and disks with similar age (i.e. flat age profiles), while late-type galaxies have disks much younger than bulges (i.e. the steepest decreasing age profiles from galaxy center to the outside).

Metallicity gradient appears to have a stronger correlation with $\sigma_{\rm e}$ rather than with galaxy age, although strong variations exists at fixed $\sigma_{\rm e}$ and varying age. The most negative metallicity gradient is also seen for galaxies in the green valley (but have larger $\sigma_{\rm e}$ than those with the lowest age gradients), which may be because of the low-metallicity gas accretion of these galaxies diluting the metallicity. Given that the stellar mass-to-light ratio $M_{\ast}/L$ depends on both age and metallicity (e.g., fig.~2 of \citealt{Ge_et_al.(2019)}), it makes sense that the distribution of gradient of $M_{\ast}/L$ on the $(\sigma_{\rm e},\mathrm{Age})$ plane is intermediate between the distributions of $\gamma_{\rm Age}$ and $\gamma_{[Z/H]}$ on this plane.

In \autoref{fig:slope_lambda_epsilon}, we present the distributions of the three stellar population gradients on the $(\epsilon,\lambda_{R_{\rm e}})$ plane (see \autoref{sec:jam_props} for definitions of the two parameters). As can be seen, the most rotational galaxies have the lowest $\gamma_{\rm Age}$ (the most negative, indicating the steepest decreasing age trend from center to the outer part of galaxies), again consistent with the results seen from \autoref{fig:slope_mass_size} and \autoref{fig:slope_age2sigma}. We note that galaxies with high $\lambda_{R_{\rm e}}$ and large $\epsilon$ (i.e. edge-on), however, have flatter age profiles (less negative $\gamma_{\rm Age}$), compared to those with similar $\lambda_{R_{\rm e}}$ but smaller $\epsilon$ (i.e. more face-on). That is because more disk component is observed for edge-on galaxies, making the averaged age younger at the center (compared to the face-on view of the galaxy) and resulting in the flatter age profiles in these galaxies. In addition, the dust attenuation effect is also significantly stronger from the edge-on view of the galaxies, and its degeneracy with stellar age also potentially contributes to the flattening of stellar population gradients.

The distribution of metallicity gradient on the $(\epsilon,\lambda_{R_{\rm e}})$ plane is more complicated. Firstly, galaxies with the highest $\lambda_{R_{\rm e}}$ ($\lambda_{R_{\rm e}}\sim 0.8$) and relatively low $\epsilon$ ($\epsilon \lesssim 0.6$) have the most negative $\gamma_{[Z/H]}$ (i.e. the steepest decreasing metallicity profiles from galaxy center to the outskirts), same as the age gradient. Secondly, the most edge-on galaxies with intermediate/large $\lambda_{R_{\rm e}}$ ($0.3\lesssim \lambda_{R_{\rm e}}\lesssim 0.7$) have the largest metallicity gradient (close to 0, i.e. flat metallicity profiles), different from the age gradients. Thirdly, the slow-rotating galaxies (within the region of black lines) appear to have intermediate $\gamma_{[Z/H]}$, indicating relatively obvious decreasing metallicity profiles of these galaxies. The distribution of gradient of stellar mass-to-light ratio again appears to be the combination of age gradient and metallicity gradient on the $(\epsilon,\lambda_{R_{\rm e}})$ plane, where galaxies with the highest $\lambda_{R_{\rm e}}$ ($\lambda_{R_{\rm e}}\sim 0.8$) and relatively low $\epsilon$ ($\epsilon \lesssim 0.6$) have the steepest decreasing stellar mass-to-light ratio profiles and the most edge-on galaxies with intermediate $\lambda_{R_{\rm e}}$ ($0.1\lesssim \lambda_{R_{\rm e}}\lesssim 0.6$) have the flattest stellar mass-to-light ratio profiles.

To further explore the interesting distribution of metallicity on the $(\epsilon,\lambda_{R_{\rm e}})$ plane, we present \autoref{fig:mass_light_lambda_epsilon}, where we show the distributions of galaxy mass and $r-$band luminosity (see \autoref{sec:jam_props} for the definitions of the two parameters) on this plane. As can be seen, galaxy mass and $r-$band luminosity show remarkably similar trends as metallicity gradient (see the middle panel of \autoref{fig:slope_lambda_epsilon}) on the $(\epsilon,\lambda_{R_{\rm e}})$ plane: the slow-rotating galaxies and the most rotation-supported galaxies appear to have the highest mass and luminosity (and the lowest metallicity gradients in the middle panel of \autoref{fig:slope_lambda_epsilon}), while the galaxies below the magenta curve (i.e. the ones being edge-on but having low $\lambda_{R_{\rm e}}$) appear to have the lowest mass and luminosity (and the flattest metallicity gradients in the middle panel of \autoref{fig:slope_lambda_epsilon}). Previously, many studies have pointed out the mass-dependence of metallicity gradient of galaxies (e.g. \citealt{Carollo_et_al.(1993),Kuntschner2010,Sanchez-Blazquez_et_al.(2014),Gonzalez_Delgado_et_al.(2015),Ho_et_al.(2015),Goddard2017,Zheng_et_al.(2017),Santucci_et_al.(2020)}), but it is the first time to see their tight correlation on a two-dimensional plane. We note that the mass distribution on the $(\epsilon,\lambda_{R_{\rm e}})$ plane is not consistent with that in \citet[][fig.~9]{Graham_et_al.(2018)}, especially for the most rotation-supported galaxies. The possible reason may be the different calculations of galaxy mass. The galaxy mass used in this work is totally based on dynamical modelling (see \autoref{sec:jam_props} and \citetalias{Zhu_et_al.(2023a)} for details), while that used in \citet{Graham_et_al.(2018)} is from the empirical correlation between dynamical mass and $K_S$ band magnitude, which is calibrated to ATLAS$^{\rm 3D}$ dynamical models for early-type galaxies only \citep{Cappellari(2013)}. The origin of the interesting distribution of metallicity gradient on the $(\epsilon,\lambda_{R_{\rm e}})$ plane and its tight correlation with galaxy mass (luminosity) may rely on the mass assembly histories of the galaxies, and is worth investigating with cosmological simulations. We plan to carry out this study in the following papers of this MaNGA DynPop series.

\section{Star-formation history}
\label{sec:sfh}

\begin{figure*}
\centering
\includegraphics[width=0.68\columnwidth]{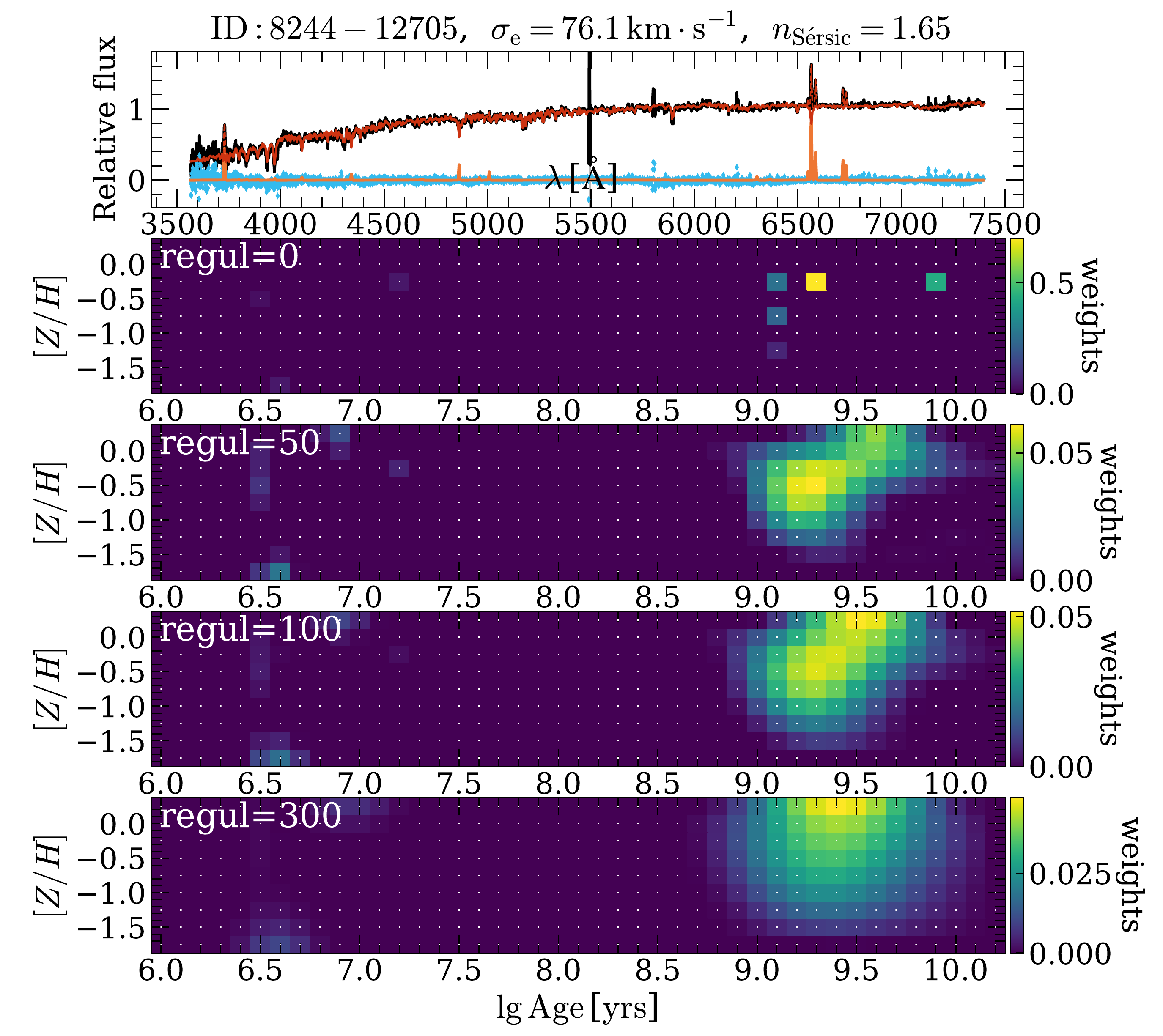}
\includegraphics[width=0.68\columnwidth]{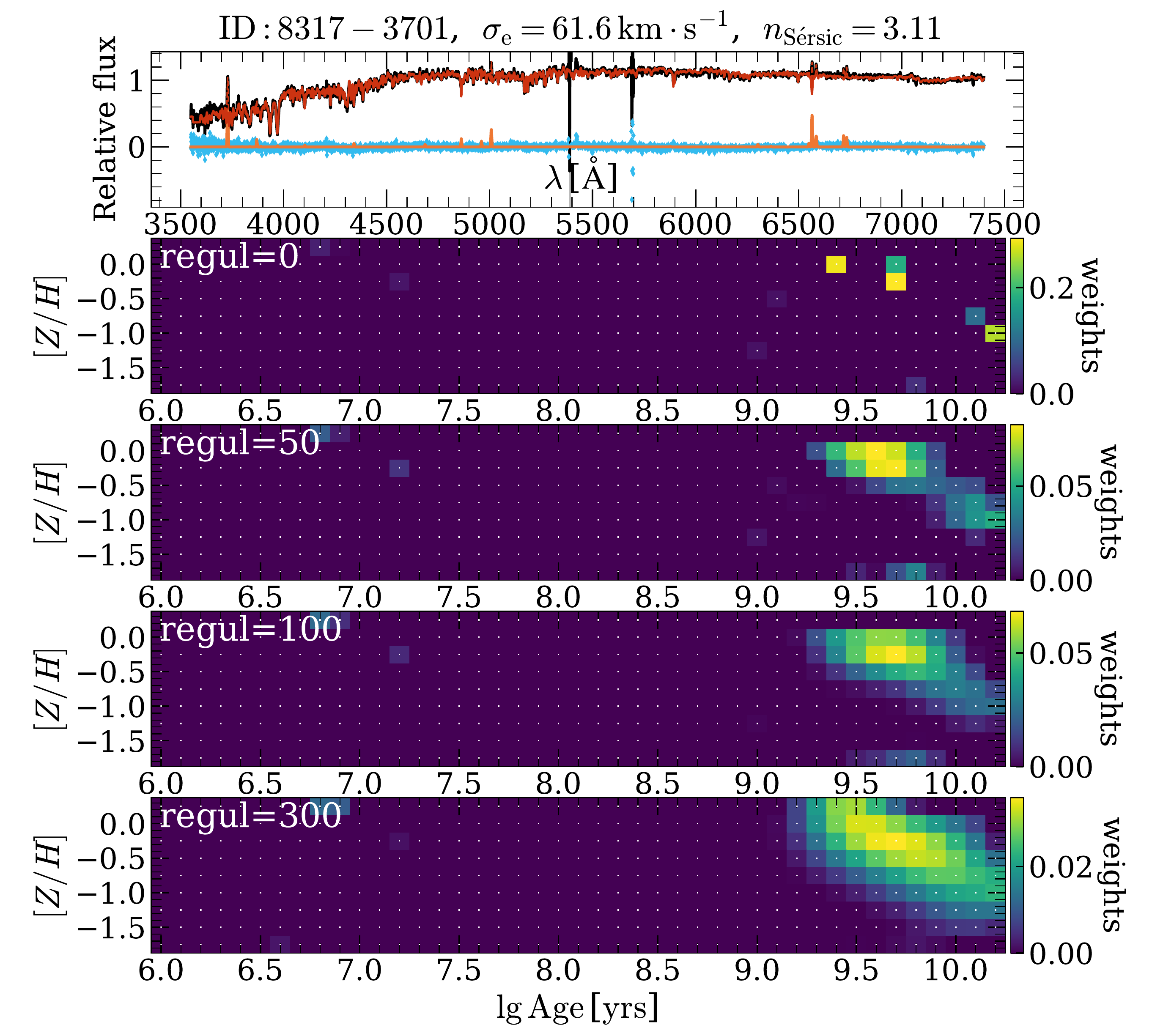}
\includegraphics[width=0.68\columnwidth]{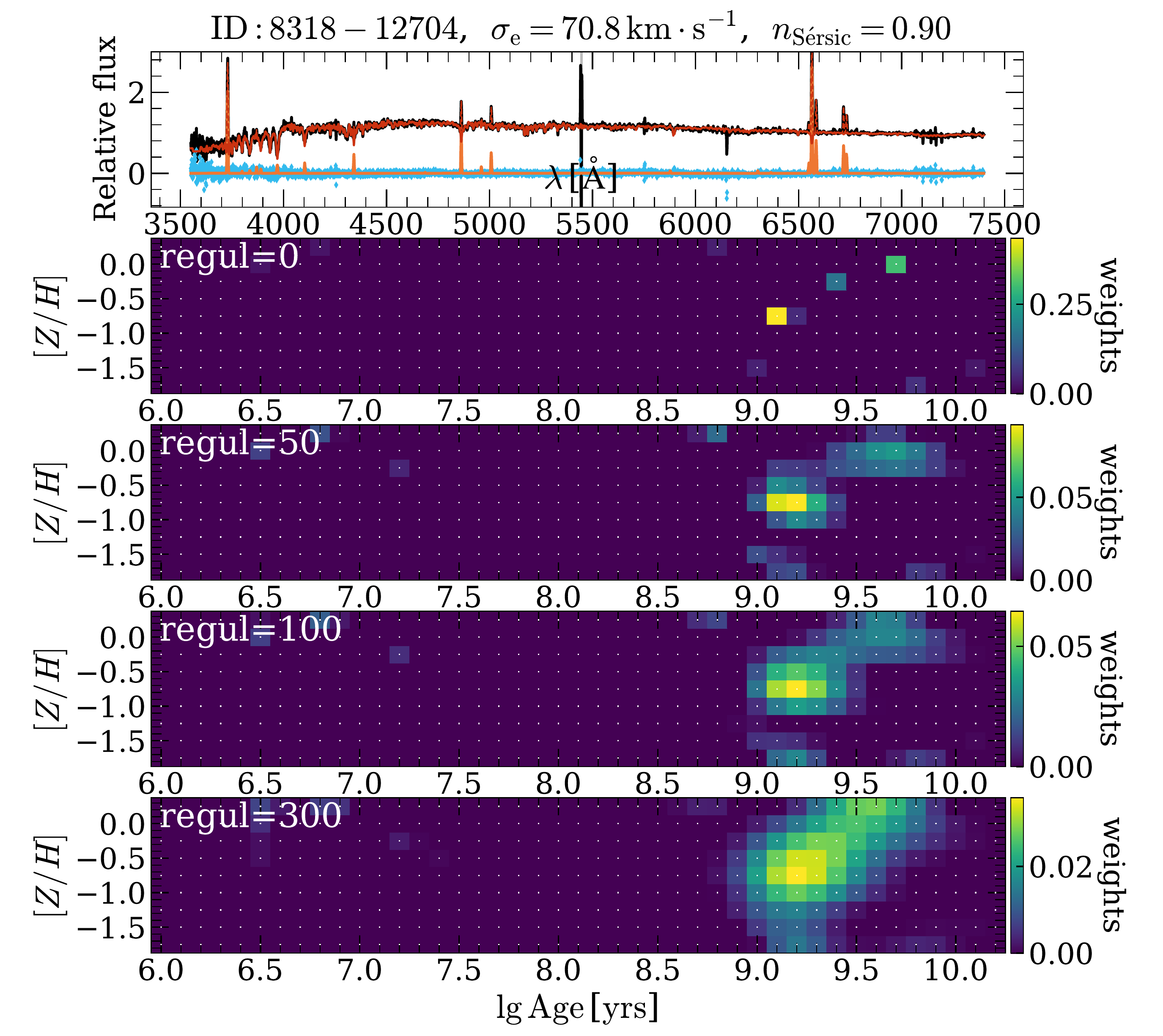}
\includegraphics[width=0.68\columnwidth]{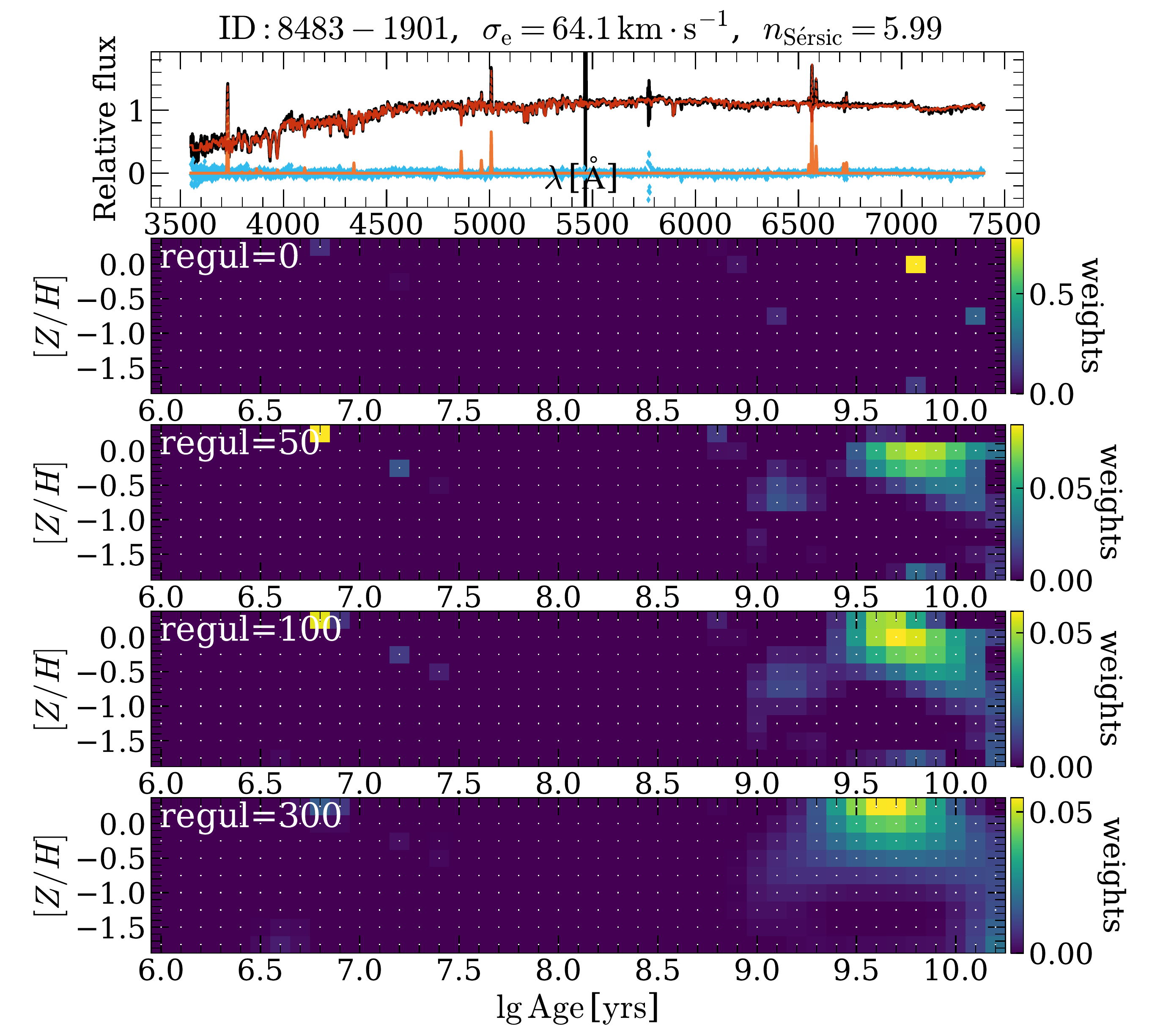}
\includegraphics[width=0.68\columnwidth]{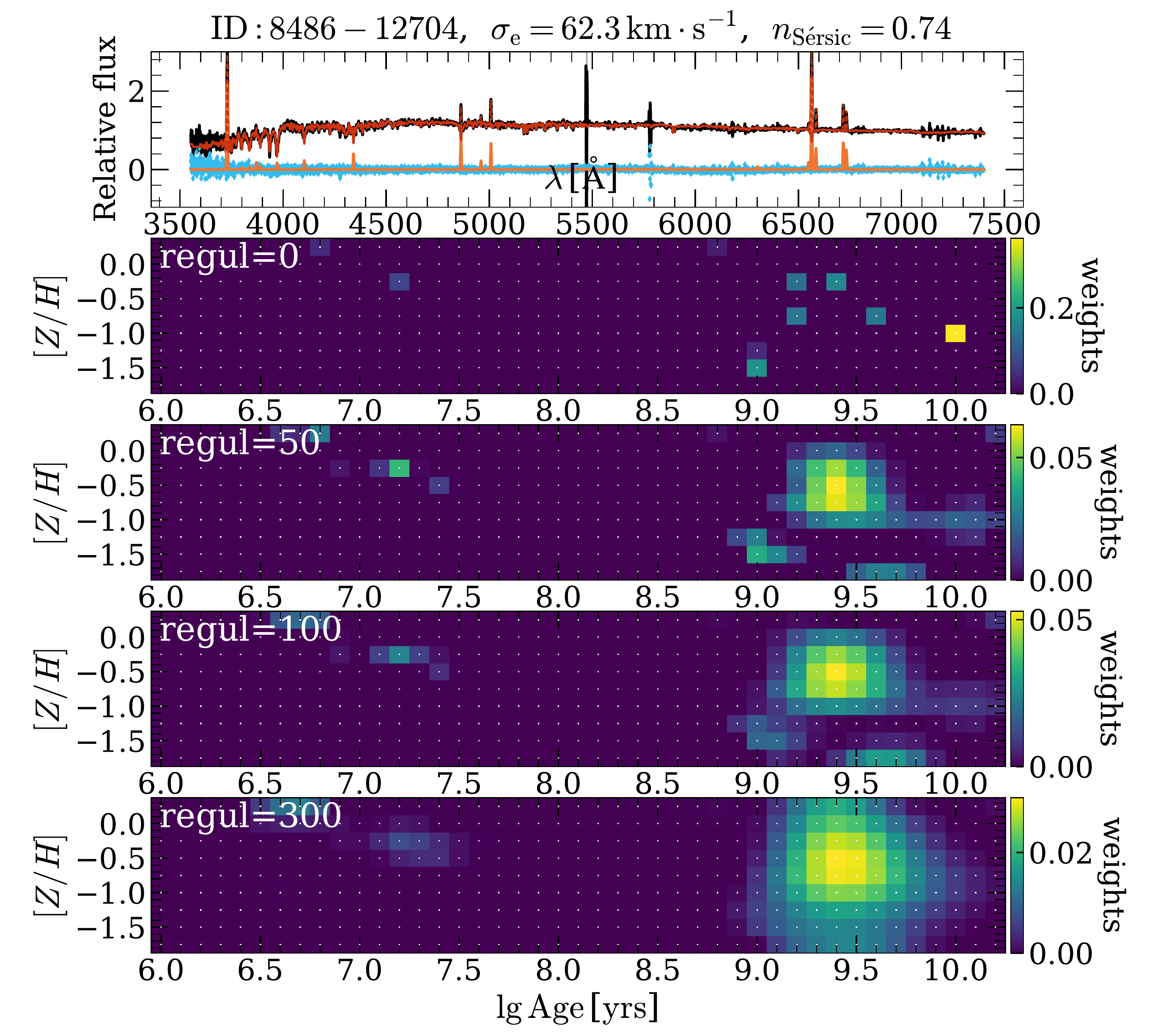}
\includegraphics[width=0.68\columnwidth]{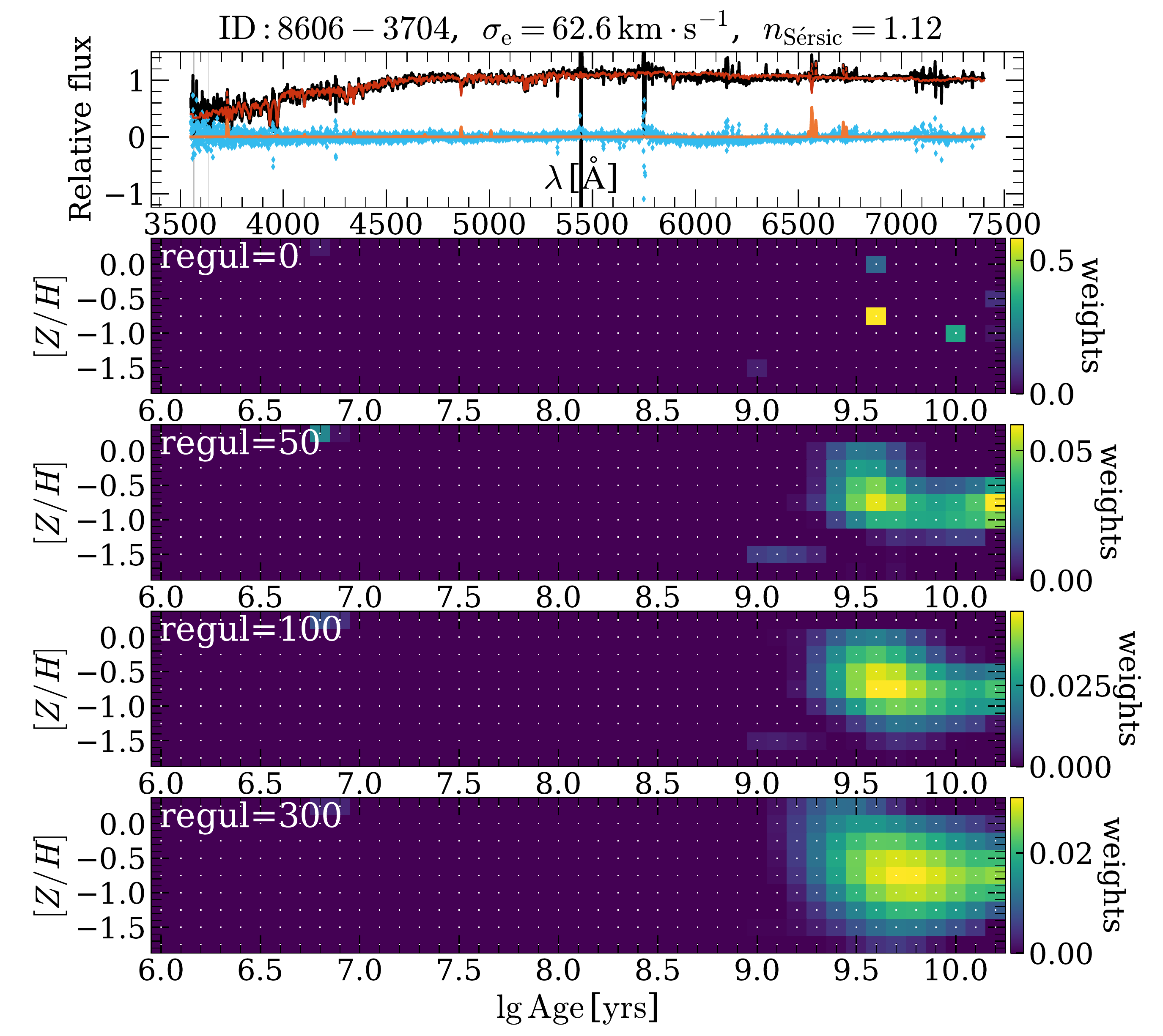}

\includegraphics[width=0.68\columnwidth]{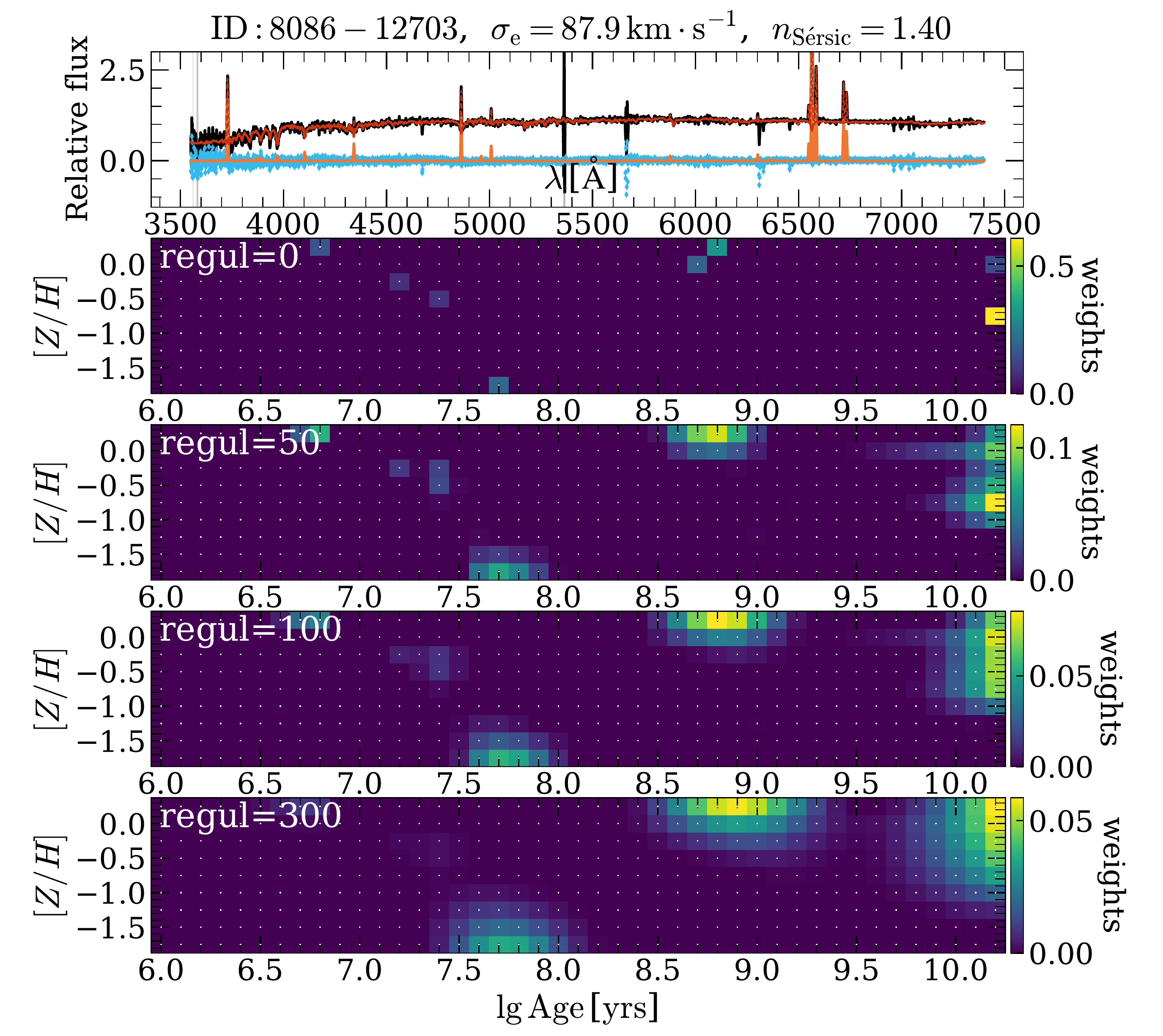}
\includegraphics[width=0.68\columnwidth]{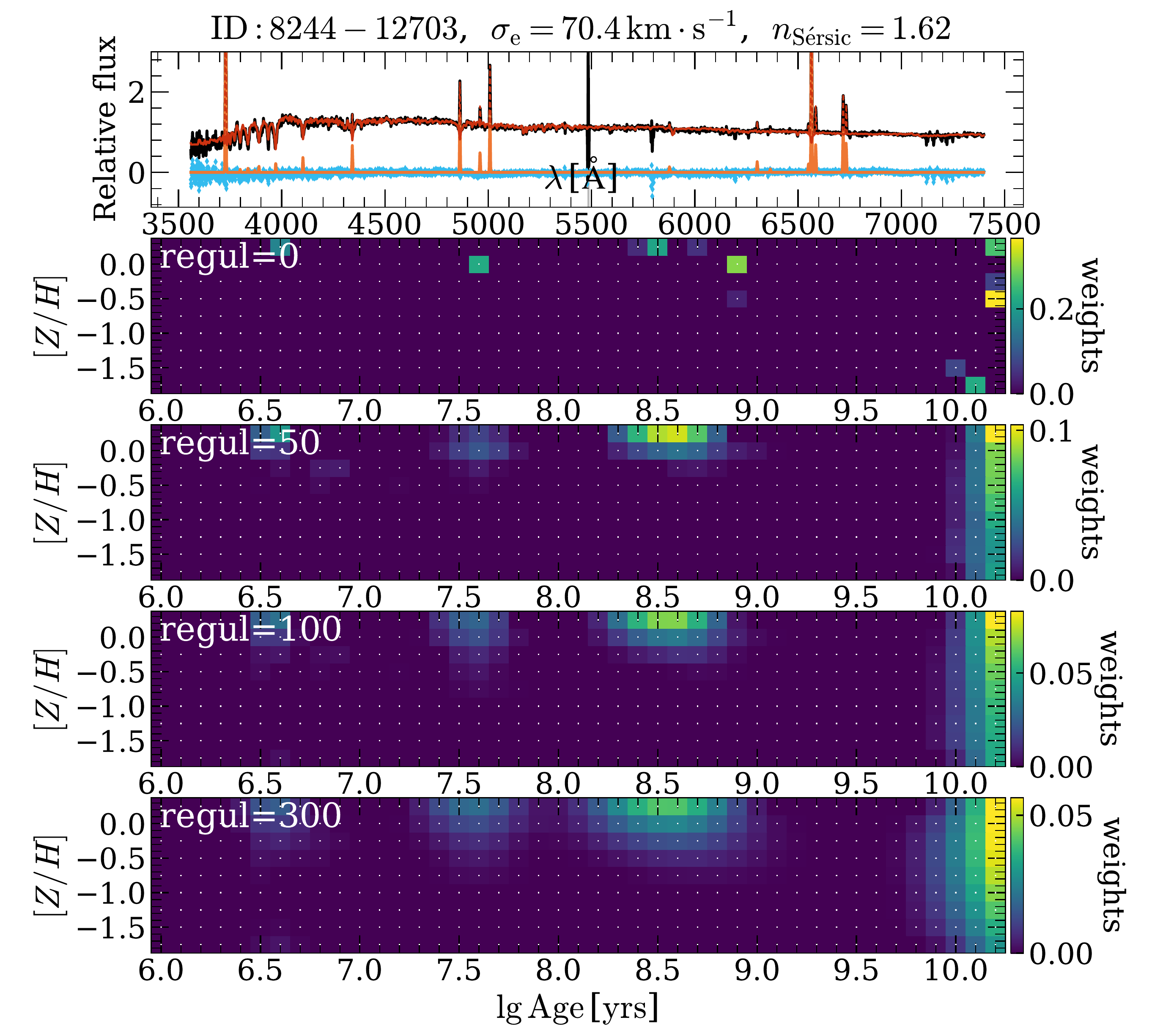}
\includegraphics[width=0.68\columnwidth]{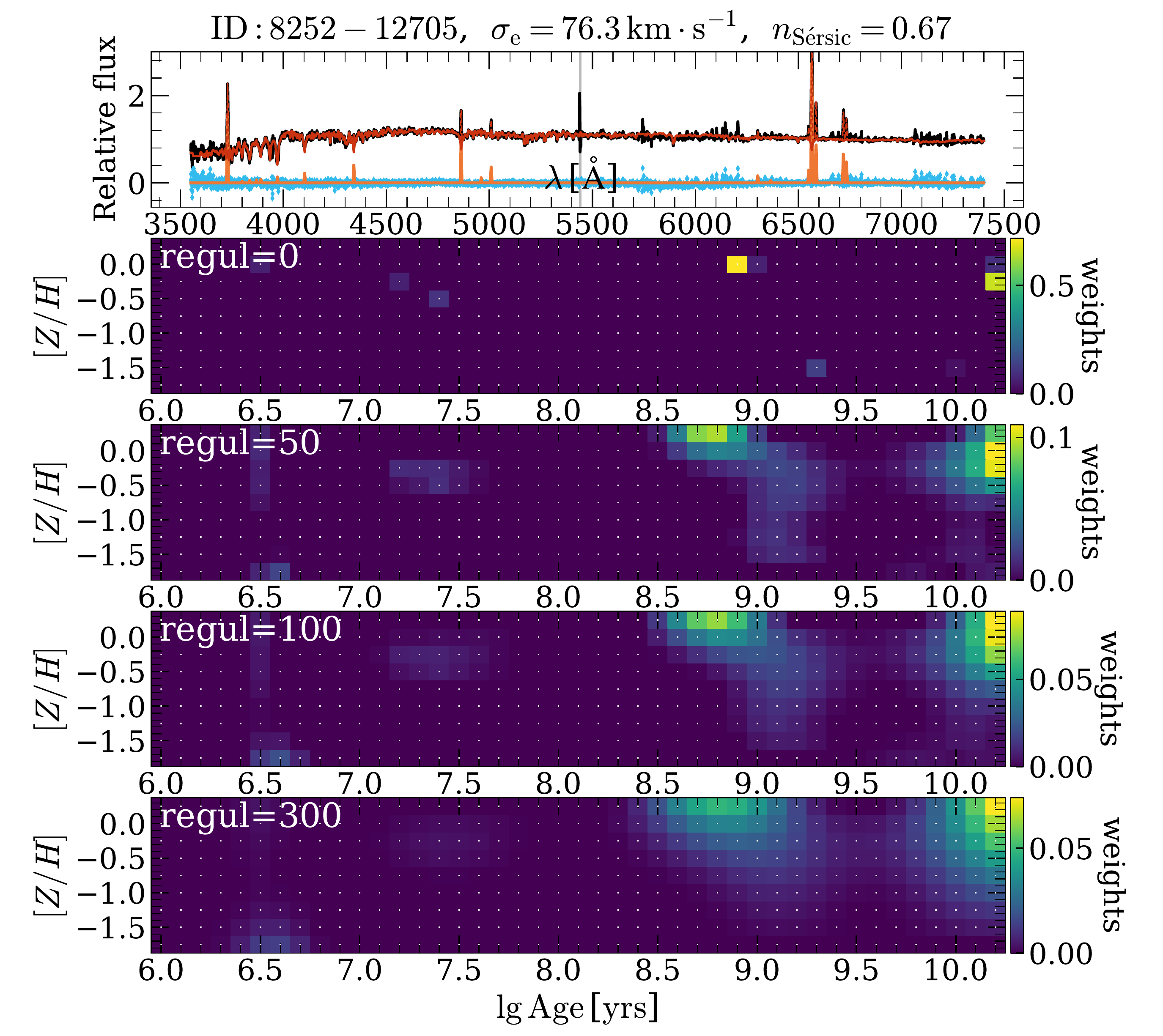}
\includegraphics[width=0.68\columnwidth]{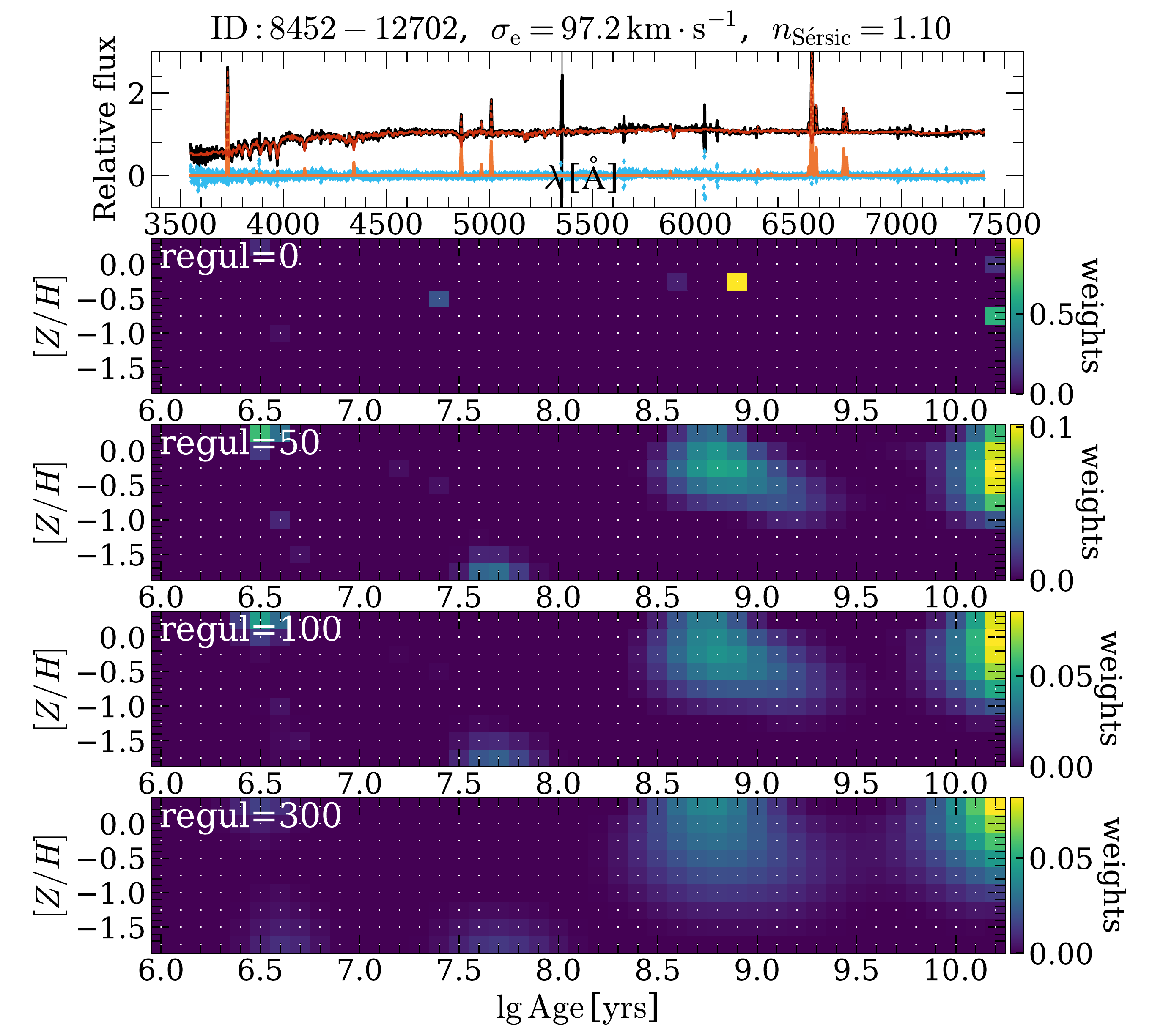}
\includegraphics[width=0.68\columnwidth]{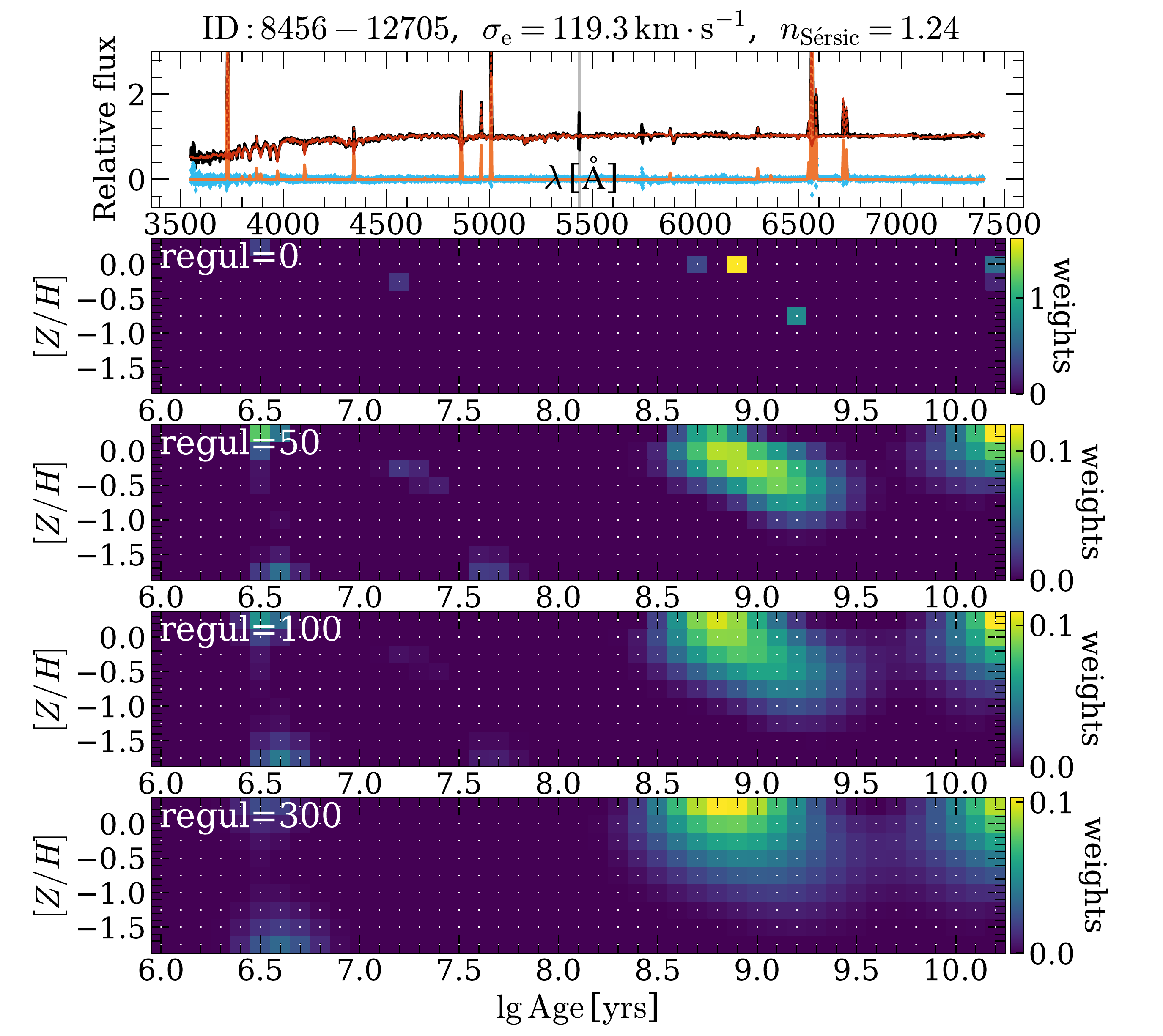}
\includegraphics[width=0.68\columnwidth]{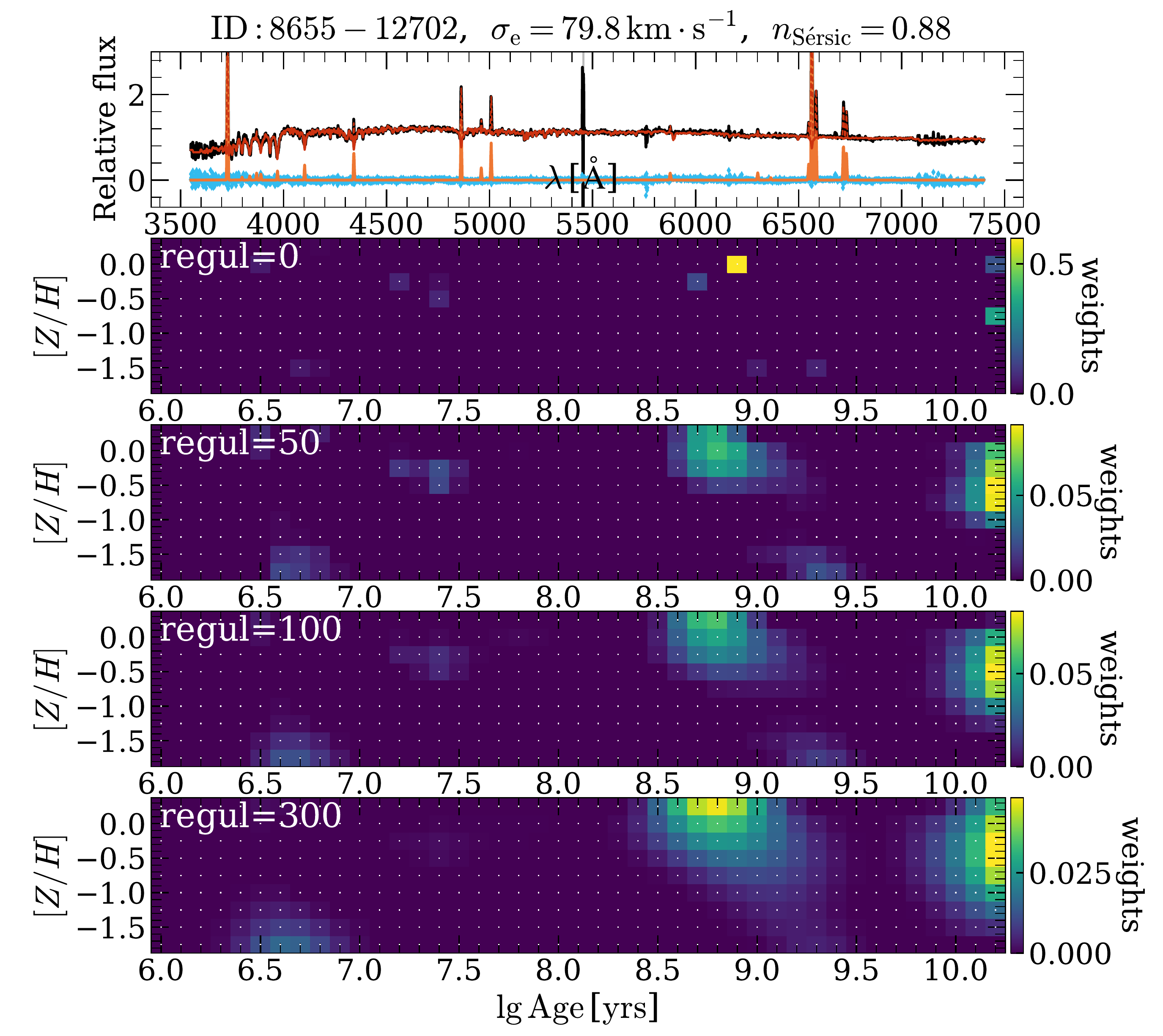}
\caption{Examples of {\sc ppxf} fitting results with different \texttt{regul} values. The top two rows are the galaxies whose star-formation history can be roughly approximated by a single star-formation burst. The bottom two rows show the galaxies with multiple star-formation bursts. For each galaxy, the fitted spectrum is the stacked spectrum within half-light isophote (see \autoref{sec:global_and_maps} for details). In each sub-figure, the observed and best-fitted spectra (without regularization, i.e. \texttt{regul}$=0$) are shown in the top panel with black and red curves, respectively. The cyan diamonds denote the fitting residuals and the orange curve represent the gas-only best-fitted spectrum (see \autoref{sec:sp} and \autoref{fig:example_spec} for more details of spectrum fitting). Template weight distributions with different \texttt{regul} values (\texttt{regul}$=0,\,50,\,100,\,300$) are shown in the panels from top to bottom. All these fittings are based on the {\sc fsps} SSP model (see \autoref{sec:libraries}). Galaxy ID, velocity dispersion, and $\rm S\acute{e}rsic$ index are shown in the top of each sub-figure. The $\rm S\acute{e}rsic$ index used here is from the NASA Sloan Atlas (NSA) catalogue~\citep{Blanton_et_al.(2011)}.}
\label{fig:reguls}
\end{figure*}

\begin{figure*}
\centering
\includegraphics[width=2\columnwidth]{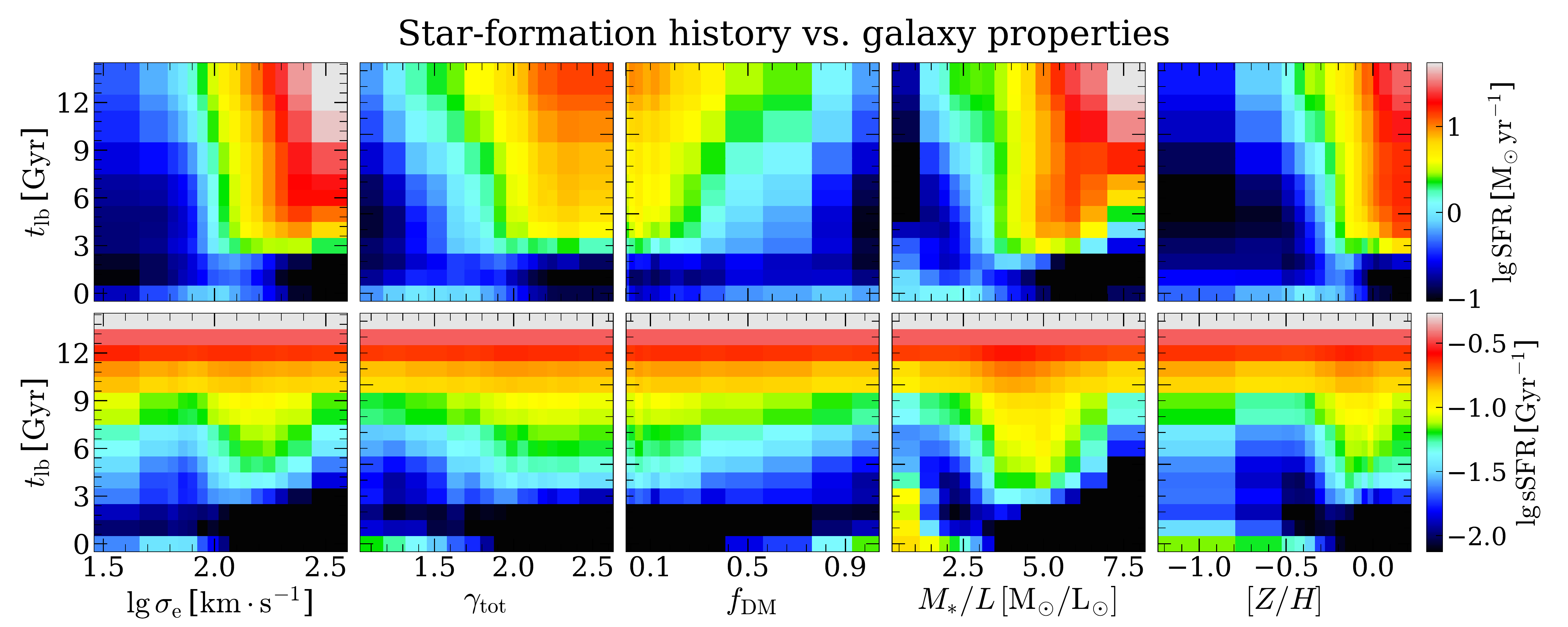}
\caption{Star-formation history (SFH; represented by star-formation rate in top panels and specific star-formation rate in bottom panels as a function of lookback time, $t_{\rm lb}$) versus galaxy properties: velocity dispersion within $R_{\rm e}$ ($\lg\,\sigma_{\rm e}$), total density slope ($\gamma_{\rm tot}$), dark matter fraction within $R_{\rm e}$ ($f_{\rm DM}$), stellar mass-to-light ratio from SPS non-regularized fitting ($M_{\ast}/L$), and global metallicity ($[Z/H]$) from left to right. In each panel, galaxies are equally divided into 15 bins according to their properties (X-axis; about 400 galaxies in each bin). For galaxies in each bin, we rebin their logarithmically spaced weights from {\sc ppxf} ($\texttt{regul=100}$) onto a linear time axis with the time step being 1 Gyr. Star-formation rate (SFR) and specific star-formation rate (sSFR) are then calculated as the averaged values of all the galaxies in each bin at each time step.}
\label{fig:sfh2paras}
\end{figure*}

\begin{figure*}
\centering
\includegraphics[width=1.5\columnwidth]{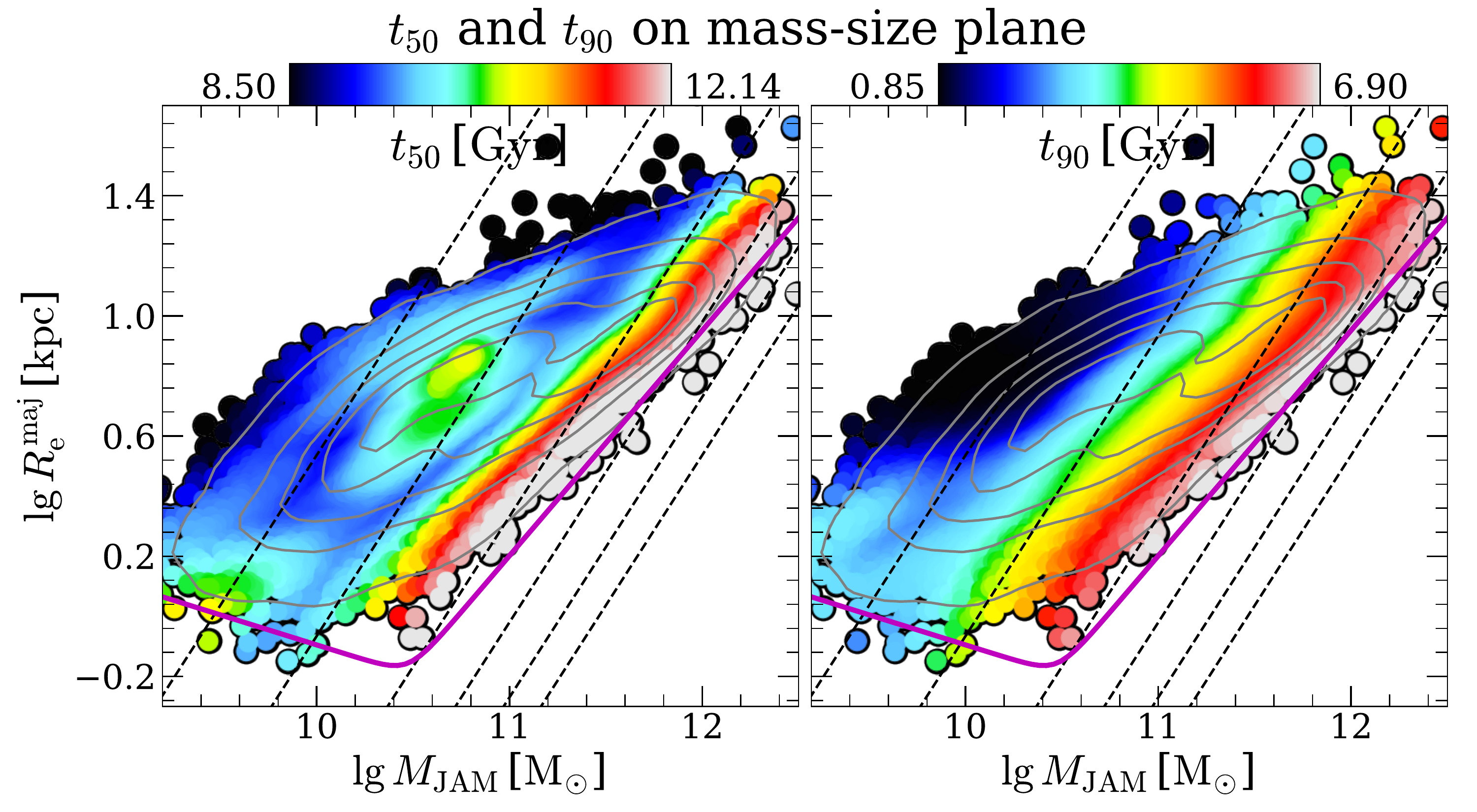}
\caption{$t_{50}$ and $t_{90}$ on the mass-size plane ($\lg\,R_{\rm e}^{\rm maj}$ versus $\lg\,M_{\rm JAM}$, see \autoref{sec:jam_props} for definitions of the two parameters), where $t_{50}$ and $t_{90}$ are the time when galaxies reach 50\% and 90\% of their present-day stellar mass, respectively. The symbols are the same as \autoref{fig:sp_mass_size}.}
\label{fig:t5090}
\end{figure*}

Star-formation history (SFH) of galaxies is one of key things to analyze in studying the evolution of galaxies. In practice, one can obtain SFH of galaxies from their spectra using either parametric methods (by assuming the specific formula of SFH; e.g. \citealt{McLure_et_al.(2018),Carnall_et_al.(2018),Zhou_et_al.(2020),Zhou_et_al.(2021)}; see \citealt{Carnall_et_al.(2019)} for the comparison between different parameterizations), or non-parametric methods (by calculating the star-formation rate from SSPs in each time bin; e.g. \citealt{Cid_Fernandes_et_al.(2005),McDermid_et_al.(2015),Leja_et_al.(2019),Cappellari2017,Cappellari(2023)}). In this work, we adopted the non-parametric method to study the star-formation history of MaNGA galaxies, using the {\sc ppxf} software. {\sc ppxf} combines the linear regularization method (e.g. \citealt{Press_et_al.(2007)}), which enables the users to obtain a smoother star-formation history while keep the best-fitted spectrum nearly unchanged within the given uncertainty (see section 3.5 of \citealt{Cappellari2017} for details of regularization).

Before applying the regularization to the whole sample, we first investigate the effect of different regularization levels on the star-formation history with a random sub-sample of MaNGA by setting the \texttt{regul} keyword to be 0, 50, 100, and 300 (\texttt{regul}$=0$ represents the non-regularized fitting which is used to derive stellar population properties and their gradients in the previous sections). In \autoref{fig:reguls}, we present several examples of {\sc ppxf} fitting with different \texttt{regul} values. We note here that all the analyses of star-formation history in this work are based on the stacked spectrum within half-light isophotes of galaxies. As can be seen, with increasing \texttt{regul} value, the distribution of SSP template weight becomes more smoothed, while the fittings roughly maintain the star-formation epochs of the galaxies, although short star-forming episodes may be hidden. We confirm that fitting with different \texttt{regul} values provides nearly the same fitting goodness of the spectrum. In this work, we apply \texttt{regul}$=100$ on the complete MaNGA sample to derive their star-formation histories and confirm that the choice of \texttt{regul} will not qualitatively change our results presented blow. We note here that due to the lack of near ultraviolet (NUV) and infrared spectrum in MaNGA survey, the star-formation history of individual galaxies may not be able to be well constrained. In particular, we are not very sensitive to small contributions of very recent star formation, which produce a black body peak in the UV due to the massive and hot stars and another black body distribution in the IR, due to the photons re-radiated by dust. However, by combining the large sample of MaNGA, we are still able to spot clues of the correlation between SFH and galaxy properties. 

In \autoref{fig:sfh2paras}, we present the correlation between SFHs and galaxy properties (see the caption of the figure). Here, we take the evolution of star-formation rate (SFR) and specific star-formation rate (sSFR) with cosmic time as the star-formation history of the galaxies. To calculate SFR at a given time, we first add up the initial stellar masses (set to be $1\mathrm{M_{\odot}}$ for each template) of all the templates with the same age (see \autoref{sec:libraries} and \autoref{fig:libs} for age grids of the templates used in this work) and then rebin the logarithmically spaced SFR on to a linear time axis (the time step is set to be 1Gyr), keeping the total newly-formed stellar mass in each time step unchanged, similar to the practice of \citet{McDermid_et_al.(2015)}. sSFR is calculated as the instantaneous SFR divided by the existing stellar mass at that time. In each panel of \autoref{fig:sfh2paras}, galaxies are first divided into 15 bins according to their properties (see the X-axes in \autoref{fig:sfh2paras}), resulting in $\sim 400$ galaxies in each bin. For each bin, we calculate the averaged SFR and sSFR at each time step. Below, we summarize the main conclusions shown in \autoref{fig:sfh2paras}.

As show in the top panels of \autoref{fig:sfh2paras}, SFR evolution tracks vary with all the 5 parameters. Specifically,
\begin{enumerate}
    \item Galaxies with high $\lg\,\sigma_{\rm e}$ ($\gtrsim 2.3$) are found to have high star-formation rate at high redshift and show obvious decreasing trend towards lower redshift. Below $t_{\rm lb}\lesssim 2.4\,\mathrm{Gyr}$, these galaxies become quenched and never form stars again, which presents a clear and sharp quenching boundary. The sharp boundary from star formation to quenching in this work is well consistent with the finding of \citet{Cappellari(2023)}, where a clear quenching boundary at $\lg\sigma_{\rm e}\sim 2.3$ is also seen (see fig.~8 of \citealt{Cappellari(2023)}). However, \citet{Cappellari(2023)} also find a decreasing trend of the time when star-formation of galaxies becomes the most active with the decreasing $\sigma_{\rm e}$ (i.e. galaxies with lower $\sigma_{\rm e}$ appear to show star-formation peak more recently), which is not seen in our results. Besides, in \citet{Cappellari(2023)}, galaxies with $\lg\,\sigma_{\rm e}\gtrsim 2.3$ are found to get quenched at least 6.8 Gyr ago (the corresponding lookback time of the median redshift of LEGA-C galaxies, i.e. $z\sim 0.76$), significantly earlier than $2.4\,\mathrm{Gyr}$ in this work. To the contrary, galaxies with low $\lg\,\sigma_{\rm e}$ ($\lesssim 2.0$) never have star-formation rate as high as the galaxies with high $\sigma_{\rm e}$, while maintain the active star-formation across their whole history, consistent with \citet{Cappellari(2023)}. The results above are also consistent with the findings of previous studies (e.g. \citealt{Heavens_et_al.(2004), Thomas2005,Panter_et_al.(2007), McDermid_et_al.(2015),Guglielmo_et_al.(2015)}), where massive galaxies are found to have decreasing star-formation activities with cosmic time and become quenched earlier (resulting in the earlier mass assembly), while the low-mass galaxies keep the star-forming status across the whole time range.

    \item A similar sharp quenching boundary is also seen in the correlation between SFR evolution and the galaxy total density slope ($\gamma_{\rm tot}$; see \autoref{sec:jam_props} for definition). Galaxies with the steepest total mass profile ($\gamma_{\rm tot}\gtrsim 2.1$) show similar trend as the galaxies with $\lg\,\sigma_{\rm e}\gtrsim 2.3$, where they are also seen to have high star-formation rate at high redshifts, while becomes quenched below $t_{\rm lb}\lesssim 2.4\,\mathrm{Gyr}$. Galaxies with flatter total density slope have continuous SFR until the present. These results indicate that galaxies with higher $\sigma_{\rm e}$ typically have steeper total density profiles, consistent with previous studies (e.g. \citealt[fig.~22c]{Cappellari(2016)}, \citealt{Li_et_al.(2019)}, and \citetalias{Zhu_et_al.(2023b)} of this series).

    \item Galaxies with low dark matter fraction ($f_{\rm DM}$; see \autoref{sec:jam_props} for definition) appear to have similar SFR evolution trend as galaxies with high $\sigma_{\rm e}$ or steep total density slope (i.e. high $\gamma_{\rm tot}$), where they have high star-formation rate at the early universe and become quenched around 2.4 Gyr ago. However, $f_{\rm DM}$ does not show clear quenching boundary as $\sigma_{\rm e}$ and $\gamma_{\rm tot}$ do, indicating that $f_{\rm DM}$ is not tightly correlated with the other two parameters (i.e. $\sigma_{\rm e}$ and $\gamma_{\rm tot}$; this will be studied in \citetalias{Zhu_et_al.(2023b)} of this series).

    \item Stellar mass-to-light ratio ($M_{\ast}/L$) and metallicity of galaxies show similar correlation of SFR evolution history with $\sigma_{\rm e}$, because they tightly correlate with $\sigma_{\rm e}$ (see \autoref{fig:sp_mass_size}; also see \citealt{McDermid_et_al.(2015),Cappellari(2016),Scott_et_al.(2017),Li_et_al.(2018)}). A clear quenching boundary is seen at $M_{\ast}/L\gtrsim 5$ and $[Z/H]\gtrsim 0$, respectively.
\end{enumerate}
We note here that our result (i.e. star-formation rate decreases with cosmic time) also disagrees with the findings of \citet{Madau_et_al.(2014)}, where the cosmic star-formation density peaks at $z\sim 2$ (see fig.~9 therein). It may be due to the lack of low mass galaxies in our sample, which contributes to the star-formation activities at lower redshifts. Also, the lack of NIR/UV constraints in our study may be another cause of the discrepancy. We will combine the NIR/UV constraints and the optical spectra to study the star-formation histories of MaNGA galaxies in a forthcoming paper.

The specific star-formation rate (sSFR) evolution tracks, however, show different correlation with galaxy properties, compared to SFR evolutions. We note here that when calculating SFR and sSFR, we assume the star-formation rate between two neighboring age grids (see \autoref{fig:libs} for age grid design of the {\sc fsps} model) to be constant. Thus, at early times (where the time step between two neighboring age grids is significantly larger than 1 Gyr), the ratio between the instantaneous SFR and the existing stellar mass (i.e. the instantaneous sSFR) is also a constant for all galaxies (see the bottom panels of \autoref{fig:sfh2paras}). This feature is also seen in \citet[][fig.~16]{McDermid_et_al.(2015)}, where galaxies in different mass bins have the same sSFR at early times. Despite the calculation issue here, we are also able to see some interesting features of the correlation between galaxy sSFR evolution trend and galaxy properties. Galaxies show clear quenching boundary for all the 5 parameters (including $f_{\rm DM}$, which does not show clear quenching boundary in its correlation with SFR evolution tracks). However, the quenching regions are not so regular as those for SFR. Specifically,  
\begin{enumerate}
    \item Galaxies with $\lg\,\sigma_{\rm e}\gtrsim 2.1$ (unlike $\lg\,\sigma_{\rm e}\gtrsim 2.3$ for SFR) appear to be quenched at the present day, while galaxies with higher $\sigma_{\rm e}$ are found to quench their star-formation slightly earlier (3.6 Gyr ago for galaxies with $\lg\,\sigma_{\rm e}\sim 2.5$, relative to 2.4 Gyr ago for galaxies with $2.1\lesssim \lg\,\sigma_{\rm e}\lesssim 2.5$). Interestingly, galaxies with low $\sigma_{\rm e}$ appear to have double peaks of sSFR, both at early times and the present.

    \item Galaxies with $\gamma_{\rm tot}\gtrsim 2.0$ and $f_{\rm DM}\lesssim 0.5$ become quenched about 2.4 Gyr ago, while galaxies with flat total density slope and high dark matter fraction appear to have double-peak sSFR evolution tracks.

    \item A clear quenching boundary is also seen at $M_{\ast}/L\sim 4$ and galaxies with high stellar mass-to-light ratio appear to be quenched earlier (6 Gyr ago for galaxies with $M_{\ast}/L\sim 7.5$, relative to 2.4 Gyr ago for those with $M_{\ast}/L\sim 5$). A double-peak sSFR evolution track for galaxies with low $M_{\ast}/L$ is also seen. 

    \item Unlike $\sigma_{\rm e}$ and $M_{\ast}/L$, metallicity of galaxies has a more regular quenching region, where galaxies with $[Z/H]\gtrsim -0.2$ are quenched 2.4 Gyr ago.
\end{enumerate}

With SFR evolution tracks of all the galaxies, we are also able to predict the mass assembly histories of the galaxies. In \autoref{fig:t5090}, we present the distribution of $t_{50}$ and $t_{90}$ on the mass-size plane, where $t_{50}$ and $t_{90}$ are the lookback times when galaxies reach $50\%$ and $90\%$ of their present-day stellar mass. As can be seen, both $t_{50}$ and $t_{90}$ show systematic variation on the mass-size plane, similar to global stellar population properties (see \autoref{fig:sp_mass_size}): galaxies with large $\sigma_{\rm e}$ tend to have larger $t_{50}$ and $t_{90}$, indicating earlier mass assembly of these galaxies, consistent with previous studies (e.g. \citealt{McDermid_et_al.(2015),Lu_et_al.(2022b),Zhou_et_al.(2021)}). Besides, the constant-$t_{50}$ (or constant-$t_{90}$) lines are also not strictly parallel with constant-$\sigma_{\rm e}$ lines, indicating that the star-formation history properties of galaxies are also not fully accounted for by $\sigma_{\rm e}$, similar to global stellar population properties (see \autoref{fig:sp_mass_size}).

\section{Conclusion}

\label{sec:conclusion}
This work is the second paper of our MaNGA DynPop (Dynamics and stellar Population) project, which aims to analyze the stellar population properties and star-formation history of the final MaNGA sample (SDSS DR17; \citealt{MaNGA_dr17}). In this paper, we make use of the stellar population synthesis (SPS) method to study the stellar population of the galaxies and use the non-parametric way to analyze their star-formation history. All these are done using the Penalized Pixel-Fitting ({\sc ppxf}; \citealt{Cappellari_et_al.(2004),Cappellari2017,Cappellari(2023)}) software with the {\sc fsps} stellar model. We summarize the main contents of this paper below:

\begin{enumerate}
    \item We provide a catalogue which contains the global stellar population properties, the stellar population radial profiles, gradients, the spatially resolved stellar population maps, as well as the star-formation histories of $\sim 10000$ MaNGA galaxies. The catalogue is available from the website of MaNGA DynPop (\url{https://manga-dynpop.github.io}). The explanations of the quantities of the catalogue are presented in \autoref{table:catalog}.

    \item With the derived global stellar population properties (i.e. age, metallicity, and stellar mass-to-light ratio), we revisit the scaling relation between stellar population properties and velocity dispersion of galaxies (\autoref{fig:sp2sigma}), the distribution of total mass density slope on the $(\sigma_{\rm e},\mathrm{Age})$ and $(\sigma_{\rm e},[Z/H])$ planes (\autoref{fig:gammatot}), the distribution of stellar population on the mass-size plane (\autoref{fig:sp_mass_size}) and on the $(\epsilon,\lambda_{R_{\rm e}})$ plane (\autoref{fig:sp_lambda_epsilon}). We confirm that the positive correlation between global stellar population properties and velocity dispersion, the systematic variation of stellar population properties on the mass-size plane, as well as the variation of age on the $(\epsilon,\lambda_{R_{\rm e}})$ plane in previous studies (e.g. \citealt{Cappellari_et_al.(2013b),McDermid_et_al.(2015),Cappellari(2016),Scott_et_al.(2017),Li_et_al.(2018),van_de_Sande_et_al.(2018)}) are all well recovered and also find some more interesting new results:
    \begin{itemize}
    \item Variation of galaxy age and metallicity is not solely driven by velocity dispersion of galaxies, as we can see clear variation trend of age (metallicity) with metallicity (age) at fixed $\sigma_{\rm e}$ (\autoref{fig:sp2sigma}). Total mass density slope show similar distributions on the $(\sigma_{\rm e},\mathrm{Age})$ and $(\sigma_{\rm e},[Z/H])$ planes as stellar population properties, indicating its tight correlation with stellar populations (\autoref{fig:gammatot}).
    
    \item Metallicity and stellar mass-to-light ratio are seen to have similar systematic variations on the $(\epsilon,\lambda_{R_{\rm e}})$ plane as galaxy age, which has not been reported before, with the less rotational galaxies (i.e. lower $\lambda_{R_{\rm e}}$) being older, more metal-rich, and having higher stellar mass-to-light ratio at fixed $\epsilon$. The most edge-on galaxies have the youngest age, the lowest metallicity, and the lowest stellar mass-to-light ratio for the reason that more disk components (which are young, metal-poor, and have low stellar mass-to-light ratio) are observed for edge-on galaxies (\autoref{fig:sp_lambda_epsilon}).
    \end{itemize}

    \item We also study the stellar population profiles of galaxies with different mass. We find that the galaxies in the lowest mass bin tend to have flatter population profiles, while the ones in the highest mass bin have older center with high $M_{\ast}/L$ and significantly young outskirts, with low $M_{\ast}/L$ (\autoref{fig:sp_profile}). This is because the highest mass bin contains not only the quenched elliptical galaxies, but also the massive disk galaxies with large quenched bulge at the center and star-forming disk at the outer part. This is consistent with the findings in \autoref{fig:slope_mass_size} and \autoref{fig:slope_age2sigma}, where the galaxies with the steepest stellar population profiles (i.e. the most negative population gradients) appear to be in the massive disk region (\autoref{fig:slope_mass_size}, see also fig.~23 of \citealt{Cappellari(2016)}) or in the green valley (\autoref{fig:slope_age2sigma}).

    \item Systematic variations of stellar population slopes are also observed on the $(\epsilon,\lambda_{R_{\rm e}})$ plane. Galaxies that are rotational and have relatively low $\epsilon$ ($\lesssim 0.6$) have the lowest gradients of age and stellar mass-to-light ratio (i.e. the steepest decreasing age and stellar mass-to-light ratio profiles from center to the outer part of galaxies). Galaxies with the flattest stellar population profiles locate in the region of largest $\epsilon$ (i.e. the most edge-on view of galaxies), because the more disk component observed from the edge-on view of galaxies contribute more younger population to the galaxy center, flattening the profiles (\autoref{fig:slope_lambda_epsilon}). Interestingly, metallicity gradient shows remarkably similar distribution as mass and luminosity of galaxies on the $(\epsilon,\lambda_{R_{\rm e}})$ plane, confirming the mass dependence of metallicity gradients. (\autoref{fig:slope_lambda_epsilon} and \autoref{fig:mass_light_lambda_epsilon}).

    \item Galaxies with different properties are found to have different star-formation histories. The galaxies with high, steep total density slope, low dark matter fraction, high stellar mass-to-light ratio, and high metallicity appear to have significantly higher star-formation rate (SFR) at early times, which decreases with cosmic time obviously, compared to other galaxies. Clear and regular quenching boundary are seen for the correlation between star-formation rate evolution and galaxy properties (galaxies with $\lg\,\sigma_{\rm e}\gtrsim 2.3$, $\gamma_{\rm tot}\gtrsim 2.1$, $M_{\ast}/L\gtrsim 5$, or $[Z/H]\gtrsim 0$ are found to be quenched $\sim 2.4\rm Gyr$ ago). For specific star-formation rate (sSFR) evolution, however, the quenched regions are still seen but appear to be more complicated than that of SFR (\autoref{fig:sfh2paras}). The lookback times when galaxies reach 50\%  and 90\% of their present-day stellar mass ($t_{50}$ and $t_{90}$) are seen to have systematic variations on the mass-size plane, similar as stellar population properties (\autoref{fig:t5090}).
\end{enumerate}

\section*{Acknowledgements}
We thank the referee for the insightful comments which improved the paper. We also acknowledge Drs. Junqiang Ge, Shuang Zhou, and Niu Li for helpful discussions on the stellar population synthesis and Paul Tol for making his colour blind friendly colour schemes publicly available. This work is partly supported by the National Key Research and Development Program of China (No. 2018YFA0404501 to SM), by the National Natural Science Foundation of China (Grant No. 11821303, 11761131004, and 11761141012). This project is also partly supported by Tsinghua University Initiative Scientific Research Program ID 2019Z07L02017. We also acknowledge the science research grants from the China Manned Space Project with No. CMS-CSST-2021-A11. KZ and RL acknowledge the support of National Natural Science Foundation of China (Nos. 11988101, 11773032, 12022306), the support from the Ministry of Science and Technology of China (Nos. 2020SKA0110100),  the science research grants from the China Manned Space Project (Nos. CMS-CSST-2021-B01, CMS-CSST-2021-A01), CAS Project for Young Scientists in Basic Research (No. YSBR-062), and the support from K.C.Wong Education Foundation.

Funding for the Sloan Digital Sky Survey IV has been provided by the Alfred P. Sloan Foundation, the U.S. Department of Energy Office of Science, and the Participating Institutions. 

SDSS-IV acknowledges support and resources from the Center for High Performance Computing  at the University of Utah. The SDSS website is www.sdss.org.

SDSS-IV is managed by the Astrophysical Research Consortium for the Participating Institutions of the SDSS Collaboration including the Brazilian Participation Group, the Carnegie Institution for Science, Carnegie Mellon University, Center for Astrophysics | Harvard \& Smithsonian, the Chilean Participation Group, the French Participation Group, Instituto de Astrof\'isica de Canarias, The Johns Hopkins University, Kavli Institute for the Physics and Mathematics of the Universe (IPMU) University of Tokyo, the Korean Participation Group, Lawrence Berkeley National Laboratory, Leibniz Institut f\"ur Astrophysik Potsdam (AIP),  Max-Planck-Institut 
f\"ur Astronomie (MPIA Heidelberg), Max-Planck-Institut f\"ur Astrophysik (MPA Garching), Max-Planck-Institut f\"ur Extraterrestrische Physik (MPE), National Astronomical Observatories of China, New Mexico State University, New York University, University of Notre Dame, Observat\'ario Nacional / MCTI, The Ohio State University, Pennsylvania State University, Shanghai Astronomical Observatory, United Kingdom Participation Group, Universidad Nacional Aut\'onoma de M\'exico, University of Arizona, University of Colorado Boulder, University of Oxford, University of Portsmouth, University of Utah, University of Virginia, University of Washington, University of Wisconsin, Vanderbilt University, and Yale University.
%%%%%%%%%%%%%%%%%%%%%%%%%%%%%%%%%%%%%%%%%%%%%%%%%%
\section*{Data Availability}
All the stellar population properties and star-formation history properties used in this work are publicly available on the website of MaNGA DynPop (\url{https://manga-dynpop.github.io}). The structural and dynamical properties can be obtained from \citetalias{Zhu_et_al.(2023a)}, which are also publicly available from both the journal website and the website of MaNGA DynPop.

\bibliographystyle{mnras}
\bibliography{ref}

\newpage
\appendix
\section{Comparison with existing catalogues}
\label{apx:comparison}
In this section, we make comparisons with the existing stellar population catalogues of the full MaNGA sample, namely the {\sc pipe3d} catalogue \citep{pipe3d} and the {\sc firefly} catalogue \citep{firefly}. Both catalogues have provided a wealth of properties of stellar populations, emission lines, as well as the dust attenuation, which contribute to the legacy of the MaNGA project. 

\begin{figure*}
\centering
\includegraphics[width=1\columnwidth]{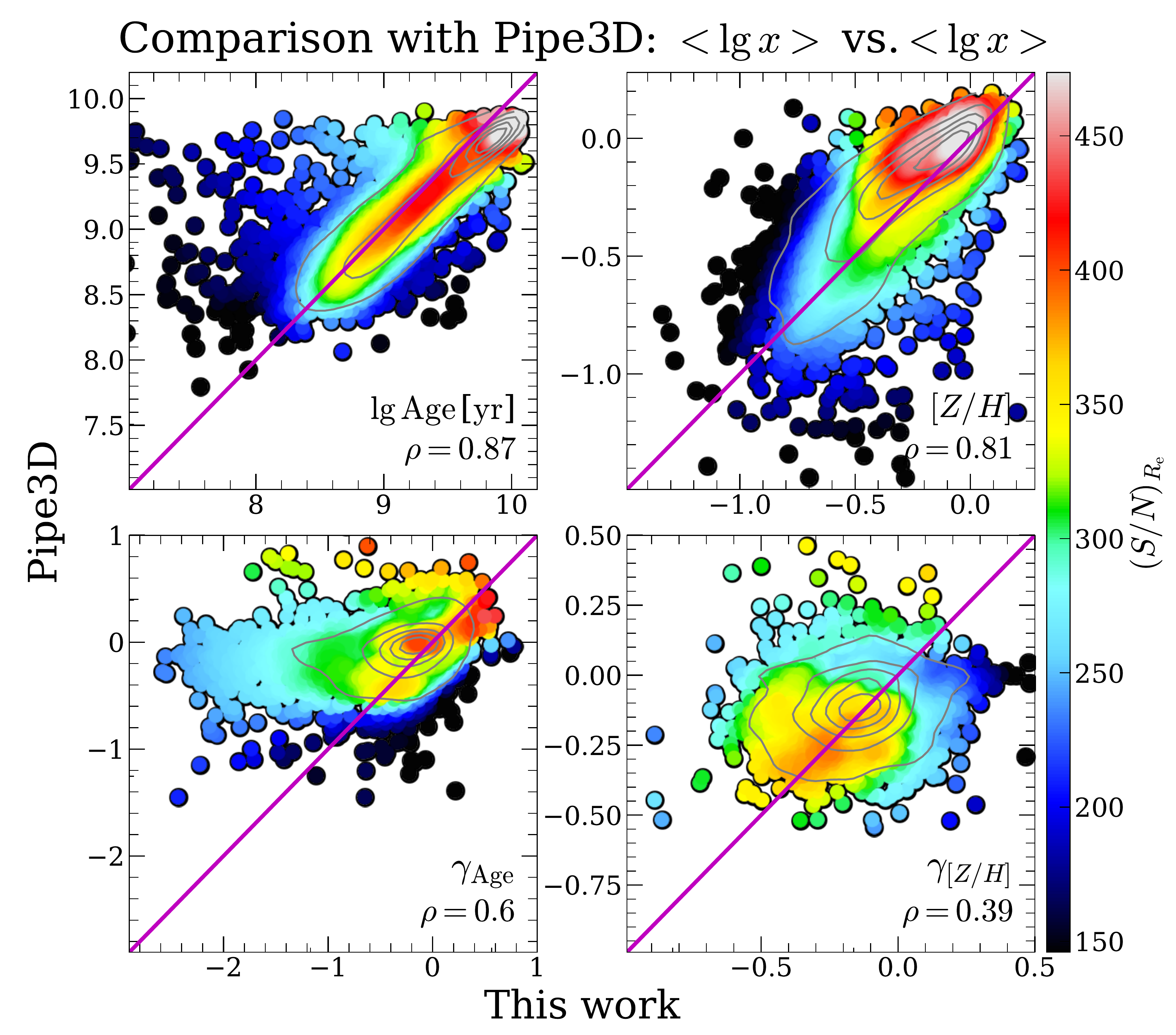}
\includegraphics[width=1\columnwidth]{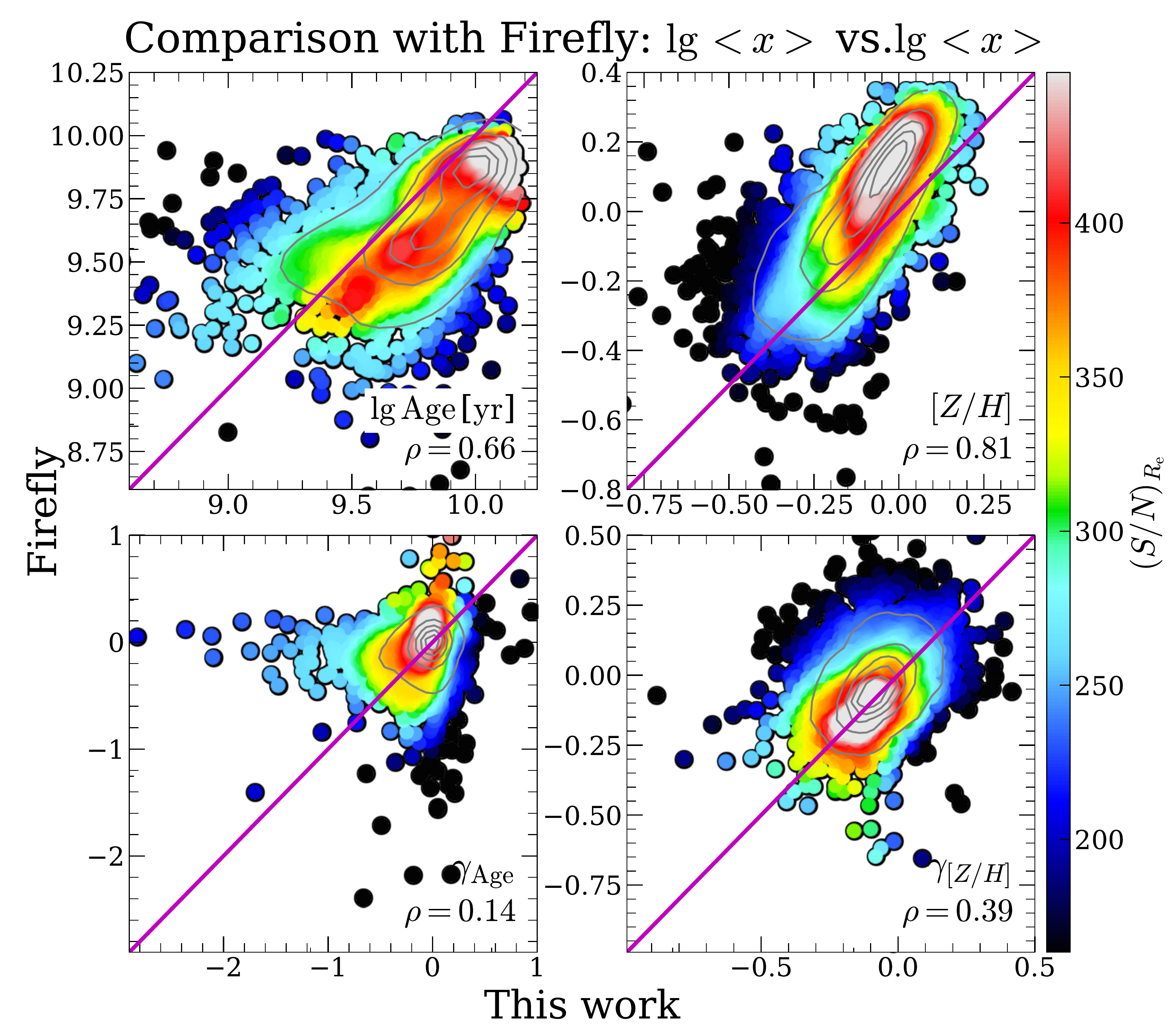}
\caption{Comparisons between the stellar population properties of this work and those from the {\sc pipe3d} (left; \citealt{pipe3d}) and the {\sc firefly} (right; \citealt{firefly}) catalogues. In both sub-figures, we present the comparisons of (1) luminosity-weighted age at effective radius (top left), (2) luminosity-weighted metallicity at effective radius (top right), (3) the slope of luminosity-weighted age profile (bottom left; calculated within $0.5-2R_{\rm e}$ for {\sc pipe3d} and $0-1.5R_{\rm e} $ for {\sc firefly}), and (4) the slope of luminosity-weighted metallicity profile (bottom right; calculated within $0.5-2R_{\rm e}$ for {\sc pipe3d} and $0-1.5R_{\rm e} $ for {\sc firefly}). For the comparison with {\sc pipe3d} catalogue (left), we adopt the geometric means (i.e. $<\lg\,x>$, where $x$ can be age and $(Z/H)/(Z_{\odot}/H_{\odot})$) when calculating the luminosity-weighted values (which is also the default setting of this work), while we adopt the arithmetic means (i.e. $\lg\,<x>$) when comparing with the {\sc firefly} catalogue (right) for consistency. In each panel, the colours indicate the signal-to-noise ratio of the stacked spectrum within the elliptical half-light isophotes and the magenta line indicates the $y=x$ correlation. The Spearman rank-order correlation coefficient, $\rho$, is shown in the bottom right of each panel. The galaxy number density is indicated by the grey contours.}
\label{fig:comparison}
\end{figure*}

In \autoref{fig:comparison}, we present the comparisons of (global) luminosity-weighted age and metallicity, as well as their radial gradients between our work and the two catalogues. We note that both the {\sc pipe3d} catalogue and the {\sc firefly} catalogue provide the stellar population properties within a ring at effective radius as the global properties of the galaxies, while in our work, the average values within the elliptical half-light isophotes are used. Thus, to make a more accurate comparison, we also calculate the stellar population properties {\em at effective radius} by simply averaging the properties within the radial range of $0.75-1R_{\rm e}$ and $1-1.25R_{\rm e}$ (already provided in the catalogue of this paper under ``Part 4: Stellar Population Radial Profiles''). When calculating the luminosity-weighted stellar populations, the {\sc pipe3d} catalogue adopted the geometric means (i.e. $<\lg\,x>$, where $x$ can be age and $(Z/H)/(Z_{\odot}/H_{\odot})$; same as this work), while the {\sc firefly} catalogue adopted the arithmetic means (i.e. $\lg\,<x>$). Thus, when making comparison with the {\sc pipe3d} catalogue, we keep our default setting (i.e. the geometric means) and when compared to the {\sc firefly} catalogue, we re-calculate the global stellar population ({\em at effective radius}) and the corresponding stellar population gradients with the same arithmetic means as {\sc firefly} for consistency. For the stellar population gradients, the two catalogues calculated them within different radial ranges ($0.5-2R_{\rm e}$ for the {\sc pipe3d} catalogue and $0-1.5R_{\rm e}$ for the {\sc firefly} catalogue). To keep the consistency, we calculate the population gradients in the same radial ranges when comparing with the two catalogues (see \autoref{sec:sp_gradients} and \autoref{fig:age_profile_calculation} for the method of gradient calculation).

In the left sub-figure of \autoref{fig:comparison}, we show the comparison between our catalogue and the {\sc pipe3d} catalogue. As can be seen, both global age and metallicity (measured {\em at effective radius}) in our work show good agreement with those from the {\sc pipe3d} catalogue, with the Spearman rank-order correlation coefficient, $\rho$, being 0.87 and 0.81, respectively. Specifically, our age is slightly ($\sim 0.1 \rm \,dex$) higher than {\sc pipe3d}, while our metallicity is slightly lower. Our age gradient also shows consistency with the {\sc pipe3d} catalogue ($\rho = 0.6$), while the galaxies with the most negative age gradient (the steepest decreasing age profiles from center to the outer part of the galaxies) in our catalogue show flatter age profiles (i.e. less negative age gradient) in the the {\sc pipe3d} catalogue. The metallicity gradient, however, show weaker agreement between {\sc pipe3d} ($\rho = 0.39$). Interestingly, we find that galaxies with higher signal-to-noise ratios show stronger agreement between the two catalogues than those with low signal-to-noise ratios. 

\begin{figure*}
\centering
\includegraphics[width=1.5\columnwidth]{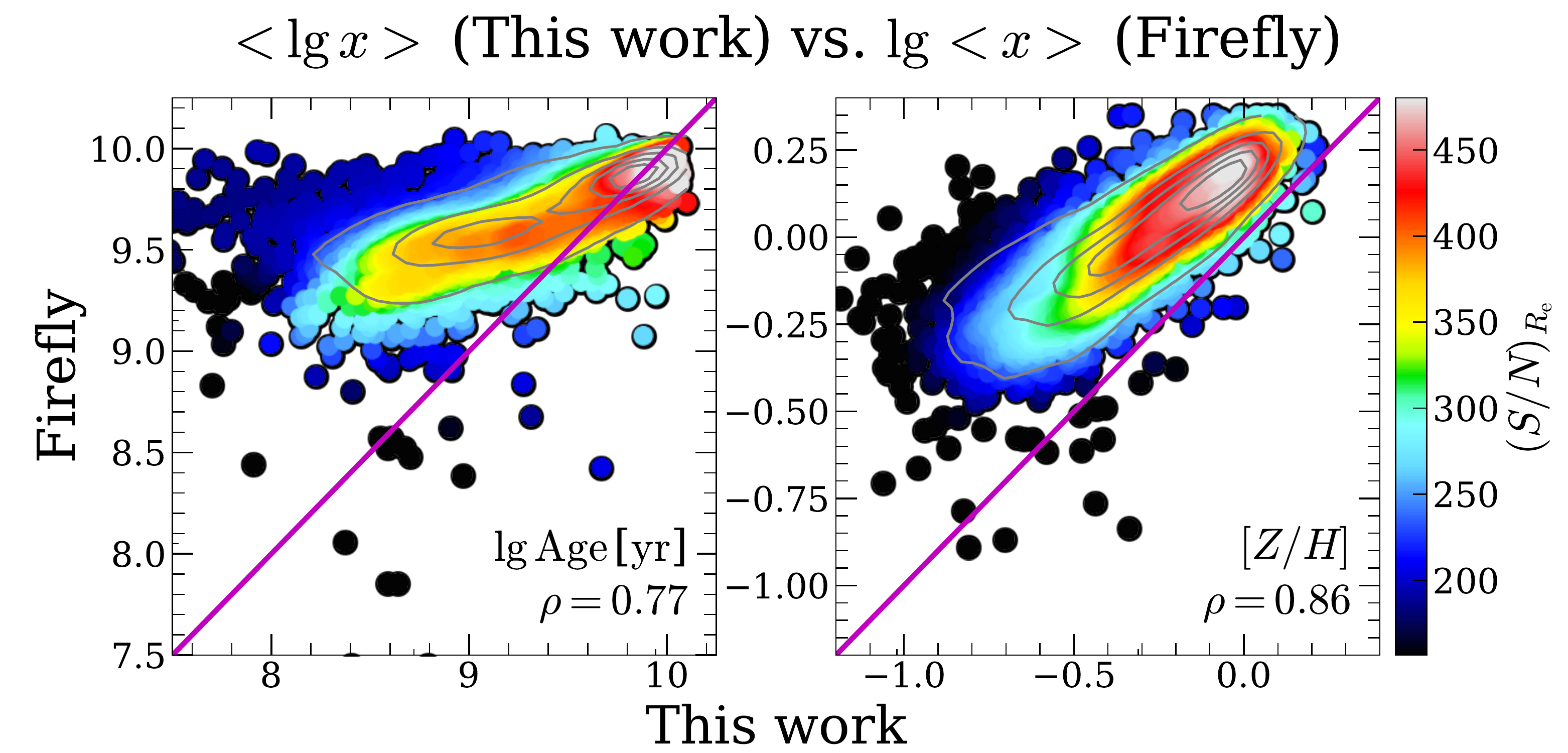}
\caption{Unfair comparison of global age and metallicity (both are measured {\em at effective radius}) between our results and {\sc firefly}. For our results, we adopt the geometric means (i.e. $<\lg\,x>$) to calculate the global stellar population properties, while {\sc firefly} adopted the arithmetic means (i.e. $\lg\,<x>$). The symbols are the same as \autoref{fig:comparison}.}
\label{fig:unfair_comparison}
\end{figure*}

For the comparison with the {\sc firefly} catalogue, which is shown in the right sub-figure of \autoref{fig:comparison}, we find that the agreement of global age and metallicity ({\em at effective radius}; arithmetic means used) between the two catalogues is still seen, with the Spearman rank-order correlation coefficients being 0.66 for age and 0.81 for metallicity. Global age of {\sc firefly} is systematically $\sim 0.3\,\rm dex$ lower than our result, while the global metallicity is systematically higher. Again, we can see that the galaxies which have the strongest disagreement between the two catalogues are typically those with low signal-to-noise ratio. \citet{Wilkinson_et_al.(2017)} pointed out the disagreement between the {\sc firefly} software and the {\sc starlight} software \citep{Cid_Fernandes_et_al.(2005)}, where the age of young galaxies in {\sc firefly} is systematically older than that from {\sc starlight} (see fig.~25 therein). \citet{firefly} also pointed out the similar trend when comparing with the {\sc pipe3d} catalogue (see fig.~13 therein). We emphasize here that these comparisons are not fair as the arithmetic means is used in {\sc firefly} (i.e. $\lg\,<x>$), while the geometric means is used in {\sc starlight} and {\sc pipe3d} (i.e. $<\lg\,x>$). In \autoref{fig:unfair_comparison}, we compare our results with geometric means (i.e. $<\lg\,x>$) to those from {\sc firefly} with arithmetic means (i.e. $\lg\,<x>$). We can see that we are also able to recover the similar trend as fig.~25 of \citet{Wilkinson_et_al.(2017)} and fig.~13 of \citet{firefly} with an independent software, {\sc ppxf}. However, when we adopt the same arithmetic means (see the right sub-figure of \autoref{fig:comparison}), the {\sc firefly} shows systematic lower age, rather than older age for young galaxies, as shown in \autoref{fig:unfair_comparison} (also see fig.~25 of \citealt{Wilkinson_et_al.(2017)} and fig.~13 of \citealt{firefly}). 

For the age and metallicity gradients, neither the age gradient, nor the metallicity gradient shows obvious agreement between our catalogue and the {\sc firefly} catalogue. This may be due to the fact that the {\sc firefly} catalogue is obtained from the ``VOR10'' DAP output of MaNGA, which is Voronoi binned \citep{Cappellari_et_al.(2003)} to $S/N\sim 10$, while we bin our data to $S/N\sim 30$. The low $S/N$ may cause large noise at outer region of galaxies in estimating stellar population properties, which cannot be fixed by averaging. We also note here that the three catalogues (i.e. the {\sc pipe3d}, the {\sc firefly}, and the catalogue provided in this work) adopt totally different stellar libraries, which may also induce differences: the {\sc pipe3d} catalogue adopted the {\sc mastar\_slog} library, which is based on the {\sc mastar} MaNGA stellar library \citep{Yan_et_al.(2019)}; the {\sc firefly} catalogue adopted the {\sc m11-miles} model templates from \citet{Maraston_et_al.(2011)}\footnote{{\sc firefly} also has a version based on the {\sc mastar} stellar library \citep{Yan_et_al.(2019)}.}; our results are based on the {\sc fsps} model (see \autoref{sec:libraries} for details). Considering the difference in the stellar libraries (and hence the boundaries of age and metallicity of the stellar templates), the fitting processes, the dust attenuation correction methods, and the definitions of effective radius, we actually do not expect all results to be consistent. We encourage the readers to make detailed comparisons when necessary.

\section{Data Explanation}
\setlength{\tabcolsep}{1mm}
\small
\vspace{10cm}{
\onecolumn
\begin{longtable}{cccp{12cm}}
\caption{Data explanations of the stellar population analysis output (based on {\sc ppxf} software with {\sc fsps} stellar model; see \autoref{sec:sp} for details). The full catalogue of all the stellar population properties and star-formation history parameters used in this paper can be obtained from the website of MaNGA DynPop (\url{https://manga-dynpop.github.io}). The order of galaxies in this catalogue corresponds to their order in the JAM catalogue of \citetalias{Zhu_et_al.(2023a)}.}
\label{table:catalog}\\
\hline \hline
Parameters & Dimensions & Units & Descriptions\\
\hline
\endfirsthead

\multicolumn{4}{c}{\autoref{table:catalog} -- continued}\\
\hline \hline
Parameters & Dimensions & Units & Descriptions\\
\hline
\endhead

\hline \hline \multicolumn{4}{c}{{Continued on next page}}
\endfoot

\hline \hline
\endlastfoot

\multicolumn{4}{c}{Part 1: General Galaxy Properties}\\
    \hline
    plateIFU & (10296,1) & & The plate ID+IFU design ID (e.g. 7443-12703; unique for each galaxy)\\
    mangaid & (10296,1) & & Unique MaNGA ID (e.g. 1-114145)\\
    obj\_ra & (10296,1) & degree & Right ascension of the science object in J2000\\
    obj\_dec & (10296,1) & degree & Declination of the science object in J2000\\
    ebvgal & (10296,1) & & E(B-V) value from SDSS dust routine for this IFU\\
    z & (10296,1) & & Redshift of the galaxy\\
    \hline
    \multicolumn{4}{c}{Part 2: Global Stellar Population Properties}\\
    \hline
    SNR\_Re  & (10296,1) & $ $ & The signal-to-noise ratio of the stacked spectrum within the elliptical half-light isophote, which is calculated as the ratio between the median values of flux and noise of the stacked spectra within the wavelength range from $4730\,\Angstrom$ to $4780\,\Angstrom$\\
    Mstar\_Re  & (10296,1) & $\lg(\mathrm{M_{\odot}})$ & Stellar mass enclosed within the elliptical half-light isophote, derived using {\sc ppxf} with a \citet{Salpeter1955} initial mass function\\
    Lr\_int\_Re  & (10296,1) & $\lg(\mathrm{L_{\odot}})$ & The {\em intrinsic} $r-$band luminosity within the elliptical half-light isophote, derived from the stacked intrinsic spectrum within the same aperture (see \autoref{sec:dust} for details)\\
    Lr\_obs\_Re  & (10296,1) & $\lg(\mathrm{L_{\odot}})$ & The {\em observed} $r-$band luminosity within the elliptical half-light isophote, derived from the stacked observed spectrum within the same aperture (see \autoref{sec:dust} for details)\\
    LW\_Age\_Re & (10296,1) & $\lg(\mathrm{yr})$ & Global $r-$band luminosity-weighted age, calculated by performing {\sc ppxf} fitting on the stacked spectrum within elliptical half-light isophote\\
    LW\_Metal\_Re & (10296,1) & & Global $r-$band luminosity-weighted $[Z/H]$\\
    MW\_Age\_Re & (10296,1) & $\lg(\mathrm{yr})$ & Global mass-weighted age\\
    MW\_Metal\_Re & (10296,1) & & Global mass-weighted $[Z/H]$\\
    ML\_int\_Re  & (10296,1) & $\lg(\mathrm{M_{\odot}/L_{\odot}})$ & Averaged {\em intrinsic} stellar mass-to-light ratio within the elliptical half-light isophote (calculated as the stellar mass  enclosed within the elliptical half-light isophote and the $r-$band luminosity derived from the intrinsic spectrum within the same aperture; see \autoref{sec:dust} and \autoref{fig:example_spec} for definition of the intrinsic spectrum)\\
    ML\_obs\_Re  & (10296,1) & $\lg(\mathrm{M_{\odot}/L_{\odot}})$ & Averaged {\em observed} stellar mass-to-light ratio within the elliptical half-light isophote (calculated as the stellar mass enclosed within the elliptical half-light isophote and the $r-$band luminosity derived from the observed spectrum within the same aperture)\\
    Av\_Re  & (10296,1) & $ $ & Best-fitted dust attenuation at $\lambda = 5500\,\Angstrom$ ($V$-band; see \citealt[][sec.~3.7]{Cappellari(2023)} for details) \\
    delta\_Re  & (10296,1) & $ $ & Best-fitted UV slope of the spectrum (see \citealt[][sec.~3.7]{Cappellari(2023)} for details) \\
    Fred\_tot\_Re  & (10296,1) & $ $ & $r-$band luminosity ratio between the observed spectrum and the intrinsic spectrum\\
    Fred\_gal\_Re  & (10296,1) & $ $ & $r-$band luminosity ratio between the observed spectrum (with the Milky Way dust attenuation corrected) and the intrinsic spectrum\\
    \hline
    \multicolumn{4}{c}{Part 3: Stellar Population Gradients}\\
    \hline
    LW\_Age\_Slope & (10296,1) & $\mathrm{dex}/R_{\rm e}$ & Gradient of $r-$band luminosity-weighted age within the elliptical half-light isophote (see \autoref{sec:sp_gradients} and \autoref{fig:age_profile_calculation} for details)\\
    LW\_Metal\_Slope & (10296,1) & $\mathrm{dex}/R_{\rm e}$ & Gradient of $r-$band luminosity-weighted $[Z/H]$ within the elliptical half-light isophote\\
    MW\_Age\_Slope & (10296,1) & $\mathrm{dex}/R_{\rm e}$ & Gradient of mass-weighted age within the elliptical half-light isophote\\
    MW\_Metal\_Slope & (10296,1) & $\mathrm{dex}/R_{\rm e}$ & Gradient of mass-weighted $[Z/H]$ within the elliptical half-light isophote\\
    ML\_int\_Slope & (10296,1) & $\mathrm{dex}/R_{\rm e}$ & Gradient of {\em intrinsic} $r-$band stellar mass-to-light ratio within the elliptical half-light isophote\\
    ML\_obs\_Slope & (10296,1) & $\mathrm{dex}/R_{\rm e}$ & Gradient of {\em observed} $r-$band stellar mass-to-light ratio within the elliptical half-light isophote\\
    \hline
    \multicolumn{4}{c}{Part 4: Stellar Population Radial Profiles}\\
    \hline
    LW\_Age\_Profile & (10296,8) & $\lg(\mathrm{yr})$ & Radial profile of $r-$band luminosity-weighted age from 0 to $2R_{\rm e}$ with the radial step being $0.25R_{\rm e}$ (i.e. eight radial bins for each galaxy; see \autoref{sec:sp_gradients} and \autoref{fig:age_profile_calculation} for details)\\
    LW\_Metal\_Profile & (10296,8) & & Radial profile of $r-$band luminosity-weighted $[Z/H]$\\
    MW\_Age\_Profile & (10296,8) & $\lg(\mathrm{yr})$ & Radial profile of mass-weighted age\\
    MW\_Metal\_Profile & (10296,8) & & Radial profile of mass-weighted $[Z/H]$\\
    ML\_int\_Profile & (10296,8) & $\lg(\mathrm{M_{\odot}/L_{\odot}})$ & Radial profile of {\em intrinsic} $r-$band stellar mass-to-light ratio\\
    ML\_obs\_Profile & (10296,8) & $\lg(\mathrm{M_{\odot}/L_{\odot}})$ & Radial profile of {\em observed} $r-$band stellar mass-to-light ratio\\
    \hline
    \multicolumn{4}{c}{Part 5: Stellar Population Maps}\\
    \hline
    BinID\_Map & (10296, $N$\footnote{$N$ is the spaxel number along X or Y-axis of this map.}, $N$) &  & IDs of Voronoi bins that the spaxels are associated with. Spaxels that have the same ID belong to the same Voronoi bin (set as -1 if a spaxel does not belong to any bins) and share the same stellar population properties (i.e. luminosity/mass-weighted age and metallicity, and stellar mass-to-light ratio)\\
    inRe\_Map & (10296, $N$, $N$) &  & 1 for spaxels within the elliptical half-light isophote and 0 for those outside the elliptical half-light isophote\\
    Mstar\_Map  & (10296, $N$, $N$) & $\lg(\mathrm{M_{\odot}})$ & Stellar mass maps\footnote{For a given Voronoi bin which consists of $N$ spaxels, the stellar mass of each spaxel is the same, given by $M_{\ast,\rm spx} = M_{\ast,\rm bin}/N$, where $M_{\ast,\rm bin}$ is the associated stellar mass of this Voronoi bin.}, derived using {\sc ppxf} with a \citet{Salpeter1955} initial mass function\\
    Lr\_int\_Map & (10296, $N$, $N$) & $\lg(\mathrm{L_{\odot}})$ & The {\em intrinsic} SDSS $r-$band luminosity maps\footnote{For a given Voronoi bin which consists of $N$ spaxels, the intrinsic $r-$band luminosity of each spaxel is the same, given by $L_{r,\rm spx}^{\rm int} = L_{r,\rm bin}^{\rm int}/N$, where $L_{r,\rm bin}^{\rm int}$ is the intrinsic $r-$band luminosity of this Voronoi bin, derived from the best-fitted intrinsic spectrum of this bin (see \autoref{sec:dust} for details).} \\
    Lr\_obs\_Map & (10296, $N$, $N$) & $\lg(\mathrm{L_{\odot}})$ & The {\em observed} SDSS $r-$band luminosity maps\footnote{Same as intrinsic luminosity maps, but for observed $r-$band luminosity.} \\
    LW\_Age\_Map & (10296, $N$, $N$) & $\lg(\mathrm{yr})$ & Spatially resolved $r-$band luminosity-weighted age maps\footnote{For a given Voronoi bin, the spaxels in this bin have the same age (and also metallicity, both luminosity- or mass-weighted), fitted from the binned spectrum.}\\
    LW\_Metal\_Map & (10296, $N$, $N$) &  & Spatially resolved $r-$band luminosity-weighted $[Z/H]$ maps\\
    MW\_Age\_Map & (10296, $N$, $N$) & $\lg(\mathrm{yr})$ & Spatially resolved stellar mass-weighted age maps\\
    MW\_Metal\_Map & (10296, $N$, $N$) &  & Spatially resolved stellar mass-weighted $[Z/H]$ maps\\
    ML\_int\_Map & (10296, $N$, $N$) & $\lg(\mathrm{M_{\odot}/L_{\odot}})$ & Spatially resolved {\em intrinsic} $r-$band stellar mass-to-light ratio maps\\
    ML\_obs\_Map & (10296, $N$, $N$) & $\lg(\mathrm{M_{\odot}/L_{\odot}})$ & Spatially resolved {\em observed} $r-$band stellar mass-to-light ratio maps\\
    Av\_Map  & (10296, $N$, $N$) & $ $ & Maps of best-fitted dust attenuation at $\lambda = 5500\,\Angstrom$ ($V$-band; see \citealt[][sec.~3.7]{Cappellari(2023)} for details) \\
    delta\_Map  & (10296, $N$, $N$) & $ $ & Maps of best-fitted UV slope of the spectrum (see \citealt[][sec.~3.7]{Cappellari(2023)} for details) \\
    Fred\_tot\_Map & (10296, $N$, $N$) & $ $ & Maps of $r-$band luminosity ratio between the observed spectrum and the intrinsic spectrum\\
    Fred\_gal\_Map  & (10296, $N$, $N$) & $ $ & Maps of $r-$band luminosity ratio between the observed spectrum (with the Milky Way dust attenuation corrected) and the intrinsic spectrum\\
    \hline
    \multicolumn{4}{c}{Part 6: Star-formation History}\\
    \hline
    T50 & (10296,1) & $\mathrm{Gyr}$ & The lookback time when galaxies reach 50\% of their present-day stellar mass\\
    T90 & (10296,1) & $\mathrm{Gyr}$ & The lookback time when galaxies reach 90\% of their present-day stellar mass\\
    SFR\_History  & (10296,15) & $\lg(\mathrm{M_{\odot}\,yr^{-1}})$ & Star-formation rate at different lookback time grids (from 0 to $14\,\rm Gyr$, with a linear time step being $1\rm \,Gyr$; see \autoref{sec:sfh} for details)\\
    sSFR\_History & (10296,15) & $\lg(\mathrm{Gyr^{-1}})$ & Specific star-formation rate at different lookback time grids (from 0 to $14\,\rm Gyr$, with a linear time step being $1\rm \,Gyr$; see \autoref{sec:sfh} for details)\\
    Mass\_Growth\_CDF & (10296,15) & $ $ & Cumulative distribution function of stellar mass growth\\
    
\end{longtable}
}

%%%%%%%%%%%%%%%%% APPENDICES %%%%%%%%%%%%%%%%%%%%%

%%%%%%%%%%%%%%%%%%%%%%%%%%%%%%%%%%%%%%%%%%%%%%%%%%

\label{lastpage}
\end{document}